\documentclass[preprint,3p,12pt]{elsarticle}

\usepackage{amssymb}
\usepackage{amsthm}

\usepackage{amsmath}
\usepackage{graphicx}
\usepackage{dcolumn}
\usepackage{bm}
\usepackage[utf8]{inputenc}
\usepackage[T1]{fontenc}
\usepackage{mathptmx}
\usepackage{xcolor}
\usepackage{hyperref}
\usepackage{subfig}
\usepackage{booktabs}
\usepackage[normalem]{ulem}

\newcommand{\er}{\mathbf{e}_R} 
\newcommand{\ez}{\mathbf{e}_Z}
\newcommand{\ep}{\mathbf{e}_\varphi}

\begin{document}

\begin{frontmatter}



\title{The GBS code for the self-consistent simulation of plasma turbulence and kinetic neutral dynamics in the tokamak boundary}

\author{M. Giacomin\corref{cor1}\fnref{aff1}}
 \ead{maurizio.giacomin@epfl.ch}
 \cortext[cor1]{Corresponding author}
\author{P. Ricci\fnref{aff1}}
\author{A. Coroado\fnref{aff1}}
\author{G. Fourestey\fnref{aff2}}
\author{D. Galassi\fnref{aff1}}
\author{E. Lanti\fnref{aff2}}
\author{D. Mancini\fnref{aff1,aff3}}
\author{N. Richart\fnref{aff2}}
\author{L. N. Stenger\fnref{aff1}}
\author{N. Varini\fnref{aff2}}

\affiliation[aff1]{organization={Ecole Polytechnique Fédérale de Lausanne (EPFL)},
            addressline={Swiss Plasma Center (SPC)}, 
            city={CH-1015 Lausanne},
            state={Switzerland}}
\affiliation[aff2]{organization={Ecole Polytechnique Fédérale de Lausanne (EPFL)},
addressline={SCITAS},
city={CH-1015 Lausanne}, 
state={Switzerland}}%
\affiliation[aff3]{organization={Università degli studi della Tuscia},
            addressline={Largo dell'Università s.n.c.}, 
            city={01100 Viterbo},
            state={Italy}}


\begin{abstract}
A new version of GBS (Ricci \emph{et al.} Plasma Phys. Control. Fusion 54, 124047, 2012; Halpern \emph{et al.} J. Comput. Phys. 315, 388-408, 2016; Paruta \emph{et al.} Phys. Plasmas 25, 112301, 2018) is described. 
GBS is a three-dimensional, flux-driven, global, two-fluid turbulence code developed for the self-consistent simulation of plasma turbulence and kinetic neutral dynamics in the tokamak boundary.
In the new version presented here, the simulation domain is extended to encompass the whole plasma volume, avoiding an artificial boundary with the core, hence retaining the core-edge-SOL interplay. 
A new toroidal coordinate system is introduced to increase the code flexibility, allowing for the simulation of arbitrary magnetic configurations (e.g. single-null, double-null and snowflake configurations), which can also be the result of the equilibrium reconstruction of an experimental discharge.  
The implementation of a new iterative solver for the Poisson and Ampère equations is presented, leading to a remarkable speed-up of the code with respect to the use of direct solvers, therefore allowing for efficient  electromagnetic simulations that avoid the use of the Boussinesq approximation. 
The self-consistent kinetic neutral model, initially developed for limited configurations, is ported to the magnetic configurations considered by the present version of GBS and carefully optimized. 
A new MPI parallelisation is implemented to evolve the plasma and neutral models in parallel, thus improving the code scalability. 
The numerical implementation of the plasma and neutral models is verified by means of the method of manufactured solutions.
As an example of the simulation capabilities of the new version of GBS, a simulation of a TCV tokamak discharge is presented.
\end{abstract}


\end{frontmatter}


\section{Introduction}

Understanding the physical phenomena taking place in the boundary of fusion plasmas is of fundamental importance for the operation and design of future devices, such as ITER and DEMO. In fact, the dynamics in the plasma boundary controls the overall performance of a fusion reactor, by setting  the boundary conditions for the core. 
In addition, by regulating the plasma-wall interaction, the boundary region controls the plasma refueling, heat exhaust and impurity
dynamics~\cite{loarte2007}.

The plasma dynamics in the boundary region is strongly nonlinear and characterized by the presence of phenomena occurring on a wide range of length and time scales, from the ion Larmor radius to the machine size, and from the gyro-motion to the time scales of a discharge. In contrast to the tokamak core, large-amplitude turbulent fluctuations do not allow decoupling the fluctuating and background  quantities. 
In addition, the plasma boundary, composed of the edge, where the magnetic field lines lie on closed and nested flux surfaces, and the scrape-off layer (SOL), where the magnetic field lines intersect the tokamak wall, is usually characterized by a complex magnetic field geometry that features one or more nulls of the poloidal magnetic field. 
The wide range of time and spatial scales, large amplitude fluctuations, and a complex magnetic geometry make the plasma boundary extremely challenging to model and simulate.

Despite recent and significant efforts to extend gyrokinetic models~\cite{qin2007,hahm2009,frei2019} and codes (e.g., XGC1~\cite{chang2017}, Gkeyll~\citep{Shi2017}, GENE~\cite{Pan2018}, ELMFIRE~\cite{chone2018}, and COGENT~\citep{dorf2020}) to simulate the SOL, fluid codes (e.g., BOUT++~\cite{Dudson2015}, GBS~\cite{Paruta2018}, GDB~\cite{Zhu2018}, GRILLIX~\cite{Stegmeir2018a}, HESEL~\cite{Nielsen2015}, and TOKAM3X~\cite{Tamain2016}) and gyro-fluid (FELTOR~\cite{wiesenberger2019}), which are computationally less demanding than gyrokinetic codes, remain the most common tools used to simulate turbulence in the tokamak boundary, taking advantage of the large collisionality of the plasma in this region. 

GBS is a first-principles, three-dimensional, flux-driven, global, turbulence code that evolves the Braginskii equations~\cite{braginskii1965} in the drift limit (see, e.g., Ref.~\cite{zeiler1997} for the derivation of the drift-reduced Braginskii equations). 
The GBS code was initially developed to study plasma turbulence in basic plasma devices. 
The initial version of GBS evolved the two-dimensional plasma dynamics in the plane perpendicular to the magnetic field~\cite{Ricci2008,ricci2009transport}, mainly studying ideal-interchange turbulence in simple magnetized plasma (SMT) configurations, such as TORPEX~\cite{fasoli2006}. 
Later, GBS was extended to include the direction parallel to the magnetic field, using a field-aligned coordinate system, and was used to perform global simulations in SMT configurations~\cite{ricci2010,ricci2011,li2009}  and linear devices~\cite{rogers2010}, such as LAPD~\cite{gekelman1991}. 
In 2012, a new version of GBS was developed to simulate plasma turbulence in the SOL of tokamak devices in limited magnetic configurations~\cite{Ricci2012}. An electrostatic model in the cold ion limit was considered. Moreover, the Boussinesq approximation was applied to simplify the numerical implementation of the divergence of the polarisation current. 

A second version of GBS was developed and presented in Ref.~\cite{halpern2016}. The physical model was improved by adding the ion temperature dynamics~\cite{mosetto2015} and electromagnetic effects~\cite{halpern2013ideal}.
The plasma model was coupled to a self-consistent kinetic neutral model, leading to the first plasma turbulence simulations of the SOL that self-consistently include the coupling to the neutral dynamics~\cite{wersal2015}.
The Boussinesq approximation was relaxed and the code parallelisation was substantially improved by means of a three-dimensional Cartesian communicator. 
The implementation of the plasma model was verified by using the method of manufactured solutions (MMS), described in Ref.~\cite{riva2014}. 

Finally, a non-field-aligned coordinate system was introduced in GBS to simulate complex magnetic geometries including one or more X-points and a third version of the code was reported in 2018, leading to the first GBS simulation of a diverted geometry~\cite{Paruta2018}.
The second-order numerical scheme was improved to a fourth-order finite difference scheme. 
Considering a simplified model with respect to the limited configuration, the diverted version of GBS was electrostatic, made use of the Boussinesq approximation, and did not include the neutral dynamics. 
Despite the use of a domain with a circular poloidal cross section that avoids the core region, limiting the choice of magnetic geometry, the version of GBS described in Ref.~\cite{Paruta2018} was used to investigate plasma turbulence in both single- and double-null magnetic configurations, shading light on the properties of blobs~\cite{paruta2019,beadle2020}.

The goal of the present paper is to describe in detail a new version of GBS that significantly improves the diverted version of GBS reported in Ref.~\cite{Paruta2018}.
First, while keeping the same fourth-order numerical scheme and non-field-aligned coordinate system, a rectangular poloidal cross section is implemented, which encompasses the whole plasma volume avoiding an artificial boundary with the core, hence retaining the core-edge-SOL turbulence interplay that is found to play a key role in determining the plasma dynamics of the tokamak boundary~\cite{Fichtmuller1998,Pradalier2017,grenfell2019}.
A new coordinate system is chosen adapted to the new geometry.
GBS simulations that leverage this new domain and coordinate system have been recently used to investigate the turbulent transport regimes of the tokamak boundary~\cite{giacomin2020} and to develop a theory-based scaling law of the pressure and density decay lengths in both the near and the far SOL~\cite{giacomin2021}.
Ultimately, the domain implemented in the present version allows for more flexibility on the choice of the magnetic configuration, which can also be loaded from an equilibrium reconstruction or a Grad-Shafranov solver. 
The possibility to simulate a magnetic configuration loaded from an equilibrium reconstruction has recently allowed for a direct comparison between GBS simulations and TCV experiments~\cite{oliveira2021}.
Moreover, leveraging the flexibility of the numerical scheme, the first turbulence simulations of snowflake magnetic configurations were performed~\cite{giacomin2020turbulence}.
Second, the physical model of Ref.~\cite{Paruta2018} is extended by adding electromagnetic effects, and the self-consistent kinetic neutral model, initially developed in the context of the limited version of GBS~\cite{halpern2016,wersal2015}, is ported to the present configuration, with an improved set of boundary conditions.
Third, both the plasma and neutral implementations are carefully refactorized and optimized.
In particular, the implementation of a new iterative solver for the Poisson and Ampère equations is introduced, leading to a remarkable speed-up of the code, therefore allowing for efficient electromagnetic simulations that avoid the use of the Boussinesq approximation. 
Fourth, the GBS parallelisation scheme is improved here by allowing the evolution in parallel of the plasma and neutral dynamics and leading to an improvement of the code scalability. 
In addition, the relatively simple numerical scheme used in GBS allows for an effective parallelisation of GBS through domain decomposition applied to all three coordinates and implemented with the Message Passing Interface (MPI).

In the present work, the numerical implementation of both the plasma and neutral models is also verified by means of the MMS for the first time.
In addition, taking advantage of the improvements of GBS brought to the present version, we present the first GBS electromagnetic simulation of a single-null TCV equilibrium, including the self-consistent evolution of the neutral dynamics.

The present paper is organized as follows. After the Introduction, Sec.~\ref{sec:physics} describes the physical model implemented in GBS. In Sec.~\ref{sec:numerics}, we summarize the numerical implementation and optimization of the plasma and neutral models. 
The code verification is reported in Sec.~\ref{sec:verification}, while parallelisation scalability tests are described in Sec.~\ref{sec:scal}. Convergence properties are presented in Sec.~\ref{sec:conv}. The first application of the new version of GBS described here is then presented in Sec.~\ref{sec:application}. The conclusions follow.

\section{Physical model}\label{sec:physics}

The aim of this section is to summarize the physical model implemented in GBS. In particular, we highlight the main differences with respect to the physical model that was evolved by the previous diverted version of GBS, described in Ref.~\cite{Paruta2018}.

\subsection{Drift-reduced Braginskii equations}

When the electron mean-free path is shorter than the parallel connection length, $\lambda_e \ll L_\parallel \sim 2\pi q R$,  and small deviations from a Maxwellian distribution function are expected, the use of the Braginskii fluid model~\cite{braginskii1965} to study the plasma dynamics is justified.
In addition, turbulence in the tokamak boundary occurs on time scales considerably longer than the ion cyclotron time scale, defined by the ion cyclotron frequency $\Omega_{ci}=e B/m_i$, and on spatial scales larger than the ion sound Larmor radius, $\rho_s=c_s/\Omega_{ci}$, with $c_s=\sqrt{T_e/m_i}$. This allows for the use of the drift-ordering approximation (see, e.g., Ref.~\cite{zeiler1997}) to simplify the Braginskii equations.

In the new version of GBS presented here, we extend the physical model implemented in Ref.~\cite{Paruta2018} by relaxing the Boussinesq approximation, adding electromagnetic effects, and coupling the Braginskii model with an improved version of the self-consistent kinetic neutral model initially developed in the limited version of GBS~\cite{halpern2016,wersal2015}.
Then, the plasma model equations take the following form:
\begin{align}
\label{eqn:density}
\frac{\partial n}{\partial t} =& -\frac{1}{B}[\phi,n]+\frac{2}{eB}\Bigl[C(p_e)-enC(\phi)\Bigr] 
-\nabla_{\parallel}(n v_{\parallel e}) + D_n\nabla_{\perp}^2 n +s_n+\nu_\text{iz}n_\text{n}-\nu_\text{rec}n\, ,\\
\label{eqn:vorticity}
\frac{\partial \Omega}{\partial t} =& -\frac{1}{B}\nabla \cdot [\phi,\boldsymbol{\omega}] - \nabla \cdot \bigl( v_{\parallel i}\nabla_\parallel \boldsymbol{\omega}\bigr) + \frac{B\Omega_{ci}}{e}\nabla_{\parallel}j_{\parallel} + \frac{2\Omega_{ci}}{e} C(p_e + p_i) \nonumber\\
&+ \frac{\Omega_{ci}}{3e}C(G_i) + D_{\Omega}\nabla_\perp^2 \Omega-\frac{n_\text{n}}{n}\nu_\text{cx}\Omega\, ,\\
\label{eqn:electron_velocity}
\frac{\partial U_{\parallel e}}{\partial t} =& -\frac{1}{B}[\phi,v_{\parallel e}] - v_{\parallel e}\nabla_\parallel v_{\parallel e} + \frac{e}{m_e}\Bigl( \frac{j_\parallel}{\sigma_\parallel}+\nabla_\parallel\phi-\frac{1}{en}\nabla_\parallel p_e-\frac{0.71}{e}\nabla_\parallel T_e -\frac{2}{3en}\nabla_\parallel G_e\Bigr) \nonumber\\
&+ D_{v_{\parallel e}}\nabla_\perp^2 v_{\parallel e}+\frac{n_\text{n}}{n}(\nu_\text{en}+2\nu_\text{iz})(v_{\parallel n}-v_{\parallel e})\,, \\
\label{eqn:ion_velocity}
\frac{\partial v_{\parallel i}}{\partial t} =& -\frac{1}{B}[\phi,v_{\parallel i}] - v_{\parallel i}\nabla_\parallel v_{\parallel i} - \frac{1}{m_i n}\nabla_\parallel(p_e+ p_i) -\frac{2}{3m_in}\nabla_\parallel G_i \nonumber\\
&+ D_{v_{\parallel i}}\nabla_\perp^2 v_{\parallel i}+\frac{n_\text{n}}{n}(\nu_\text{iz}+\nu_\text{cx})(v_{\parallel n}-v_{\parallel i})\, ,\\
\label{eqn:electron_temperature}
\frac{\partial T_e}{\partial t} =& -\frac{1}{B}[\phi,T_e] - v_{\parallel e}\nabla_\parallel T_e 
+ \frac{2}{3}T_e\Bigl[0.71\frac{\nabla_\parallel j_\parallel}{en} - \nabla_\parallel v_{\parallel e}\Bigr] + \frac{4}{3}\frac{T_e}{eB}\Bigl[\frac{7}{2}C(T_e)+\frac{T_e}{n}C(n)-eC(\phi)\Bigr] 
 \nonumber \\
&+\nabla_\parallel (\chi_{\parallel e}\nabla_\parallel T_e) + D_{T_e}\nabla_\perp^2 T_e  + s_{T_e}-\frac{n_\text{n}}{n}\nu_\text{en}m_e\frac{2}{3}v_{\parallel e}(v_{\parallel n}-v_{\parallel e}) \nonumber\\
&- \frac{4}{3}\frac{m_e}{m_i}\frac{1}{\tau_e} (T_e- T_i) + \frac{n_\text{n}}{n}\nu_\text{iz}\biggl[-\frac{2}{3}E_\text{iz}-T_e+m_e v_{\parallel e}\Bigl(v_{\parallel e}-\frac{4}{3}v_{\parallel n}\Bigr)\biggr]  \,,\\
\label{eqn:ion_temperature}
\frac{\partial T_i}{\partial t} =& -\frac{1}{B}[\phi,T_i] - v_{\parallel i}\nabla_\parallel T_i 
+ \frac{4}{3}\frac{T_i}{eB}\Bigl[C(T_e)+\frac{T_e}{n}C(n)-eC(\phi)\Bigr] - \frac{10}{3}\frac{T_i}{eB}C(T_i) \nonumber \\ 
&+ \frac{2}{3}T_i\Bigl[(v_{\parallel i}-v_{\parallel e})\frac{\nabla_\parallel n}{n} -\nabla_\parallel v_{\parallel e}\Bigr] 
 + \nabla_\parallel (\chi_{\parallel i}\nabla_\parallel T_i) + D_{T_i}\nabla_\perp^2 T_i + s_{T_i} \nonumber\\
 &+ \frac{4}{3} \frac{m_e}{m_i}\frac{1}{\tau_e} (T_e- T_i) + \frac{n_\text{n}}{n}(\nu_\text{iz}+\nu_\text{cx})\Bigl[T_n-T_i+\frac{1}{3}(v_{\parallel n}-v_{\parallel i})^2\Bigr]\,,
\end{align}
and are coupled to the Poisson and Ampère equations,
\begin{align}
\label{eqn:poisson}
\nabla \cdot \bigl( n \nabla_\perp \phi\bigr) =&\ \Omega-\frac{\nabla_\perp^2 p_i}{e}\,,\\
\label{eqn:ampere}
\biggl( \nabla_\perp^2 - \frac{e^2\mu_0}{m_e}n\biggr)v_{\parallel e}=&\ \nabla_\perp^2 U_{\parallel e} - \frac{e^2\mu_0}{m_e}n v_{\parallel i} + \frac{e^2\mu_0}{m_e} \overline{j}_\parallel\,.
\end{align}
In Eqs.~\eqref{eqn:density}--\eqref{eqn:ampere}, $\Omega = \nabla\cdot\boldsymbol{\omega} = \nabla \cdot (n \nabla_\perp\phi + \nabla_\perp p_i/e)$ is the scalar vorticity, while $U_{\parallel e} = v_{\parallel e} +  e \psi/m_e$ is the sum of the electron inertia and the electromagnetic induction contributions.
Ampère law, Eq.~\eqref{eqn:ampere}, is solved for magnetic field fluctuations  perpendicular to the equilibrium magnetic field. 
Precisely, the magnetic field perturbation, $\delta\mathbf{B}$, is written in terms of the fluctuating parallel component of the vector potential, $\psi$, i.e. $\delta\mathbf{B}=-\nabla\times (\psi \mathbf{b})$, where $\mathbf{b}=\mathbf{B}/B$ is the unit vector of the unperturbed magnetic field and $\delta\mathbf{B}$ results in a vector perpendicular to $\mathbf{B}$ (neglecting the parallel component of $\delta \mathbf{B}$ is equivalent to exclude the fast compressional Alfvén wave from the dynamics~\cite{zeiler1997}).
In addition, we avoid to evolve the externally imposed equilibrium magnetic field, which would require to couple the plasma model to a Grad-Shafranov solver and to self-consistently simulate the plasma current.
For this reason, the contribution of the equilibrium parallel current $\overline{j}_\parallel$, computed by averaging the parallel current in the toroidal direction, is subtracted to the total parallel current in Eq.~\eqref{eqn:ampere}. 
We therefore neglect large-scale variation of the equilibrium magnetic field and we focus on the electromagnetic effects rising from small-scale, small-amplitude magnetic perturbations~\cite{zeiler1997}. 
   
The spatial operators appearing in Eqs.~\eqref{eqn:density}--\eqref{eqn:ampere} are the $\mathbf{E}\times\mathbf{B}$ convective term,
\begin{equation}
\label{eqn:pb}
    [\phi,f]=\mathbf{b}\ \cdot\ \bigl(\nabla \phi \times \nabla f\bigr)\,,\\
\end{equation}
the curvature operator,
\begin{equation}
\label{eqn:curv}
    C(f)=\frac{B}{2}\Bigl(\nabla \times \frac{\mathbf{b}}{B}\Bigr)\cdot \nabla f\,,\\
\end{equation}
the parallel gradient, which includes the electromagnetic flutter contribution, 
\begin{equation}
\label{eqn:gradpar}
    \nabla_\parallel f=\mathbf{b}\cdot\nabla f + \frac{1}{B}[\psi,f]\,,\\
\end{equation}
and the perpendicular Laplacian,
\begin{equation}
\label{eqn:lapl}
    \nabla_\perp^2 f=\nabla\cdot\bigl[(\mathbf{b}\times\nabla f)\times\mathbf{b}\bigl]\,,\\
\end{equation}
where $f$ is a general scalar function. 
The implementation of these operators in the GBS coordinate system is outlined in Sec.~\ref{sec:diff_op}.

The source terms in the density and temperature equations, $s_n$ and $s_T$, are added to fuel and heat the plasma.
The gyroviscous terms are defined as
\begin{align}
\label{eqn:ion_gyro}
    G_i&=-\eta_{0i}\Bigl[2\nabla_\parallel v_{\parallel i}+\frac{1}{B}C(\phi)+\frac{1}{en B}C(p_i)\Bigr]\,,\\
    \label{eqn:ele_gyro}
    G_e&=-\eta_{0e}\Bigl[2\nabla_\parallel v_{\parallel e} +\frac{1}{B}C(\phi)-\frac{1}{enB}C(p_e)\Bigr]\,,
\end{align}
where $\eta_{0i}=0.96 n T_{i}\tau_i$ and $\eta_{0e}=0.73 n T_{e}\tau_e$.
The numerical diffusion terms, $D_f\nabla_{\perp}^2 f$, are added for numerical stability.

Plasma and neutral dynamics are coupled through the ionization, recombination and charge-exchange processes as well as elastic electron-neutral collisions. These are described by the Krook operators with collision frequencies defined as
\begin{align}
\label{eqn:rate_first}
    \nu_\text{iz} = n\langle v_e \sigma_\text{iz}(v_e)\rangle\,,\\
    \nu_\text{rec} = n\langle v_e \sigma_\text{rec}(v_e)\rangle\,,\\
    \nu_\text{en} = n\langle v_e \sigma_\text{en}(v_e)\rangle\,\\
\end{align}
and
\begin{equation}
    \label{eqn:rate_last}
    \nu_\text{cx} = n\langle v_i \sigma_\text{iz}(v_i)\rangle\,,
\end{equation}
where $v_e$ and $v_i$ are the electron and ion velocities, $\sigma_\text{iz}$, $\sigma_\text{rec}$, $\sigma_\text{en}$ and $\sigma_\text{cx}$ are the ionization, recombination, elastic electron-neutral, and charge-exchange collision cross sections. The effective reaction rates, $\langle v\sigma\rangle$, are taken from the OpenADAS database~\citep{summers2006}.  In Eq.~\eqref{eqn:electron_temperature}, $E_\text{iz}$ is the ionization energy. 

While Eqs.~\eqref{eqn:density}--\eqref{eqn:ampere} are presented in physical units, they are implemented in GBS in dimensionless form. In the following, we use physical units, but we present the simulation results using GBS dimensionless units. 
The density, $n$, is normalized to the reference density $n_0$. 
The electron and ion temperatures, $T_e$ and $T_i$, are normalized to the reference values $T_{e0}$ and $T_{i0}$, respectively.
The electron and ion parallel velocities, $v_{\parallel e}$ and $v_{\parallel i}$, are normalized to the reference sound speed $c_{s0}=\sqrt{T_{e0}/m_i}$. The magnetic field is normalized to its modulus $B_0$ at the tokamak magnetic axis. The electrostatic potential, $\phi$, is normalized to $T_{e0}/e$ and $\psi$ is normalized to $\rho_{s0} B_0$, with $\rho_{s0}= c_{s0}/\Omega_{ci}$ the reference ion sound Larmor radius.
Perpendicular lengths are normalized to $\rho_{s0}$ and parallel lengths are normalized to the tokamak major radius, $R_0$. Time is normalized to $R_0/c_{s0}$.
By normalizing Eqs.~\eqref{eqn:density}--\eqref{eqn:ampere}, the following dimensionless parameters that regulate the system dynamics are identified: the normalized ion sound Larmor radius, $\rho_* = \rho_{s0}/R_0$, the ion to electron reference temperature ratio, $\tau = T_{i0}/T_{e0}$, the normalized electron and ion parallel thermal conductivities,
\begin{equation}
    \chi_{\parallel e} = \biggl(\frac{1.58}{\sqrt{2\pi}}\frac{m_i}{\sqrt{m_e}}\frac{(4\pi\epsilon_0)^2}{e^4}\frac{c_{s0}}{R_0}\frac{T_{e0}^{3/2}}{\lambda n_0}\biggr)T_e^{5/2}
\end{equation}
and
\begin{equation}
    \chi_{\parallel i} =\biggl(\frac{1.94}{\sqrt{2\pi}}\sqrt{m_i}\frac{(4\pi\epsilon_0)^2}{e^4}\frac{c_{s0}}{R_0}\frac{T_{e0}^{3/2}\tau^{5/2}}{\lambda n_0}\biggr)T_i^{5/2}\,,
\end{equation}
the reference electron plasma $\beta$, $\beta_{e0}=2\mu_0 n_0 T_{e0}/B_0^2$, and the normalized Spitzer resistivity, $\nu = e^2n_0R_0/(m_ic_{s0}\sigma_\parallel) = \nu_0 T_e^{-3/2}$, with 
\begin{equation}
\sigma_\parallel = \biggl(1.96\frac{n_0 e^2 \tau_e}{m_e}\biggr)n=\biggl(\frac{5.88}{4\sqrt{2\pi}}\frac{(4\pi\epsilon_0)^2}{e^2}\frac{ T_{e0}^{3/2}}{\lambda\sqrt{m_e}}\biggr)T_e^{3/2}
\label{eqn:resistivity}
\end{equation}
and
\begin{equation}
\nu_0 =\frac{4\sqrt{2\pi}}{5.88}\frac{e^4}{(4\pi\epsilon_0)^2}\frac{\sqrt{m_e}R_0n_0\lambda}{m_i c_{s0} T_{e0}^{3/2}}\,,
\end{equation}
where $\lambda$ is the Coulomb logarithm.

\subsection{Kinetic model for neutral atoms}\label{sec:neutrals}

The effect of neutral particles on plasma turbulence has been approached by modelling the neutral dynamics with fluid and kinetic models.
For example, two-dimensional fluid models that include the neutral gas continuity equations, accounting for neutral gas ionization and charge exchange processes, have been used to investigate the role of the neutral gas in the SOL~\cite{bisai2005,thrysoe2018}. 
More recently, three-dimensional turbulent simulations of the tokamak boundary, coupling fluid plasma and fluid neutral models, have been carried out with the BOUT++ code~\cite{leddy2017}.
However, since the neutral mean-free path in the SOL can be large, a kinetic approach is often preferred for the evolution of neutral dynamics. 
The first attempt to couple a kinetic neutral model to a plasma turbulence model was presented in Ref.~\cite{marandet2013}, where the EIRENE kinetic code~\cite{reiter2005} based on a Monte Carlo algorithm was coupled to the two-dimensional turbulence code TOKAM2D. 
The first three-dimensional turbulent simulation of the SOL with the self-consistent evolution of a kinetic neutral model was carried out by using GBS in limited geometry~\cite{wersal2015}.
Recently, the three-dimensional turbulence code TOKAM3X has been coupled to the EIRENE code to simulate plasma turbulence and kinetic neutrals in a diverted geometry~\cite{fan2018}.

We summarize here the kinetic model for neutral atoms implemented in GBS. The model is described in detail in Ref.~\cite{wersal2015} and extended here to the simulation of diverted configurations to include the contribution of the $\mathbf{E}\times\mathbf{B}$ and diamagnetic fluxes to the ion flux to the wall.
We consider a single mono-atomic neutral species represented by a distribution function $f_\text{n}$ with its dynamics being described by the following kinetic equation,
\begin{equation}
\label{eqn:kinetic_neutrals}
    \frac{\partial f_\text{n}}{\partial t} + \mathbf{v}\cdot\nabla f_\text{n} = -\nu_\text{iz} f_\text{n} -\nu_\text{cx}\Bigl(f_\text{n}-\frac{n_\text{n}}{n_i}f_i\Bigr)+\nu_\text{rec}f_i\,,
\end{equation}
where $f_i$ and $n_i$ are the ion distribution function and the ion density, respectively. The neutral-neutral collisions, which have a lower reaction rate than the charge-exchange and ionization processes in medium size tokamaks or attached conditions, are neglected. The elastic electron-neutral collisions, retained in Eq.~\eqref{eqn:electron_velocity}, are neglected in Eq.~\eqref{eqn:kinetic_neutrals} because of the electron-to-neutral mass ratio. 

The boundary conditions for $f_\text{n}$ at the wall are derived under the assumption that the impacting neutrals and ions are either reflected or absorbed. If absorbed, the neutral particle is immediately released with a velocity that depends on the wall properties and is independent of the impacting particle velocity. 
The distribution function of the neutrals flowing from wall to the plasma volume (i.e. neutrals with velocity such that $\mathbf{v}\cdot \mathbf{\hat{n}}>0$, with $\mathbf{\hat{n}}$ the unit vector normal to the wall) is therefore given by
\begin{equation}
\label{eqn:boundary_kinetic}
    f_\text{n}(\mathbf{x}_b,\mathbf{v}) = (1-\alpha_\text{refl})\Gamma_\text{out}(\mathbf{x}_b)\chi_\text{in}(\mathbf{x}_b,\mathbf{v},T_b)+\alpha_\text{refl}[f_\text{n}(\mathbf{x}_b,\mathbf{v}-2\mathbf{v}_p)+f_i(\mathbf{x}_b,\mathbf{v}-2\mathbf{v}_p)]\,,
\end{equation}
where $\alpha_\text{refl}$ is the reflection fraction, assumed the same for neutrals and ions, $\mathbf{x}_b$ indicates the boundary position, $\Gamma_\text{out}=\Gamma_\text{out,n}+\Gamma_\text{out,i}$ is the sum of the neutral and ion fluxes to the wall and projected in the direction perpendicular to it, $\mathbf{v}_p=\text{v}_p \mathbf{\hat{n}}$ is the velocity perpendicular to the boundary, with  $\text{v}_p=\mathbf{v}\cdot \mathbf{\hat{n}}$, and $\chi_\text{in}$ is the inflowing velocity distribution function given by the Knudsen cosine law,
\begin{equation}
    \chi_\text{in}(\mathbf{x}_b,\mathbf{v},T_b)=\frac{3}{4\pi}\frac{m^2}{T_b^2}\cos\theta\exp\Bigl(-\frac{m \text{v}^2}{2 T_b}\Bigr)\,,
\end{equation}
with $\theta=\arccos(\text{v}_p/\text{v})$ and $T_b$ the wall temperature~\cite{wersal2015}. 

In the limit where the turbulent time scale is much longer than the typical time of flight of neutrals, $\tau_\text{turb}\gg \tau_n$, the neutral adiabatic approximation can be applied. This corresponds to impose $\partial_t f_\text{n} = 0$ in Eq.~\eqref{eqn:kinetic_neutrals}.
Moreover, we assume that the neutral mean free path is shorter than typical parallel lengths of the plasma structures. 
Under these assumptions, the formal solution of Eq.~\eqref{eqn:kinetic_neutrals} can be obtained by using the method of characteristics~\cite{wersal2015},
\begin{equation}
\label{eqn:dist_fun}
    \begin{split}
    f_\text{n}(\mathbf{x}_\perp,x_\parallel,\mathbf{v},t)=\int_0^{r_{\perp b}} \biggl[\frac{S(\mathbf{x}'_\perp,x_\parallel,\mathbf{v},t)}{v_\perp}+\delta(r_\perp'-r_{\perp b})f_\text{n}(\mathbf{x}_{\perp b}',x_\parallel,\mathbf{v},t)\biggr]\\
    \times \exp\Bigl[-\frac{1}{v_\perp}\int_0^{r_\perp'}\nu_\text{eff}(\mathbf{x}_\perp'',x_\parallel,t)\mathrm{d}r_\perp''\Bigr]\mathrm{d}r_\perp'\,,
    \end{split}
\end{equation}
where we express a position $\mathbf{x}$ in terms of $\mathbf{x}_\perp$, the coordinate on the plane perpendicular to $\mathbf{B}$, and $x_\parallel$, the coordinate parallel to $\mathbf{B}$, $r_\perp'$ is the coordinate along the neutral characteristic defined by $\mathbf{x}_\perp'=\mathbf{x}_\perp-r_\perp'\mathbf{v}_\perp/\text{v}_\perp$, $r_{\perp b}$ denotes the distance along the characteristic from the position $\mathbf{x}$ and the wall, $\mathbf{v}_\perp$ is the component of the velocity perpendicular to $\mathbf{B}$, and $\nu_\text{eff}=\nu_\text{iz}+\nu_\text{cx}$ is the effective collision frequency for neutral loss. In the following, we drop the parametric dependencies on $t$ and $x_\parallel$ to simplify the notation.

The volumetric source term in Eq.~\eqref{eqn:dist_fun} results from charge-exchange and recombination processes and is given by~\cite{wersal2015}
\begin{equation}
\label{eqn:neutral_source}
    S(\mathbf{x}_\perp',\mathbf{v}) = \nu_\text{cx}(\mathbf{x}_\perp')n_\text{n}(\mathbf{x}_\perp')\Phi_i(\mathbf{x}_\perp',\mathbf{v})+\nu_\text{rec}(\mathbf{x}_\perp')f_i(\mathbf{x}_\perp',\mathbf{v})\,,
\end{equation}
where $\Phi_i = [m_i/(2\pi T_i)]^{3/2}\exp[-m_i \text{v}^2/(2T_i)]$ is the ion velocity distribution. The source of neutrals at the wall, $\delta(r_\perp'-r_{\perp b})f_\text{n}(\mathbf{x}_{\perp b}',x_\parallel,\mathbf{v},t)$, is given by the boundary conditions in Eq.~\eqref{eqn:boundary_kinetic}.

The ion recycling term present in the boundary conditions and the recombination term appearing in $S(\mathbf{x}_\perp',\mathbf{v})$ (see Eqs.~\eqref{eqn:boundary_kinetic}~and~\eqref{eqn:neutral_source}) do not depend on the neutral distribution function and can be computed directly from the plasma quantities. 
On the other hand, the charge-exchange term in $S(\mathbf{x}_\perp',\mathbf{v})$ as well as the reflected and re-emitted neutrals in the boundary conditions (see Eqs.~\eqref{eqn:boundary_kinetic}~and~\eqref{eqn:neutral_source}) depend on $n_\text{n}(\mathbf{x}_\perp) = \int f_\text{n} \mathrm{d}\mathbf{v}$, the neutral density. 
This suggests to integrate Eq.~\eqref{eqn:dist_fun}. In fact, Ref.~\cite{wersal2015} shows that, by integrating Eq.~\eqref{eqn:dist_fun}, a linear integral equation for $n_\text{n}(\mathbf{x}_\perp)$ is obtained, 
\begin{align}
\label{eqn:neutral_density}
    n_\text{n}(\mathbf{x}_\perp) =& \int_D n_\text{n}(\mathbf{x}_\perp')\nu_\text{cx}(\mathbf{x}_\perp')K_{p\rightarrow p}(\mathbf{x}_\perp,\mathbf{x}_\perp')\mathrm{d}A'\nonumber\\
    &+\int_{\partial D} (1-\alpha_\text{refl})\Gamma_\text{out,n}(\mathbf{x}_{\perp b}')K_{b\rightarrow p}(\mathbf{x}_\perp,\mathbf{x}_{\perp b}',T_b)\mathrm{d}a_b'\nonumber\\
    &+ n_\text{n[out,i]}(\mathbf{x}_\perp) +n_{\text{n[rec]}}(\mathbf{x}_\perp) \,,
\end{align}
where $\mathrm{d}A'$ is the infinitesimal area in the poloidal plane $D$,  $\mathrm{d}a_b'$ is the infinitesimal length along the boundary $\partial D$, and $\Gamma_\text{out,n}(\mathbf{x}_{\perp b})$ is the neutral flux towards the wall,
\begin{align}
\label{eqn:neutral_flux}
\Gamma_\text{out,n}(\mathbf{x}_{\perp b})=&\int_D n_\text{n}(\mathbf{x}_\perp')\nu_\text{cx}(\mathbf{x}_\perp')K_{p\rightarrow b}(\mathbf{x}_{\perp b},\mathbf{x}_\perp')\mathrm{d}A'\nonumber\\
&+\int_{\partial D}(1-\alpha_\text{refl})\Gamma_\text{out,n}(\mathbf{x}_{\perp b}')K_{b\rightarrow b}(\mathbf{x}_{\perp b},\mathbf{x}_{\perp b}',T_b)\mathrm{d}a_b'\nonumber\\
&+\Gamma_\text{out,n[out,i]}(\mathbf{x}_{\perp b}) +\Gamma_\text{out,n[rec]}(\mathbf{x}_{\perp b})\,.
\end{align}
The contribution to the neutral density due to the ion recycling at the wall and recombination events in Eq.~\eqref{eqn:neutral_density}, $n_\text{n[out,i]}$ and $n_{\text{n[rec]}}$, as well as the corresponding contribution to the neutral flux in Eq.~\eqref{eqn:neutral_flux}, $\Gamma_\text{out,n[out,i]}$ and $\Gamma_\text{out,n[rec]}$, are defined as
\begin{align}
    n_\text{n[out,i]}(\mathbf{x}_\perp)=&\int_{\partial D} \Gamma_\text{out,i}(\mathbf{x}_{\perp b}')\bigl[ (1-\alpha_\text{refl}) K_{b\rightarrow p}(\mathbf{x}_\perp,\mathbf{x}_{\perp b}',T_b)\nonumber\\
    &+ \alpha_\text{refl} K_{b\rightarrow p}(\mathbf{x}_\perp,\mathbf{x}_{\perp b}',T_i)\bigr]\mathrm{d}a_b'\,,\\
    n_{\text{n[rec]}}(\mathbf{x}_\perp)=&\int_{D} n_i(\mathbf{x}_\perp')\nu_\text{rec}(\mathbf{x}_\perp')K_{p\rightarrow p}(\mathbf{x}_{\perp},\mathbf{x}_\perp')\mathrm{d}A'\,,\\
    \Gamma_\text{out,n[out,i]}(\mathbf{x}_{\perp b})=&\int_{\partial D}\Gamma_\text{out,i}(\mathbf{x}_{\perp b}')\bigl[(1-\alpha_\text{refl}) K_{b\rightarrow b}(\mathbf{x}_{\perp b},\mathbf{x}_{\perp b}',T_b)\nonumber\\
    &+\alpha_\text{refl}K_{b\rightarrow b}(\mathbf{x}_{\perp b},\mathbf{x}_{\perp b}',T_i)\bigr] \mathrm{d}a_b'\,,\\
    \Gamma_\text{out,n[rec]}(\mathbf{x}_{\perp b}) =& \int_{D} n_i(\mathbf{x}_\perp')\nu_\text{rec}(\mathbf{x}_\perp')K_{p\rightarrow b}(\mathbf{x}_{\perp b},\mathbf{x}_\perp')\mathrm{d}A'\,.
\end{align}
Generalizing Ref.~\cite{wersal2015}, the ion flux projected in the direction perpendicular to the wall, $\Gamma_\text{out,i}$, is evaluated by considering the ion parallel flux and the perpendicular fluxes due to the $\mathbf{E}\times\mathbf{B}$ and the diamagnetic drifts, $\mathbf{v}_E=\mathbf{B}\times\nabla\phi/B^2$ and $\mathbf{v}_{di}=\mathbf{B} \times \nabla p_i/nB^2$, that is
\begin{equation}
    \Gamma_\text{out,i}(\mathbf{x}_{\perp b}) =  -n_i(\mathbf{x}_{\perp b})\bigl[ v_{\parallel i}(\mathbf{x}_{\perp b})\mathbf{b} + \mathbf{v}_E(\mathbf{x}_{\perp b}) + \mathbf{v}_{di}(\mathbf{x}_{\perp b}) \bigr]\cdot\mathbf{\hat{n}}\,.
\end{equation}
The kernel functions, appearing in Eqs.~\eqref{eqn:neutral_density}~and~\eqref{eqn:neutral_flux}, involve integrals in the velocity space,
\begin{align}
\label{eqn:kernel_first}
    K_{p\rightarrow p}(\mathbf{x}_\perp,\mathbf{x}_\perp')&=K^\text{dir}_{p\rightarrow p}(\mathbf{x}_\perp,\mathbf{x}_\perp')+\alpha_\text{refl} K^\text{refl}_{p\rightarrow p}(\mathbf{x}_\perp,\mathbf{x}_\perp')\,,\\
    K_{b\rightarrow p}(\mathbf{x}_\perp,\mathbf{x}_{\perp b}',T)&=K^\text{dir}_{b\rightarrow p}(\mathbf{x}_\perp,\mathbf{x}_{\perp b}',T)+\alpha_\text{refl}K^\text{refl}_{b\rightarrow p}(\mathbf{x}_\perp,\mathbf{x}_{\perp b}',T)\,,\\
    K_{p\rightarrow b}(\mathbf{x}_{\perp b},\mathbf{x}_\perp')&=K^\text{dir}_{p\rightarrow b}(\mathbf{x}_{\perp b},\mathbf{x}_\perp')+\alpha_\text{refl}K^\text{refl}_{p\rightarrow b}(\mathbf{x}_{\perp b},\mathbf{x}_\perp')\,,\\
    \label{eqn:kernel_last}
    K_{b\rightarrow b}(\mathbf{x}_{\perp b},\mathbf{x}_{\perp b}',T)&=K^\text{dir}_{b\rightarrow b}(\mathbf{x}_{\perp b},\mathbf{x}_{\perp b}',T)+\alpha_\text{refl}K^\text{refl}_{b\rightarrow b}(\mathbf{x}_{\perp b},\mathbf{x}_{\perp b}',T)\,,
\end{align}
where we consider the direct path between two points as well as the paths that include one reflection at the wall (we neglect paths with multiple reflections as $\lambda_n/L<1$, with $\lambda_n$ the averaged neutral mean free path and $L$ the typical machine size in poloidal plane) and we define, for path~=~\{dir, refl\},
\begin{align}
\label{eqn:kpp}
    K_{p\rightarrow p}^\text{path}(\mathbf{x}_\perp,\mathbf{x}_\perp')&=\int_0^\infty \frac{1}{r_\perp'}\Phi_{\perp i}(\mathbf{x}_\perp',\mathbf{v}_\perp)\exp\biggl[-\frac{1}{v_\perp}\int_0^{r_\perp'}\nu_\text{eff}(\mathbf{x}_\perp'')\mathrm{d}r_\perp''\biggr]\mathrm{d}v_\perp\,,\\
    \label{eqn:kbp}
    K_{b\rightarrow p}^\text{path}(\mathbf{x}_\perp,\mathbf{x}_{\perp b}',T)&=\int_0^\infty \frac{v_\perp}{r_\perp'}\cos\theta'\chi_{\perp \text{in}}(\mathbf{x}_{\perp b}',\mathbf{v}_\perp,T)\exp\biggl[-\frac{1}{v_\perp}\int_0^{r_\perp'}\nu_\text{eff}(\mathbf{x}_\perp'')\mathrm{d}r_\perp''\biggr]\mathrm{d}v_\perp\,,\\
    \label{eqn:kpb}
    K_{p\rightarrow b}^\text{path}(\mathbf{x}_{\perp b},\mathbf{x}_\perp')&=\int_0^\infty \frac{v_\perp}{r_\perp'}\cos\theta\Phi_{\perp i}(\mathbf{x}_\perp',\mathbf{v}_\perp)\exp\biggl[-\frac{1}{v_\perp}\int_0^{r_\perp'}\nu_\text{eff}(\mathbf{x}_\perp'')\mathrm{d}r_\perp''\biggr]\mathrm{d}v_\perp\,,\\
    \label{eqn:kbb}
    K_{b\rightarrow b}^\text{path}(\mathbf{x}_{\perp b},\mathbf{x}_{\perp b}',T)&=\int_0^\infty \frac{v_\perp^2}{r_\perp'}\cos\theta\cos\theta'\chi_{\perp \text{in}}(\mathbf{x}_b',\mathbf{v}_\perp,T)\exp\biggl[-\frac{1}{v_\perp}\int_0^{r_\perp'}\nu_\text{eff}(\mathbf{x}_\perp'')\mathrm{d}r_\perp''\biggr]\mathrm{d}v_\perp\,,
\end{align}
with
$\Phi_{\perp i} (\mathbf{x}_\perp,\mathbf{v}_\perp)=\int \Phi_i (\mathbf{x}_\perp,\mathbf{v}) \mathrm{d}v_\parallel = m_i/(2\pi T_i)\exp[-m_i v_\perp^2/(2T_i)]$ and $\chi_{\perp \text{in}} (\mathbf{x}_\perp,\mathbf{v}_\perp)=\int \chi_{\text{in}} (\mathbf{x}_\perp,\mathbf{v}) \mathrm{d}v_\parallel = 3m_i^2/(4\pi T_i^2)v_\perp\cos\theta\exp[-m_iv_\perp^2/(4 T_i)] \mathcal{K}_0[m_i v_\perp^2/(4 T_i)]$, being $\mathcal{K}_0(x)$ the modified Bessel function of the second kind.
The vector $\mathbf{x}_\perp''$ indicates a position along the path from the source to the target points.
The four kernels represent the four possibilities for neutral particles of being generated within the plasma, $p$, or at the boundary, $b$, and reach a point also in the plasma or in the boundary.  

We note that a neutral flux can be externally imposed by means of a gas puff that introduces a localized source of neutrals. 
Similarly, a pumping region on the wall can also be considered.
This can be simply implemented by multiplying the kernel functions, $K_{b \rightarrow p}$ and $K_{b \rightarrow b}$, by a recycling coefficient smaller than one.

\subsection{Differential operators}\label{sec:diff_op}

\begin{figure}
    \centering
    \includegraphics[scale=0.4]{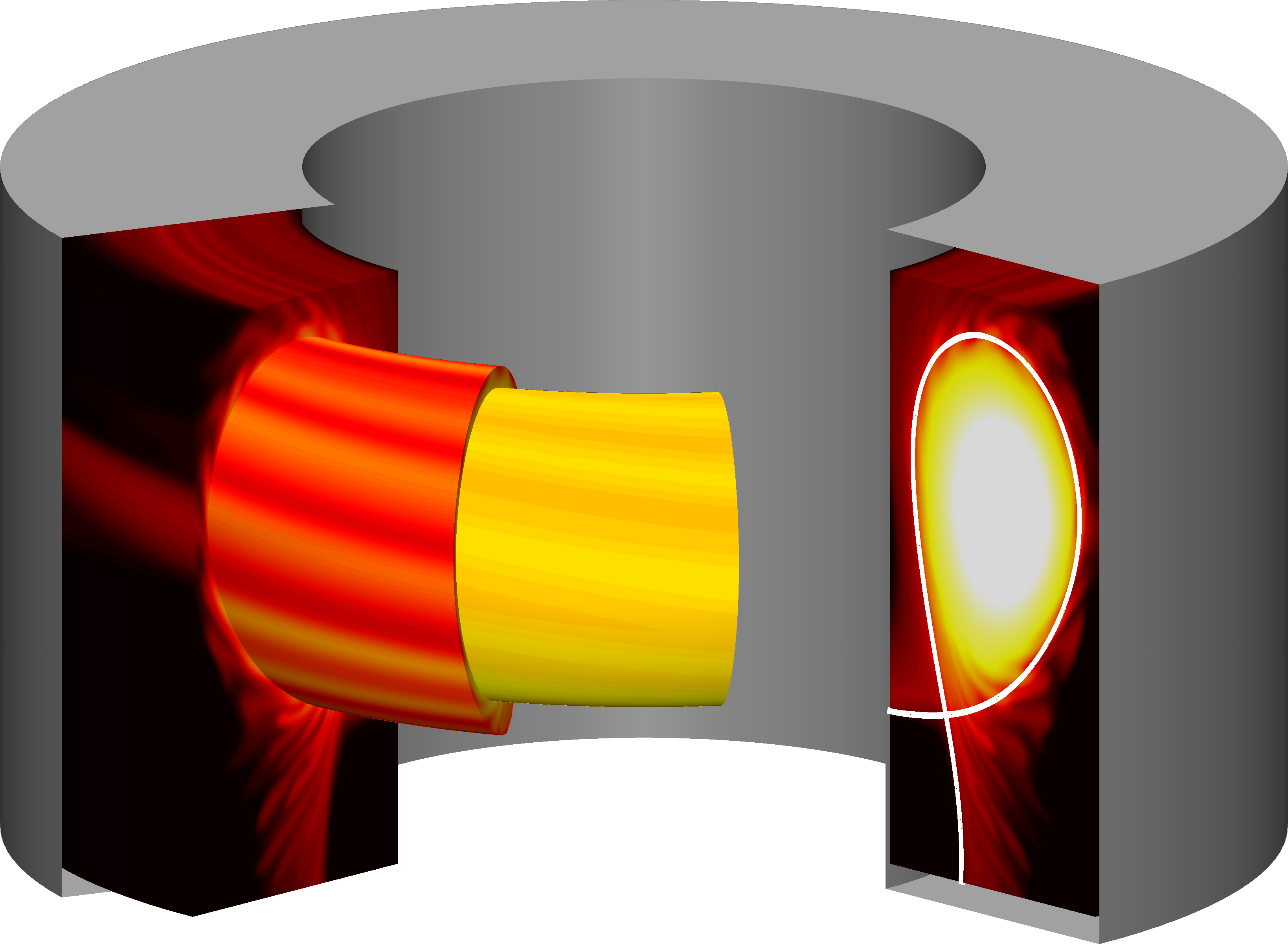}
    \caption{Representation of a three dimensional time snapshot of the plasma electron pressure, $p_e=n T_e$, from a GBS simulation. The domain encompasses the whole tokamak volume, and develops over the complete toroidal angle and poloidal cross section. The white line denotes the separatrix. The magnetic equilibrium is given by the equilibrium reconstruction of the TCV discharge \#65402 at time 1.0~s.}
    \label{fig:tcv_3d}
\end{figure}

The differential operators in Eqs.~\eqref{eqn:pb}--\eqref{eqn:lapl} are written in the $(R,\varphi,Z)$ cylindrical non-field-aligned coordinate system, where $R$ is the distance from the axis of symmetry of the torus, $Z$ is the vertical coordinate, and $\varphi$ is the toroidal angle. The poloidal cross section has a rectangular shape, particularly suitable for the simulation of the TCV tokamak,  and the domain encompasses the whole plasma volume.
A representation of the GBS domain is shown in Fig.~\ref{fig:tcv_3d}.

The toroidally symmetric equilibrium magnetic field, used to compute the GBS differential operators, is written in terms of the poloidal magnetic flux $\Psi$, as
\begin{equation}
\label{eqn:magfield}
    \mathbf{B}=RB_\varphi\nabla\varphi+ \nabla\varphi\times\nabla\Psi\,.
\end{equation}
The poloidal magnetic flux is a function of $R$ and  $Z$, and can be provided as an analytical function, an equilibrium reconstruction, or as the result of a Grad-Shafranov solver. As an example, the magnetic separatrix of a TCV equilibrium reconstruction is shown in Fig.~\ref{fig:tcv_3d}.  

The spatial differential operators in Eqs.~\eqref{eqn:density}--\eqref{eqn:ampere} are written in $(R, \varphi, Z)$ coordinates. They are expanded in the large aspect ratio limit, $\epsilon\sim r/R_0 \ll 1$, assuming the poloidal component of the magnetic field smaller than the toroidal one, $\delta \sim B_p/B_\varphi \ll 1$, with a safety factor at the midplane $q\sim\epsilon/\delta$ of order unity. Only the leading order terms in $\epsilon$ and $\delta$ are retained.

In the following, we derive the expression of the operators as they are implemented in GBS.
As a first step, we note that, at zeroth-order in $\epsilon$ and $\delta$, the modulus of the magnetic field is constant, 
\begin{equation}
\label{eqn:modB}
    \frac{B^2}{B_0^2}= \frac{B_R^2}{B_0^2}+\frac{B_Z^2}{B_0^2}+\frac{B_\varphi^2}{B_0^2}= \frac{B_\varphi^2}{B_0^2} + O(\delta^2)=1 + O(\epsilon,\delta^2)\,,
\end{equation}
where the magnetic field is written as
\begin{equation}
    \mathbf{B} = B_R \mathbf{e}_R + B_Z \mathbf{e}_Z + B_\varphi \mathbf{e}_\varphi\,,
\end{equation}
with $\mathbf{e}_R$, $\mathbf{e}_Z$, and $\mathbf{e}_\varphi$ the basis vectors.
Hence, the Poisson brackets, Eq.~\eqref{eqn:pb}, can be written as 
\begin{align}
    [\phi,f] &= \mathbf{b}\cdot \nabla\phi \times \nabla f \nonumber\\
    &= \mathbf{b}\cdot\Bigl[\bigl(\partial_R\phi\, \er + \partial_Z\phi\, \ez + \frac{1}{R}\partial_\varphi\phi\,\ep\bigr)\times \bigl(\partial_R f\, \mathbf{e}_R + \partial_Z f\, \mathbf{e}_Z + \frac{1}{R}\partial_\varphi f\, \mathbf{e}_\varphi\bigr)\Bigr]\\
    &= \frac{1}{R}\frac{B_R}{B} (\,\partial_\varphi\phi\,\partial_Z f - \partial_Z\phi\,\partial_\varphi f) + \frac{1}{R}\frac{B_Z}{B} (\partial_R\phi\,\partial_\varphi f - \partial_\varphi\phi\,\partial_R f) + \frac{B_\varphi}{B} (\partial_Z\phi\,\partial_R f - \partial_R\phi\,\partial_Z f)\nonumber\,,
\end{align}
which, in dimensionless units, leads to
\begin{align}
    [\phi, f] &=\frac{\rho_{s0}}{R}\frac{B_R}{B} (\,\partial_\varphi\phi\,\partial_Z f - \partial_Z\phi\,\partial_\varphi f) + \frac{\rho_{s0}}{R}\frac{B_Z}{B} (\partial_R\phi\,\partial_\varphi f - \partial_\varphi\phi\,\partial_R f) + \frac{B_\varphi}{B} (\partial_Z\phi\,\partial_R f - \partial_R\phi\,\partial_Z f)\nonumber\\
    &=\frac{B_\varphi}{B} (\partial_Z\phi\,\partial_R f - \partial_R\phi\,\partial_Z f)+O(\epsilon,\delta)\,,
\end{align}
since $B_Z/B\sim B_R/B \sim \delta$ and $\rho_{s0}/R\sim \rho_*\ll \epsilon$.

Neglecting local current and therefore assuming $\nabla\times\mathbf{B} = 0$, the curvature operator, Eq.~\eqref{eqn:curv}, is expanded in $\epsilon$ and $\delta$ as 
\begin{equation}
      C(f) = \frac{1}{2 B} \biggl( \frac{B_\varphi}{B^2}\partial_Z B^2 \partial_R f - \frac{B_\varphi}{B^2} \partial_R B^2\partial_Z f \biggr) + O(\epsilon,\delta) \,.
\end{equation}
The spatial derivatives $\partial_Z B^2$ and $\partial_R B^2$ can be determined by using  Eq.~\eqref{eqn:modB}, 
\begin{align}
    \frac{\partial_Z B^2}{B_0^2} &= \frac{\partial}{\partial Z} \Bigl(\frac{R_0}{R}\Bigr)^2 + O(\delta^2) = 0 + O(\delta^2)\\
    \frac{\partial_R B^2}{B_0^2} &= \frac{\partial}{\partial R} \Bigl(\frac{R_0}{R}\Bigr)^2 + O(\delta^2) =  -2\frac{R_0^2}{R^3} + O(\delta^2) = -\frac{2}{R} + O(\epsilon,\delta^2)\,,
\end{align}
where we use $B_\varphi^2/B_0^2 = R_0^2/R^2 = 1 + O(\epsilon)$.
Finally, by keeping only the leading order terms in $\epsilon$ and $\delta$, the curvature operator, normalized to $1/(R_0\rho_{s0})$, becomes
\begin{equation}
    \label{eqn:curv_final}
    C(f)  = \frac{B_\varphi}{B_0} \partial_Z f + O(\epsilon,\delta) \,.
\end{equation}

The parallel gradient, Eq.~\eqref{eqn:gradpar}, is normalized to $1/R_0$ and, in dimensionless units, is given by
\begin{align}
    \label{eqn:gradpar_final}
    \nabla_\parallel f &= \frac{\mathbf{B}}{B} \cdot \nabla f = \rho_*^{-1}\Bigl(\frac{B_R}{B} \partial_R f + \frac{B_Z}{B} \partial_Z f + \frac{B_\varphi}{B} \frac{\rho_{s0}}{R}\partial_\varphi f\Bigr)\nonumber\\
    &= \partial_Z\Psi\,\partial_R f - \partial_R\Psi\,\partial_Z f + \frac{B_\varphi}{B_0}\partial_\varphi f + O(\epsilon,\delta)\,,
\end{align}
where $B_R=\partial_Z \Psi/R$, $B_Z=-\partial_R \Psi/R$.

The perpendicular laplacian, Eq.~\eqref{eqn:lapl}, can be developed as
\begin{equation}
\label{eqn:lapl_int}
    \nabla_\perp^2 f = \nabla \cdot \Bigl[ \frac{1}{B^2} (\mathbf{B}\times \nabla f)\times \mathbf{B}\Bigr] = \frac{1}{B^2}  \nabla \cdot \bigl[ (\mathbf{B}\times \nabla f)\times \mathbf{B}\bigr] - \frac{\nabla B^2}{B^4}\cdot \bigl[ (\mathbf{B}\times \nabla f)\times \mathbf{B}\bigr]\,.
\end{equation}
The second term on the right-hand side of Eq.~\eqref{eqn:lapl_int} is one order $\epsilon$ smaller than the first one, which can be written in cylindrical coordinates as
\begin{equation}
\begin{split}
    (\mathbf{B}\times\nabla f)\times\mathbf{B} &=\Bigl(B_Z^2\partial_R f -B_Z B_R \partial_Z f -  \frac{B_R B_\varphi}{R}\partial_\varphi f + B_\varphi^2 \partial_R f\Bigr)\er\\
    &+\Bigl(B_\varphi^2\partial_Z f -  \frac{B_Z B_\varphi}{R}\partial_\varphi f - B_Z B_R \partial_R f +B_R^2\partial_Z f\Bigr)\ez\\
    &+\Bigl(\frac{B_R^2}{R}\partial_\varphi f - B_\varphi B_R \partial_R f -B_\varphi B_Z \partial_Z f + \frac{B_Z^2}{R}\partial_\varphi f\Bigr) \ep\,,
\end{split}
\end{equation}
and expanded in $\epsilon$ and $\delta$,
\begin{equation}
\label{eqn:lapl_intermidiate}
 (\mathbf{B}\times\nabla f)\times\mathbf{B}\simeq \frac{B_\varphi^2}{B^2}\partial_R f\,\er + \frac{B_\varphi^2}{B^2}\partial_Z f\,\ez + O(\epsilon,\delta)\,.   
\end{equation}
Thus, the leading order terms of Eq.~\eqref{eqn:lapl_intermidiate} can be expressed as
\begin{equation}
    \nabla_\perp^2 f = \partial_{RR}^2 f + \partial_{ZZ}^2 f + O(\epsilon,\delta)\,.
\end{equation}

In addition to the differential operators given by Eqs.~\eqref{eqn:pb}--\eqref{eqn:lapl}, there are three operators appearing in Eqs.~\eqref{eqn:density}--\eqref{eqn:ion_temperature} that can be derived from Eqs.~\eqref{eqn:pb}--\eqref{eqn:lapl} by applying the same ordering in $\epsilon$ and $\delta$: the parallel laplacian, $\nabla_\parallel^2 f$, the curvature of the parallel gradient, $C(\nabla_\parallel f)$, and the parallel gradient of the curvature, $\nabla_\parallel [C(f)]$.
In dimensionless units, the parallel laplacian is given by
\begin{equation}
    \begin{split}
        \nabla_\parallel^2 f &= \bigl(\partial_Z\Psi\partial_{RZ}^2\Psi-\partial_R\Psi\partial_{ZZ}^2\Psi\bigr)\partial_R f + \bigl(\partial_R\Psi\partial_{RZ}^2\Psi-\partial_Z\Psi\partial_{RR}^2\Psi\bigr)\partial_Z f \\
        &+\bigl[(\partial_Z\Psi)^2\partial_{RR}^2 f+(\partial_R\Psi)^2\partial_{ZZ}^2f-2\partial_Z\Psi\partial_R\Psi\partial_{RZ}^2f\bigr]\\
        &+ 2\frac{B_\varphi}{B}\partial_Z\Psi\partial_{\varphi R}^2 f - 2\frac{B_\varphi}{B}\partial_R\Psi\partial_{\varphi Z}^2 f + \partial_{\varphi\varphi}^2 f + O(\epsilon,\delta)\,,
    \end{split}
\end{equation}
the curvature of the parallel gradient by
\begin{equation}
    C(\nabla_\parallel f) = \frac{B_\varphi}{B}\bigl(\partial_{ZZ}\Psi\partial_R f+\partial_Z\Psi\partial_{RZ}f-\partial_{RZ}\Psi\partial_Z f-\partial_R\Psi\partial_{ZZ}f\bigr)+\partial_{Z\varphi}f+ O(\epsilon,\delta)\,,
\end{equation}
and the parallel gradient of the curvature by
\begin{equation}
    \nabla_\parallel [C(f)] = \frac{B_\varphi}{B}\partial_Z\Psi \partial_{RZ} f -\frac{B_\varphi}{B}\partial_R\Psi \partial_{ZZ} f+ \partial_{Z\varphi} f + O(\epsilon,\delta)\,.
\end{equation}

\subsection{Boundary and initial conditions}\label{sec:boundary}

The large electric field established on the $\rho_s$ scale in the magnetic pre-sheath violates the hypothesis behind the drift-approximation. Consequently, the boundary conditions of GBS are applied at the magnetic pre-sheath entrance. 
The set of generalized Bohm-Chodura sheath boundary conditions implemented in GBS was originally derived in Ref.~\cite{Loizu2012} in the cold ion limit and extended in Ref.~\cite{mosetto2015} to include warm ions. In this version of GBS, we neglect the correction terms arising from the variation of the density and electrostatic potential in the direction tangent to the wall, and we apply the magnetic pre-sheath boundary conditions at the walls containing the strike points, i.e. the walls where the divertor legs terminate. 
For instance, in the case of the TCV magnetic equilibrium shown in Fig.~\ref{fig:tcv_3d}, magnetic pre-sheath boundary conditions are applied at the bottom and inner walls. 
The boundary conditions assume the following form: 
\begin{align}
    \label{eqn:boundary_first}
    v_{\parallel i}=& \pm c_s\sqrt{1+ \frac{T_i}{T_e}}\,,\\
    \label{eqn:boundary_second}
    v_{\parallel e}=& \pm c_s\  \exp\Bigl(\Lambda_0-\frac{e\phi}{T_e}\Bigr)\,,\\
    \partial_s n =& \mp \frac{n}{c_s\sqrt{1+ \frac{T_i}{T_e}}}\partial_s v_{\parallel i}\,,\\
    \partial_s T_e =&\ \partial_s T_i=\ 0\,,\\
    \label{eqn:boundary_omega}
    \Omega =& \mp \frac{m_i n}{e} c_s\sqrt{1+ \frac{T_i}{T_e}}\partial_{ss}^2 v_{\parallel i}\,,\\
    \label{eqn:boundary_pot}
    \partial_s \phi =& \mp \frac{m_i c_s}{e\sqrt{1+\frac{T_i}{T_e}}} \partial_s v_{\parallel i}\,,
\end{align}
where $\Lambda_0=\log\sqrt{m_i/(2\pi m_e)}\simeq 3$ for hydrogen plasmas, and $s$ denotes the derivative in the direction perpendicular to the wall ($s=Z$ for the top and bottom walls, and $s=R$ for the inner and outer walls).
The top (bottom) sign refers to the magnetic field pointing towards (away from) the target plate.
In addition to the boundary conditions in Eqs.~\eqref{eqn:boundary_first}--\eqref{eqn:boundary_pot}, we simply consider $\psi=0$ at the magnetic pre-sheath entrance.
In order to avoid the discontinuity of the parallel velocities at the location where the magnetic field is tangent to the wall, a smoothing function from $+c_s$ to $-c_s$ is applied to guarantee that $v_{\parallel i}$ and $v_{\parallel e}$  vary without strong discontinuities.
In addition, we impose $\partial_s |v_{\parallel i}|=0$ and $\partial_{ss}^2 |v_{\parallel i}|=0$ in Eqs.~\eqref{eqn:boundary_first}-\eqref{eqn:boundary_pot} in these smoothing regions where the normal derivative to the wall of $|v_{\parallel i}|$ can become negative, leading to unphysical boundary conditions for $n$, $\phi$ and $\Omega$.  
We highlight that the smoothing affects less than 2\% of the wall surface and is at certain distance from the strike points, therefore it has a negligible effect on the overall dynamics.

The boundary conditions in Eqs.~\eqref{eqn:boundary_first}-\eqref{eqn:boundary_omega} are also applied at the walls that do not contain strike points, while the Dirichlet boundary condition $\phi=\Lambda T_e/e$ is used for $\phi$, with ${\Lambda = \Lambda_0 - \log\sqrt{1+T_i/T_e}}$. 
In fact, Poisson equation is ill-defined if a Neumann boundary condition for the electrostatic potential is applied at the four walls of the domain.

As long as the quasi-steady state that is established after a transient is statistically independent of the initial conditions, the initial profiles do not present physical interest.
As initial condition for our simulations, we impose $\phi = \Lambda T_e/e$, with $T_e$ the initial electron temperature, which is uniform over the entire GBS domain. Also $\Omega$, $n$, and $T_i$ have initial uniform profiles. The electron and ion parallel velocities are properly designed functions to satisfy the boundary conditions $v_{\parallel e,i}|_\text{wall}=\pm c_s\sqrt{1+ T_i/T_e}$ with $\partial_s v_{\parallel e,i}|_\text{wall} = 0$, being $s$ the coordinate perpendicular to the wall. The  constraint $\partial_s v_{\parallel i}|_\text{wall} = 0$ guarantees vanishing Neumann boundary conditions for $n$, $\phi$, $\omega$, in agreement with the uniform initial profiles of these quantities. Random noise is added to all fields to trigger the turbulent dynamics.

\section{Implementation and optimization}\label{sec:numerics}

While the time evolution is provided by a fourth-order Runge-Kutta algorithm and the spatial discretization by a fourth-order centered finite differences scheme, similarly to the GBS diverted version of Ref.~\cite{Paruta2018},  GBS now relies on an optimized iterative solver for the Poisson and Ampère equations. In addition, the kinetic neutral model, initially developed in the limited version of GBS~\cite{halpern2016,wersal2015} and ported here to the diverted geometry, is significantly optimized.

\begin{figure}
    \centering
    \includegraphics[scale=0.4]{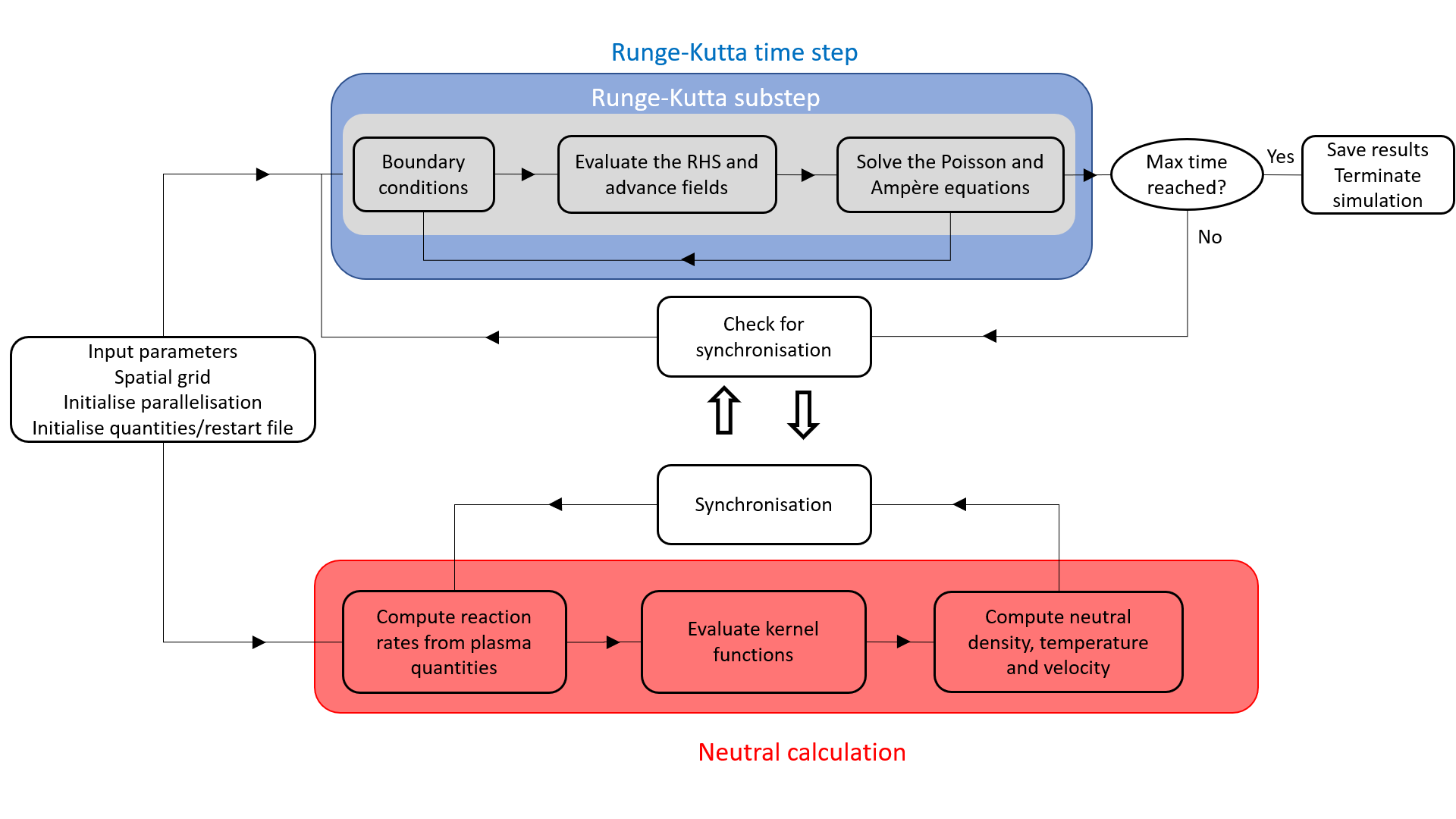}
    \caption{Workflow of the GBS code. The grey and blue areas highlight a Runge-Kutta substep and step, respectively, for the solution of the plasma equations. After every Runge-Kutta time step the plasma module checks if updated values of the neutral quantities are available. The workflow of a neutral calculation is highlighted by the red area. The synchronization is performed at the end of each neutral calculation. }
    \label{fig:workflow}
\end{figure}

We show in Fig.~\ref{fig:workflow} the workflow of GBS.
Following the initialization of the simulation, plasma and neutral modules are run simultaneously. 
Each Runge-Kutta substep, used to advance the GBS plasma equations, performs three tasks (see the grey area of Fig.~\ref{fig:workflow}). First, the boundary conditions are applied to every plasma quantity and the necessary spatial operators are applied to the plasma fields. 
Then, the right-hand side of the drift-reduced Braginskii equations is evaluated and Eqs.~\eqref{eqn:density}--\eqref{eqn:ion_temperature} are advanced, updating the values of $n$, $T_e$, $T_i$, $\Omega$, $U_{\parallel e}$ and $v_{\parallel i}$. 
Finally, the Poisson and Ampère equations, Eqs.~\eqref{eqn:poisson}~and~\eqref{eqn:ampere}, are numerically solved to update $\phi$ and $\psi$. 
After every Runge-Kutta time step (see the blue area of Fig.~\ref{fig:workflow}), the plasma module checks if the neutral calculation is completed and updates the value of the neutral-related terms in Eqs.~\eqref{eqn:density}--\eqref{eqn:ion_temperature} with the result of the new neutral calculation, if available.  
In parallel to the plasma evolution, the neutral density, temperature and velocity are computed by the neutral module (red area in Fig.~\ref{fig:workflow}). At the beginning of each neutral calculation, GBS computes the reaction rates, Eqs.~\eqref{eqn:rate_first}--\eqref{eqn:rate_last},  by using the current plasma quantities. Then, it evaluates the kernel functions, Eqs.~\eqref{eqn:kernel_first}--\eqref{eqn:kernel_last}, and finally it computes the neutral density, temperature and velocity by solving Eq.~\eqref{eqn:neutral_density}.
After every neutral calculation, the plasma and neutral modules synchronize, and the values of the plasma density, electron and ion temperatures used to compute the reaction rates is updated. 
A minimum neutral calculation frequency can be imposed in order to guarantee a good convergence of the simulation results, as detailed in Sec.~\ref{sec:conv}.

The remainder of the chapter is structured as follows. First, we describe the spatial discretization of the plasma equations. Then, we focus on the implementation and optimization of the Poisson and Ampère laws. Finally, we report on the implementation and optimization of the neutral model.

\subsection{Spatial discretization of the plasma equations}

All plasma quantities are evaluated on a uniform Cartesian grid, which discretizes the $R$, $\varphi$ and $Z$ coordinates, with a size $N_R\times N_Z\times N_\varphi$, where $N_R$, $N_Z$ and $N_\varphi$ are the number of grid points in the radial, vertical and toroidal direction, respectively. The neutral quantities are also discretized on a uniform grid in the $(R, \varphi, Z)$ cylindrical coordinate system, $N_R'\times N_Z'\times N_\varphi'$, that is in general coarser than the plasma grid in the poloidal plane ($N_R'<N_R$ and $N_Z'<N_Z$), while the neutral and plasma grid resolutions along the toroidal direction are the same ($N_\varphi'=N_\varphi$). 
The grid spacing is denoted as $\Delta R$, $\Delta Z$ and $\Delta \varphi$.
The grid for $n$, $T_{e}$, $T_{i}$, $\Omega$, and $\phi$, denoted as $\phi$-grid, is staggered in the vertical and toroidal directions with respect to the grid for $v_{\parallel e}$, $v_{\parallel i}$ and $\psi$, which is denoted as $v$-grid. 
More precisely, the $\phi$-grid is shifted by $\Delta Z/2$  and $\Delta\varphi/2$, along the vertical and toroidal directions, respectively, with respect to the $v$-grid.
The use of staggered grids prevents the formation of checkerboard patterns that can appear when treating an advection problem with a finite centered difference scheme, as shown in Ref.~\cite{patnakar1980}.
In the following, we denote as $f_{i,j,k}$ a quantity at the position $(R_i,\varphi_j,Z_k)$ in the $v$-grid and as $f_{i,j+\frac{1}{2},k+\frac{1}{2}}$ a quantity at the position $(R_i,\varphi_{j+\frac{1}{2}},Z_{k+\frac{1}{2}})$ in $\phi$-grid.

The differential operators in Eqs.~\eqref{eqn:curv}--\eqref{eqn:lapl} are computed as a linear combination of first and second derivatives along $R$, $Z$ and $\varphi$, with subsequent interpolation between staggered grids, if needed. 
For instance, the first derivative along $Z$ is discretized at the fourth-order by means of a 5-point stencil as
\begin{equation}
\label{eqn:der_first}
    (\partial_Z f)_{i,j,k} =\frac{1}{\Delta Z}\Bigl(\frac{1}{12}f_{i,j,k-2}-\frac{2}{3}f_{i,j,k-1}+\frac{2}{3}f_{i,j,k+1}-\frac{1}{12}f_{i,j,k+2}\Bigr)\,,
\end{equation}
if both the field and its derivative are evaluated on the $v$-grid. On the other hand, we have
\begin{equation}
\label{eqn:der_stag}
    (\partial_Z f)_{i,j+\frac{1}{2},k+\frac{1}{2}} =\frac{1}{\Delta Z}\Bigl(\frac{1}{24}f_{i,j,k-1}-\frac{9}{8}f_{i,j,k}+\frac{9}{8}f_{i,j,k+1}-\frac{1}{24}f_{i,j,k+2}\Bigr)\,,
\end{equation}
if the field is evaluated on the $v$-grid and its derivative on the $\phi$-grid. Analogous expressions to Eqs.~\eqref{eqn:der_first}~and~\eqref{eqn:der_stag} hold for fields evaluated on the $\phi$-grid, if their derivatives are evaluated on the $\phi$-grid or $v$-grid, respectively.
A quantity evaluated on the $v$-grid is interpolated to the $\phi$-grid by using a fourth-order interpolation,
\begin{equation}
    (I_Z f)_{i,j+\frac{1}{2},k+\frac{1}{2}} = -\frac{1}{16}f_{i,j,k-1}+\frac{9}{16}f_{i,j,k}+\frac{9}{16}f_{i,j,k+1}-\frac{1}{16}f_{i,j,k+2}\,,
\end{equation}
and analogous expression is used to interpolate from the $\phi$-grid to the $v$-grid.
The second derivative along Z is given by
\begin{equation}
    (\partial_{ZZ} f)_{i,j,k} = \frac{1}{\Delta Z^2}\Bigl(-\frac{1}{12}f_{i,j,k-2}+\frac{4}{3}f_{i,j,k-1}-\frac{5}{2}f_{i,j,k}+\frac{4}{3}f_{i,j,k+1}-\frac{1}{12}f_{i,j,k+2}\Bigr)\,,
\end{equation}
if both the field and its derivative are evaluated on the $v$-grid. Analogous expression holds if both the field and its derivative are evaluated on the $\phi$-grid.
The Poisson brackets, Eq.~\eqref{eqn:pb},  are discretized by means of a fourth-order Arakawa scheme~\cite{peterson2013}.
We note that the discretization at fourth-order of the curvature-related contributions in the gyroviscous terms, Eqs.~\eqref{eqn:ion_gyro}~and~\eqref{eqn:ele_gyro},  requires a 7-point stencil because of the presence of second derivatives on the $\phi$-grid of a quantity evaluated on the $v$-grid and vice versa. In order to use a 5-point stencil for all GBS operators, these derivatives are implemented at the second-order. For example, the second derivative on the $\phi$-grid of a field evaluated on the $v$-grid is given by
\begin{equation}
    (\partial_{ZZ} f)_{i,j+\frac{1}{2},k+\frac{1}{2}}=\frac{1}{\Delta Z^2}\Bigl(\frac{1}{2}f_{i,j,k-1}-\frac{1}{2}f_{i,j,k}-\frac{1}{2}f_{i,j,k+1}+\frac{1}{2}f_{i,j,k+2}\Bigr)\,.
\end{equation}

\subsection{Implementation and optimization of the Poisson and Ampère equations}
\label{sec:petsc}

In Ref.~\cite{Paruta2018}, the Boussinesq approximation is applied, thus considerably simplifying the implementation of the Poisson equation, and the electrostatic potential is computed by a direct inversion of the perpendicular laplacian operator, Eq.~\eqref{eqn:lapl}, through LU factorization  using the external MUMPS library~\cite{MUMPS}. 
Despite being computationally demanding, the LU factorization  is carried out once for all at the beginning of the simulation and therefore does not significantly impact the cost of the simulations. 

In the version of GBS described here, the Boussinesq approximation is relaxed, requiring the implementation at fourth-order of the $\nabla \cdot n \nabla_\perp\phi = \partial_R(n\partial_R\phi) + \partial_Z(n\partial_Z\phi)$ operator, i.e.
\begin{align}
    \bigl[\partial_R (n\partial_R \phi)\bigr]_{i,j,k}=&\frac{1}{\Delta R^2}\Bigl(\delta_{i-2,j,k}\phi_{i-2,j,k}+\delta_{i-1,j,k}\phi_{i-1,j,k}+\delta_{i,j,k} \phi_{i,j,k}\nonumber\\
    &+\delta_{i+1,j,k}\phi_{i+1,j,k} + \delta_{i+2,j,k}\phi_{i+2,j,k}\Bigr),
\end{align}
with
\begin{align}
    \delta_{i-2,j,k} =&\frac{n_{i-2,j,k}}{144}-\frac{n_{i-1,j,k}}{18}-\frac{n_{i,j,k}}{12}+\frac{n_{i+1,j,k}}{18}-\frac{n_{i+2,j,k}}{144}\,,\nonumber\\
    \delta_{i-1,j,k} =&-\frac{n_{i-2,j,k}}{18}+\frac{4n_{i-1,j,k}}{9}+\frac{4n_{i,j,k}}{3}-\frac{4n_{i+1,j,k}}{9}+\frac{n_{i+2,j,k}}{18}\,,\nonumber\\
    \delta_{i,j,k} =&-\frac{5n_{i,j,k}}{2}\,,\\
    \delta_{i+1,j,k} =&\frac{n_{i-2,j,k}}{18}-\frac{4n_{i-1,j,k}}{9}+\frac{4n_{i,j,k}}{3}+\frac{4n_{i+1,j,k}}{9}-\frac{n_{i+2,j,k}}{18}\,,\nonumber\\
    \delta_{i+2,j,k} =&-\frac{n_{i-2,j,k}}{144}+\frac{n_{i-1,j,k}}{18}-\frac{n_{i,j,k}}{12}-\frac{n_{i+1,j,k}}{18}+\frac{n_{i+2,j,k}}{144}\,,\nonumber
\end{align}
and similarly for $\bigl[\partial_Z(n\partial_Z\phi)\bigr]_{i,j,k}$.
Since $n$ depends on time, the $\delta$ coefficients vary in time, therefore the matrix that discretizes the Laplacian operator has to be assembled and factorized every time step, leading to a dramatic increase of the computational effort, which becomes prohibitively large already for a medium size grid if a LU factorization is used to solve Poisson equation. 
In fact, an initial profiling of GBS, carried out on a medium size poloidal grid ($N_R\times N_Z = 300\times 600$), corresponding to the one used to simulate a medium size tokamak such as TCV, shows that, if the MUMPS direct solver is used, more than 50\% of the time used to advance the plasma equations is spent in the factorization of the matrix that discretizes Poisson equation.  
We note that a similar matrix needs to be assembled and inverted every time step to solve the Ampère equation (see Eq.~\eqref{eqn:ampere}).
However,  the factorization of the Poisson matrix is computationally more expensive than the factorization of the matrix that discretizes Ampère equation.

In order to avoid the LU factorization  and optimize GBS, an iterative solver is used. 
We rely on the Data Management for Structured Grids (DMDA) of the PETSc library~\cite{petsc-web-page} that provides a flexible framework to implement a large number of solvers and preconditioners. 
By carrying out a scan on different solvers and preconditioners, analysing more than 2000 combinations, we find that the best performance is achieved with the algebraic multigrid preconditioner \texttt{boomerAMG} provided by the external package HYPRE~\cite{falgout2002} and the deflated \texttt{GMRES} solver~\cite{wakam2011}. 
In order to fine tune the preconditioner parameters, we focus on the main options of \texttt{boomerAMG}, i.e. the strong threshold, the maximum number of levels, the coarsen type, and the interpolation type (see Ref.~\cite{falgout2002} for details). The best performance is achieved by using a strong threshold of 0.25, a maximum of levels of 30, the \texttt{Falgout} coarsen type, and the \texttt{ext+i} interpolation type.  

\begin{figure}
    \centering
    \includegraphics[scale=0.65]{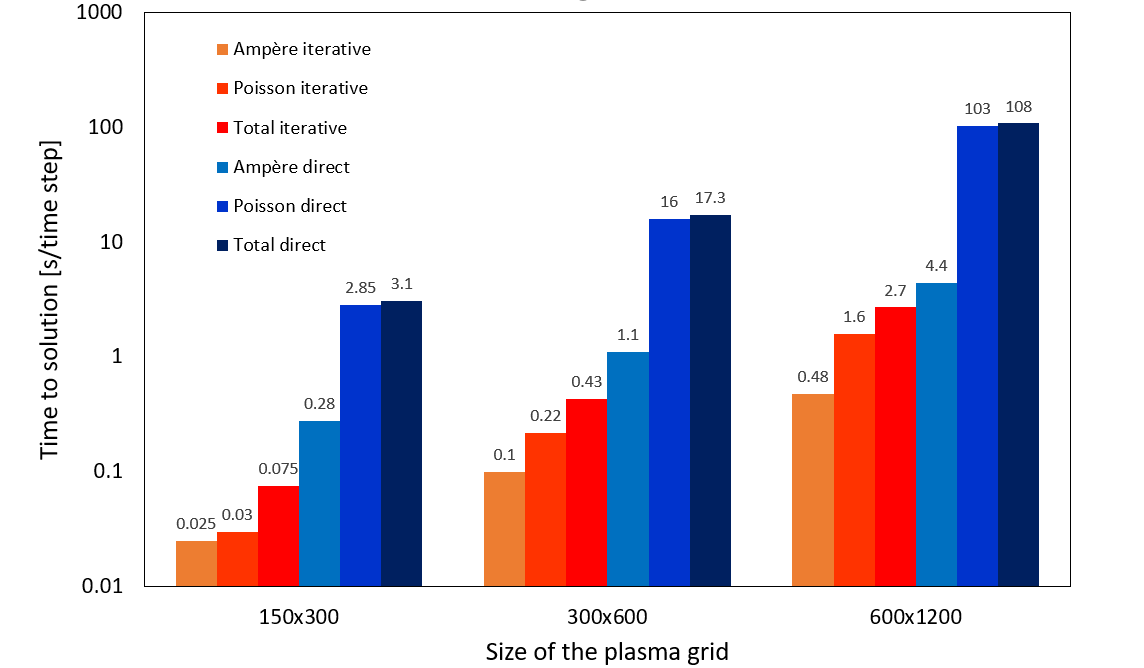}
    \caption{Time to solution per time step averaged from simulations of 10 time steps that consider different poloidal grid sizes. No coupling to neutral dynamics is considered here. The simulations are carried out on one node (36 cores) of the multi-core partition of Piz Daint (Cray XC40 equipped with two 18-core Intel Xeon E5-2695 v4 CPUs at 2.10GHz) and are performed taking as initial conditions the results of a simulation in turbulent state. This allows us to compare the solvers in typical working conditions. The iterative and direct solvers are based on the PETSc and MUMPS libraries, respectively. The time of MUMPS includes the factorization.}
    \label{fig:petsc_vs_mumps}
\end{figure}

In order to evaluate the performance gain arising from the iterative solver implemented with the PETSc library with this set of optimized parameters, we run a set of GBS simulations considering the $N_R\times N_Z = 150\times 300$, $N_R\times N_Z = 300\times 600$ and $N_R\times N_Z = 600\times 1200$ grids. 
We refer to these grids as the half-TCV, TCV and double-TCV grids, since $N_R\times N_Z = 300\times 600$ is the typical poloidal grid used for a simulation of a TCV discharge with toroidal magnetic field of 0.9~T.
The time to solution per time step is shown in Fig.~\ref{fig:petsc_vs_mumps} for the different grid sizes.  
Across all sizes considered here, we observe a speed-up of, approximately, a factor of 40 when the iterative solver is used, with respect to a direct solver, reducing considerably the cost of our simulations and making the simulation at the TCV scales possible, otherwise prohibitively expensive. 
As an aside, we also note that the solution of the Poisson and Ampère equations using the iterative solver and the evaluation of the right-hand side of Eqs.~\eqref{eqn:density}-\eqref{eqn:ion_temperature} have approximately the same computational cost in the half-TCV and TCV simulations.
On the other hand, more than half of the computational time is spent to solve the Poisson equation in the double-TCV simulation, even using the iterative solver (see Fig.~\ref{fig:petsc_vs_mumps}). 
Simulations at scales larger than TCV require therefore further optimization. 

\subsection{Implementation and optimization of neutral model in diverted geometry}\label{sec:neutrals_impl}

We summarize here the implementation and the optimization of the neutral model.
The discretization of the kernel functions, Eqs.~\eqref{eqn:kernel_first}--\eqref{eqn:kernel_last}, allows us to write the equations for the neutral density, Eq.~\eqref{eqn:neutral_density}, and neutral flux to the wall, Eq.~\eqref{eqn:neutral_flux},  as a linear system
\begin{equation}
\label{eqn:neutral_system}
    \begin{bmatrix}
    n_\text{n}\\
    \Gamma_\text{out,n}
    \end{bmatrix}=
    \begin{bmatrix}
    \nu_\text{cx} K_{p\rightarrow p} & (1-\alpha_\text{refl}) K_{b\rightarrow p}\\
    \nu_\text{cx} K_{p\rightarrow b} & (1-\alpha_\text{refl}) K_{b\rightarrow b}
    \end{bmatrix}
    \begin{bmatrix}
    n_\text{n}\\
    \Gamma_\text{out,n}
    \end{bmatrix}+
    \begin{bmatrix}
    n_{\text{n[rec]}}+n_\text{n[out,i]}\\
    \Gamma_\text{out,n[rec]}+\Gamma_\text{out,n[out,i]}
    \end{bmatrix}\,.
\end{equation}
The linear system of Eq.~\eqref{eqn:neutral_system} can then be solved by inverting the kernel matrix,
\begin{equation}
\label{eqn:k_matrix}
    K=\begin{bmatrix}
    \nu_\text{cx} K_{p\rightarrow p} & (1-\alpha_\text{refl}) K_{b\rightarrow p}\\
    \nu_\text{cx} K_{p\rightarrow b} & (1-\alpha_\text{refl}) K_{b\rightarrow b}
    \end{bmatrix}\,,
\end{equation}
which, in the limit of large aspect ratio considered here, has $[N_R'N_z'+2(N_R'+N_Z')]^2$ elements and is inverted for each poloidal plane at every neutral calculation.
Since our simulation domain encompasses the whole plasma volume, contrary to the previous neutral version of GBS, the $K$ matrix is a dense matrix with, in principle, all its elements strictly positive. 
However, since the value of the matrix elements decays exponentially with the distance between the two connected points, becoming negligible when their distance is several neutral mean free paths, we introduce a threshold value below which the matrix element is considered as vanishing. 
Since the evolution of the neutral dynamics is computationally more expensive than the evolution of the plasma dynamics, we recalculate the neutral quantities over a time interval longer than the time step used to evolve the plasma equation, but still smaller than the turbulence timescale. 

The solution of the kinetic equation is a two steps process. First, the elements of the $K$ matrix are computed, requiring the evaluation of the kernel functions, Eqs.~\eqref{eqn:kernel_first}--\eqref{eqn:kernel_last}. 
In the present version of GBS, the evaluation of $K$ is optimized by improving the code vectorisation and avoiding unnecessary computation.  Then, the $K$ matrix is inverted to compute the neutral density by using Eq.~\eqref{eqn:neutral_system}.
The matrix inversion is improved by avoiding an expensive LU factorization. 

We first focus on the evaluation of the $K$ matrix. This requires to compute $\left[N_R'N_Z' + 2(N_R'+N_Z')\right]^2$ elements, all of them involving the evaluation of one of the four kernel functions defined in Eqs.~\eqref{eqn:kernel_first}--\eqref{eqn:kernel_last}, a computationally expensive evaluation.
In fact, approximately 65\% of a neutral step is spent in the computation of the $K$ matrix elements (the remaining time is spent in the solution of the associated linear system). 
In the version of GBS presented in Ref.~\cite{wersal2015}, each element of the $K$ matrix was evaluated by using the same routine that included logic conditions used to identify points belonging to the plasma volume or the wall.
Here, the logic conditions are moved outside the routine that evaluates the elements of $K$, which is split into four specialized routines, each of them computing one of the kernel functions in Eqs.~\eqref{eqn:kernel_first}-\eqref{eqn:kernel_last}. 
Therefore, points belonging to the plasma and the wall are separated at the beginning of the simulation and a whole subset of $K$ is evaluated instead of computing each single element.
In addition, the modified Bessel function of second kind involved in the $K_{b\rightarrow p}$ and $K_{b\rightarrow p}$ kernels is pre-computed at the beginning of the simulation and later re-used at the cost of a memory access. 
Finally, the computation of the integral  $\int_0^{r_\perp'}\nu_\text{eff}(\mathbf{x}_\perp'')\mathrm{d}r_\perp''$ appearing  in the four kernel functions, Eqs.~\eqref{eqn:kernel_first}--\eqref{eqn:kernel_last}, is improved. This requires first the interpolation of the plasma quantities necessary to compute $\nu_\text{eff}$ along the neutral trajectory and then the numerical evaluation of the integral.
The interpolation is performed by using a second-order method that is carefully implemented to enable the compiler auto-vectorisation. 
In contrast to Ref.~\cite{wersal2015}, here we use the same routine to integrate along the neutral trajectory and over the velocity space. Moreover, integrals can be evaluated by either the (left) rectangular or midpoint rules. 

We focus now on the inversion of the $K$ matrix in Eq.~\eqref{eqn:neutral_system}.
We note that the $K$ matrix evolves in time, similarly to the matrix associated with the Poisson and Ampère equations.
While a direct solver based on the MUMPS library was used in Ref.~\cite{wersal2015}, requiring an expensive LU factorization, here we implement an iterative solver. Also in this case, we choose to use the framework provided by the PETSc library, opting for the use of the \texttt{GMRES} solver without preconditioner.

\begin{figure}
    \centering
    \includegraphics[scale=0.6]{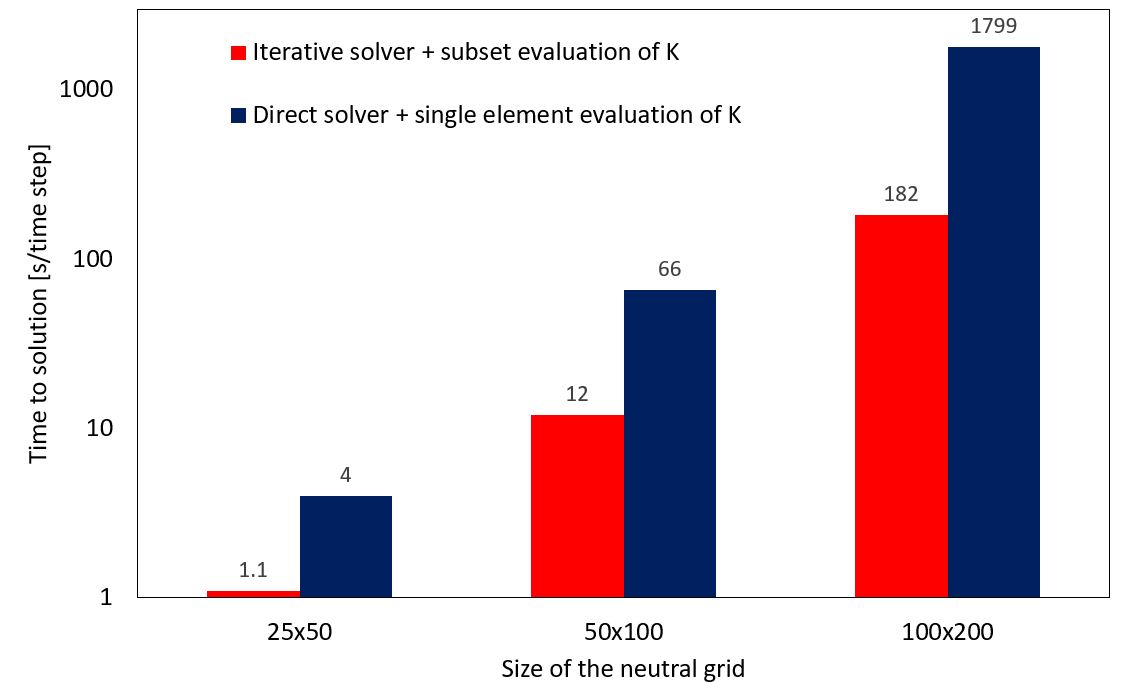}
    \caption{Time to solution per neutral calculation of the implementation based on an iterative solver and the evaluation of a subset of the $K$ matrix, compared to the implementation based on a direct solver and the evaluation of single elements of the $K$ matrix. Tests are carried out for different neutral grid sizes (one poloidal plane) on one computing node of the multi-core partition of Piz Daint (Cray XC40 equipped with two 18-core Intel Xeon E5-2695 v4 CPUs at 2.10 GHz). The implementation of the iterative and direct solvers are based on the PETSc and MUMPS libraries, respectively.}
    \label{fig:neutral_opt}
\end{figure}

In order to evaluate the improvement of performance arising from the new implementation, we analyze the time to carry out one neutral step with various neutral grid sizes. We consider a coarse neutral grid of size $N_R'\times N_Z'=25\times50$, a medium neutral grid of size $N_R'\times N_Z'=50\times 100$, and a fine neutral grid of size $N_R'\times N_Z'=100\times 200$, which is typical of a simulation of a TCV discharge. 
In Fig.~\ref{fig:neutral_opt}, the time to solution per neutral time step of the implementation that evaluates single elements of the $K$ matrix and uses the direct solver is compared to the one of the implementation that evaluates a whole subset of the $K$ matrix and uses the iterative solver.
We note that the speed-up arising from the new implementation increases with the size of the system, ranging from a factor of four to a factor of ten.

\section{Verification of GBS implementation}\label{sec:verification}

We report here on the verification of the plasma and neutral implementation. We verify the two modules separately, as presented below.

\subsection{Verification of the plasma model}

Similarly to the past versions, we verify the plasma model implementation in the present version of GBS by using the method of manufactured solutions.
Here, we briefly describe the MMS. A detailed description can be found in Ref.~\cite{riva2014} where the MMS was applied for the first time to verify a plasma turbulence code. 

In order to verify the numerical implementation of a model $M$, the analytical solution $s$ of the model $M$, which satisfies $M(s)=0$, is compared to the numerical solution $s_h$ of the discretized model $M_h$, where $h$ is the discretization parameter. The code is successfully verified if the discretization error $e_h=||s-s_h||\to h^p$ as $h\to 0$, with $p$ the order of the adopted discretization scheme.
Since the analytical solution of a model equations is in general unknown, the main idea behind the MMS is to manufacture an arbitrary analytical function, $u$, and evaluate $S=M(u)$.
We remark that, in general, $S\neq 0$, since $u$ is not the solution of $M$. On the other hand, the arbitrary function $u$ is solution of the model $N$, defined as $N(u)=M(u)-S=0$. Since $S$ can be analytically evaluated, $N_h$ and $M_h$ are affected by the same discretization error. Therefore, verifying $N_h$ is equivalent to verify $M_h$, with the discretization error given by $e_h=||u-u_h||$.

In the following, we report for the first time the verification of the implementation in GBS of the plasma model that avoids the Boussinesq approximation and includes electromagnetic effects. 
We consider a simulation domain with a rectangular poloidal cross section of size $L_R=37.5\,\rho_{s0}$ and $L_Z=50\,\rho_{s0}$, in the radial and vertical directions, respectively. The verification is carried out with $\rho_*^{-1}=100$, $\nu=1$, $\beta_{e0}=10^{-4}$, $\tau=1$ and $m_i/m_e=1$. 
The dimensionless parameters are chosen so that the terms on the right-hand side of Eqs.~\eqref{eqn:density}--\eqref{eqn:ion_temperature} are all of the same order of magnitude. 

The magnetic field considered for the verification is analytically obtained by solving the Biot-Savart law in the infinite aspect ratio limit for a current density with a Gaussian distribution inside the simulation domain, which guarantees continuity and derivability of the magnetic field, and an additional current filament outside the simulation domain, which produces the X-point. This leads to 
\begin{equation}
\label{eqn:psi}
    \begin{split}
    \Psi(R,Z)=\frac{I_0}{2}\biggl\{\log\biggl[\frac{(R-R_1)^2+(Z-Z_1)^2}{\rho_{s0}^2}\biggr]+\mathrm{EI}\biggl[\frac{(R-R_1)^2+(Z-Z_1)^2}{\sigma_0^2}\biggr]\\+\log\biggl[\frac{(R-R_1)^2+(Z-Z_2)^2}{\rho_{s0}^2}\biggr]\biggr\}\,,
    \end{split}
\end{equation}
where $\mathrm{EI}(x)$ is the exponential integral function, with $I_0=40\,\rho_{s0}^2B_0$, $\sigma_0=6.25\,\rho_{s0}$, $R_1=100\,\rho_{s0}$, $Z_1= 0$, and $Z_2=-40\,\rho_{s0}$.
The manufactured solutions for the evolved scalar quantities $u=n, T_{e}, T_{e}, v_{\parallel e}, v_{\parallel i}, \Omega, \phi, \psi$ are chosen as
\begin{equation}
\label{eqn:manu_sol}
    u_M(R,Z,\varphi,t) = A_u \bigl[ B_u + \sin(C_u Z +\alpha_u)\sin(D_u \varphi + \beta_u)\sin(E_u t + F_u R + \gamma_u )\bigr]\,,
\end{equation}
where $A_u$, $B_u$, $C_u$, $D_u$, $E_u$, $F_u$, $\alpha_u$, $\beta_u$, and $\gamma_u$ are arbitrary constants whose value is chosen to excite all terms in the right-hand side of Eqs.~\eqref{eqn:density}--\eqref{eqn:ampere}, ensuring that none of them provides a dominating contribution to the numerical error.
The source terms can be computed by substituting Eq.~\eqref{eqn:manu_sol} into Eqs.~\eqref{eqn:density}--\eqref{eqn:ampere}. This process is carried out by using the symbolic calculation offered by the Mathematica software package~\citep{Mathematica} and the analytical expressions of the source terms are directly converted to the Fortran language of the GBS code.  

In order to decouple the tolerance error associated with the iterative solver from the discretization error, we first consider the verification of GBS where Poisson and Ampère equations are solved with a direct method. 
As a second step, we discuss the verification of the iterative solver.
We do not include the curvature-related contributions appearing in the gyroviscous terms, Eqs.~\eqref{eqn:ion_gyro}~and~\eqref{eqn:ele_gyro}, since they are implemented at second-order.
These terms are verified independently. 

\begin{figure}
    \centering
    \subfloat[]{\includegraphics[width=0.46\textwidth]{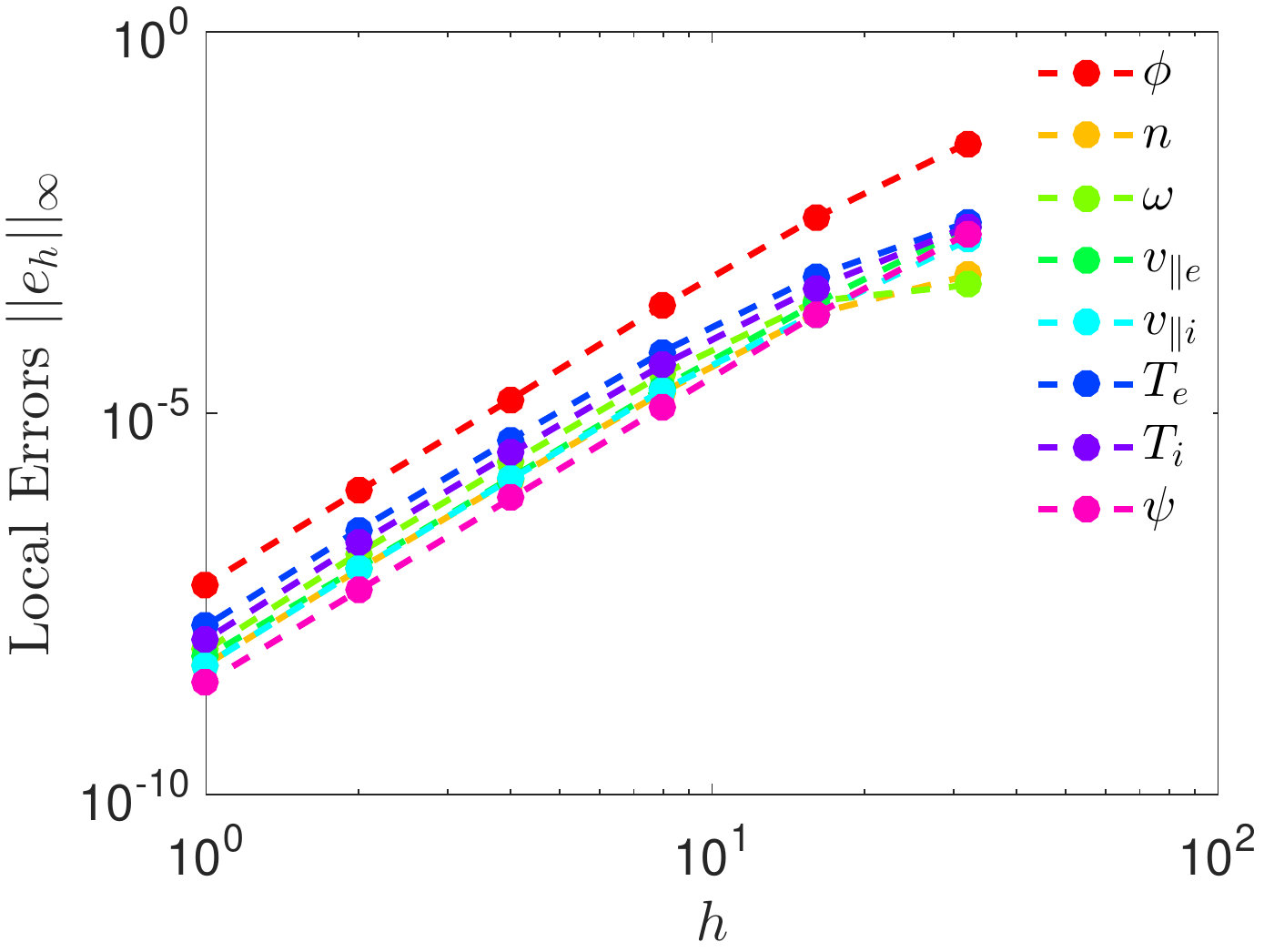}}\quad
    \subfloat[]{\includegraphics[width=0.46\textwidth]{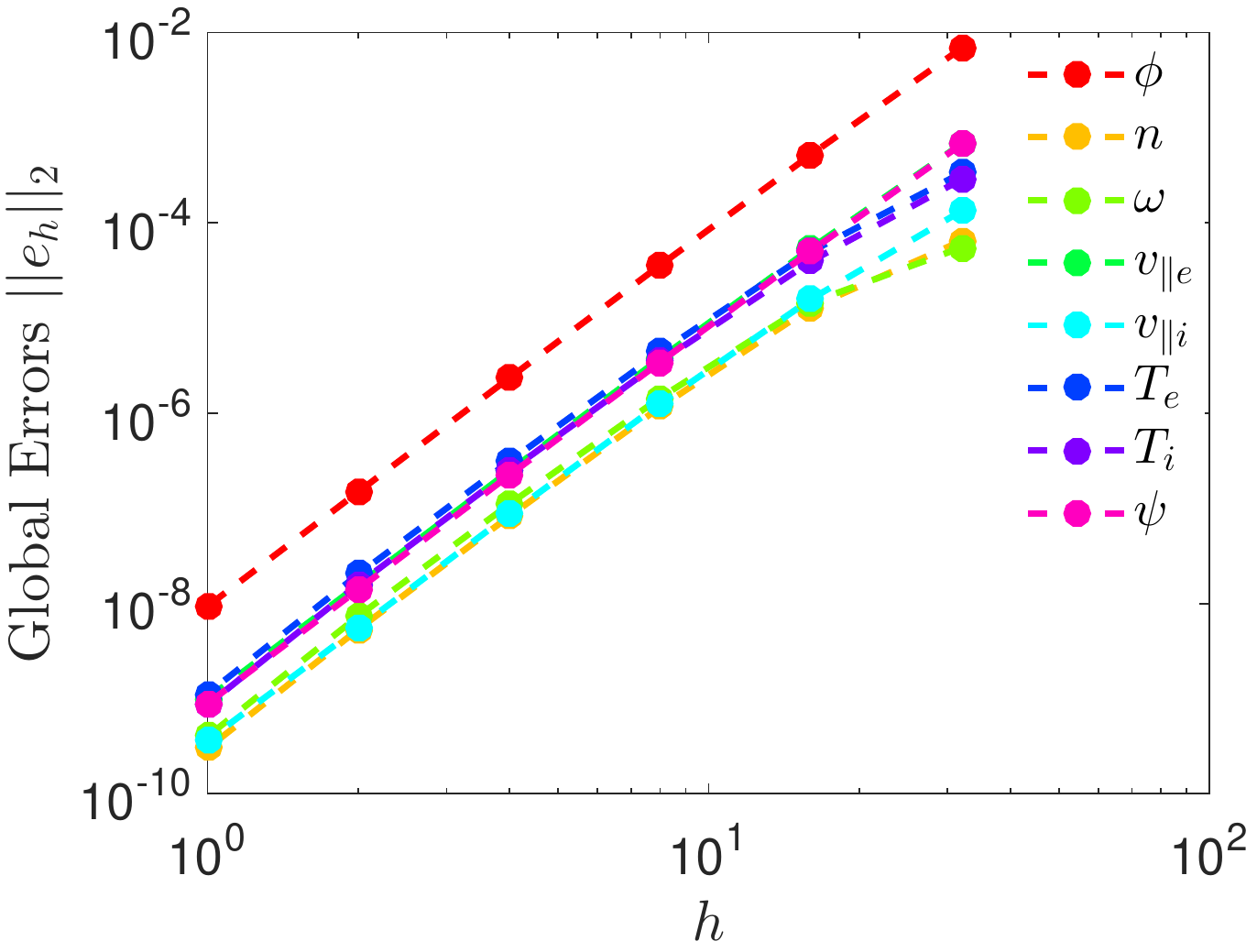}}\\
    \subfloat[]{\includegraphics[width=0.45\textwidth]{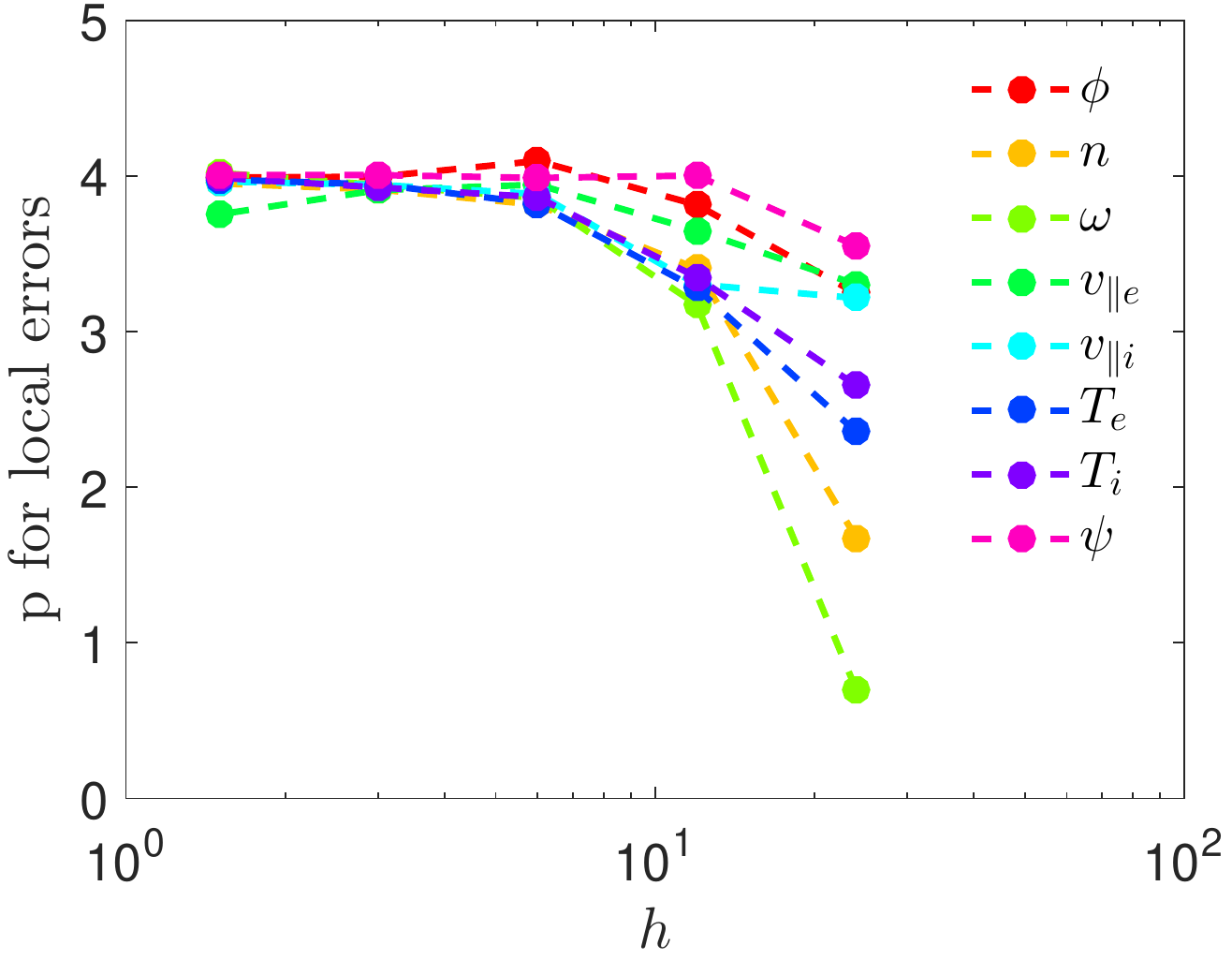}}\quad
    \subfloat[]{\includegraphics[width=0.45\textwidth]{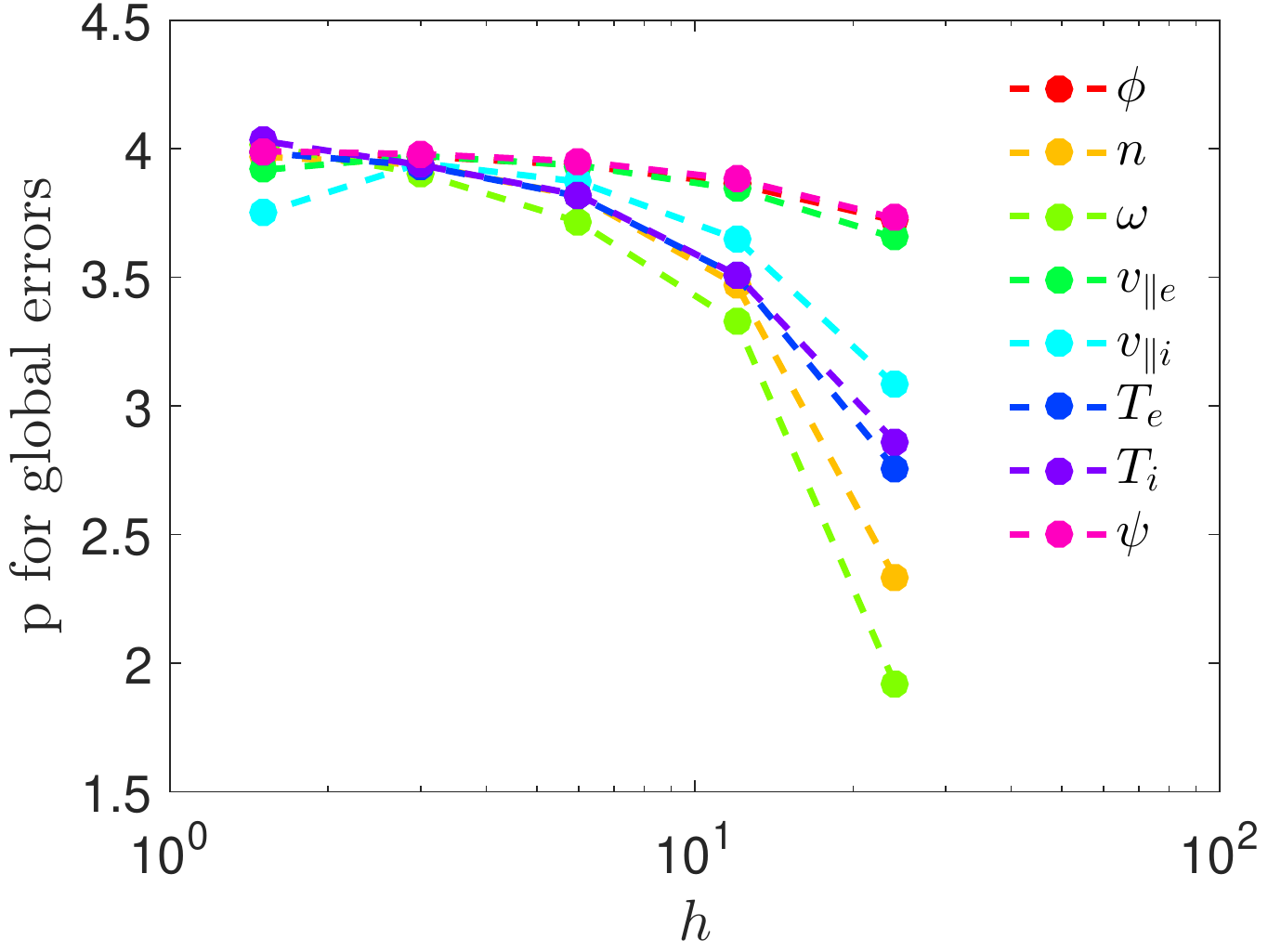}}
    \caption{Local, $L_\infty$, (a) and global, $L_2$, (b) norms of the discretization error as a function of the grid resolution parameter $h= \Delta R/\Delta R_0 = \Delta Z/\Delta Z_0 = \Delta \varphi/\Delta \varphi_0 = \Delta t/\Delta t_0$, where $\Delta R_0$, $\Delta Z_0$, and $\Delta \varphi_0$ is the grid spacing for the $N_R\times N_Z\times N_\varphi = 256 \times 256 \times 256$ grid with time step $\Delta t_0 = 6.25 \times 10^{-6}$.
    The order  of convergence $p$ is also shown for the local (c) and the global (d) norms of the discretization errors. As expected from the order of convergence of the numerical scheme, $p$ tends to 4 as $h$ decreases. Results are shown in dimensionless units. }
    \label{fig:verification}
\end{figure}

Figs.~\ref{fig:verification}~(a)~and~(b) show the $L_2$ and $L_\infty$ norms of the discretization errors at $t=0.01$. We consider various grid refinements as a function of the discretization parameter $h= \Delta R/\Delta R_0 = \Delta Z/\Delta Z_0 = \Delta \varphi/\Delta \varphi_0 = \Delta t/\Delta t_0$, where $\Delta R_0$, $\Delta Z_0$, and $\Delta \varphi_0$ is the spacing of the $N_R\times N_Z\times N_\varphi = 256 \times 256 \times 256$ grid, with time step $\Delta t_0 = 6.25 \times 10^{-6}$. The coarsest grid considered for the verification is $N_R\times N_Z\times N_\varphi = 8 \times 8 \times 8$ with $\Delta t = 2 \times 10^{-4}$. 
The order of accuracy, 
\begin{equation}
    p= \frac{\log(e_{rh}/e_h)}{\log{r}}\,,
\end{equation}
with $rh$ indicating the coarsening of the temporal and spatial grids by a factor of $r$, is shown in Figs.~\ref{fig:verification}~(c)~and~(d).
Since a fourth-order numerical scheme is used to discretize both space and time, we expect the discretization error to decrease as $h^4$ when $h\to 0$, in good agreement with the result of Fig.~\ref{fig:verification} that shows the convergence to $p=4$ for both the $L_2$ and $L_\infty$ norms as $h$ decreases. 

\begin{figure}
    \centering
    \subfloat[]{\includegraphics[width=0.45\textwidth]{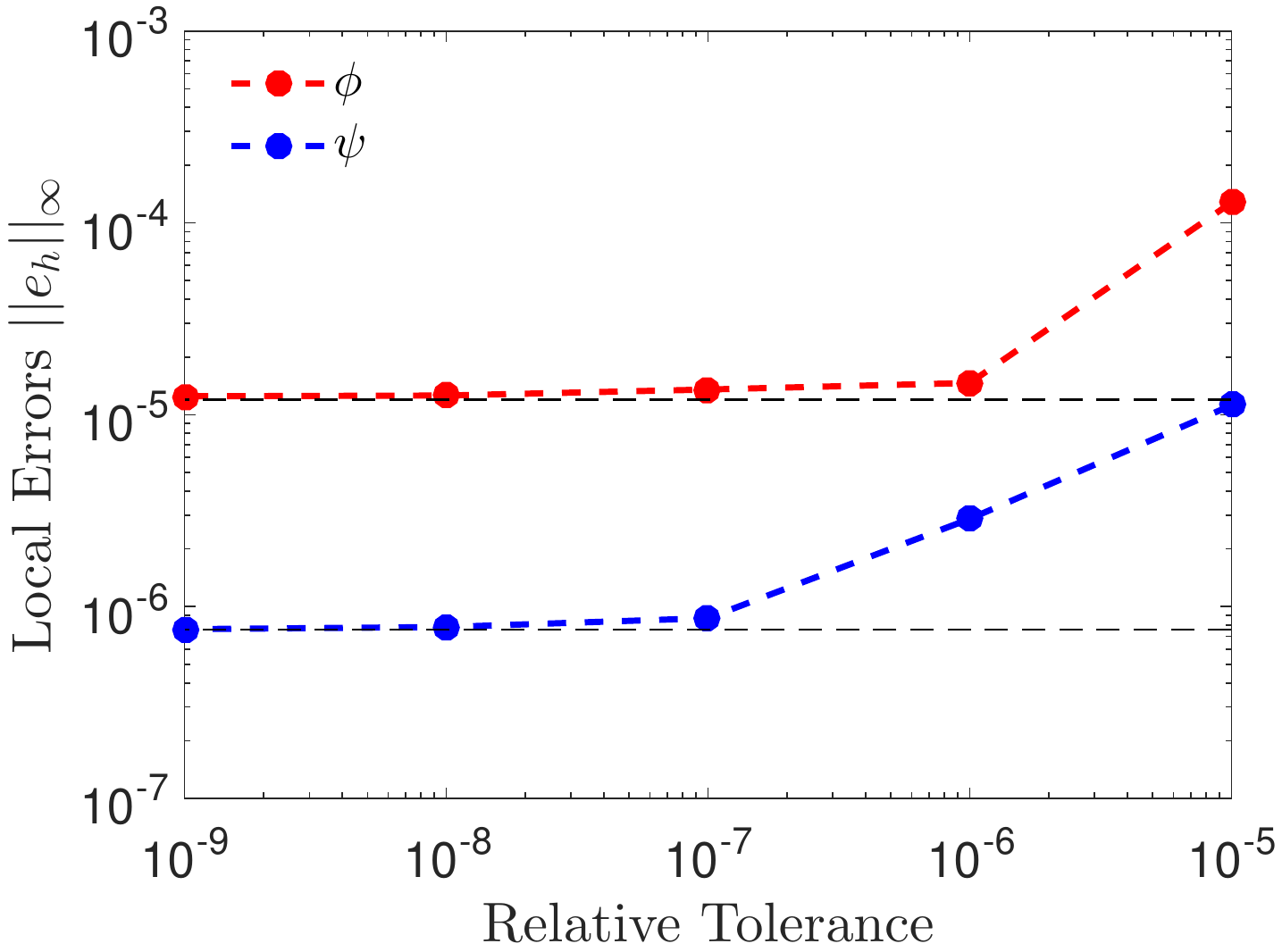}}\quad
    \subfloat[]{\includegraphics[width=0.45\textwidth]{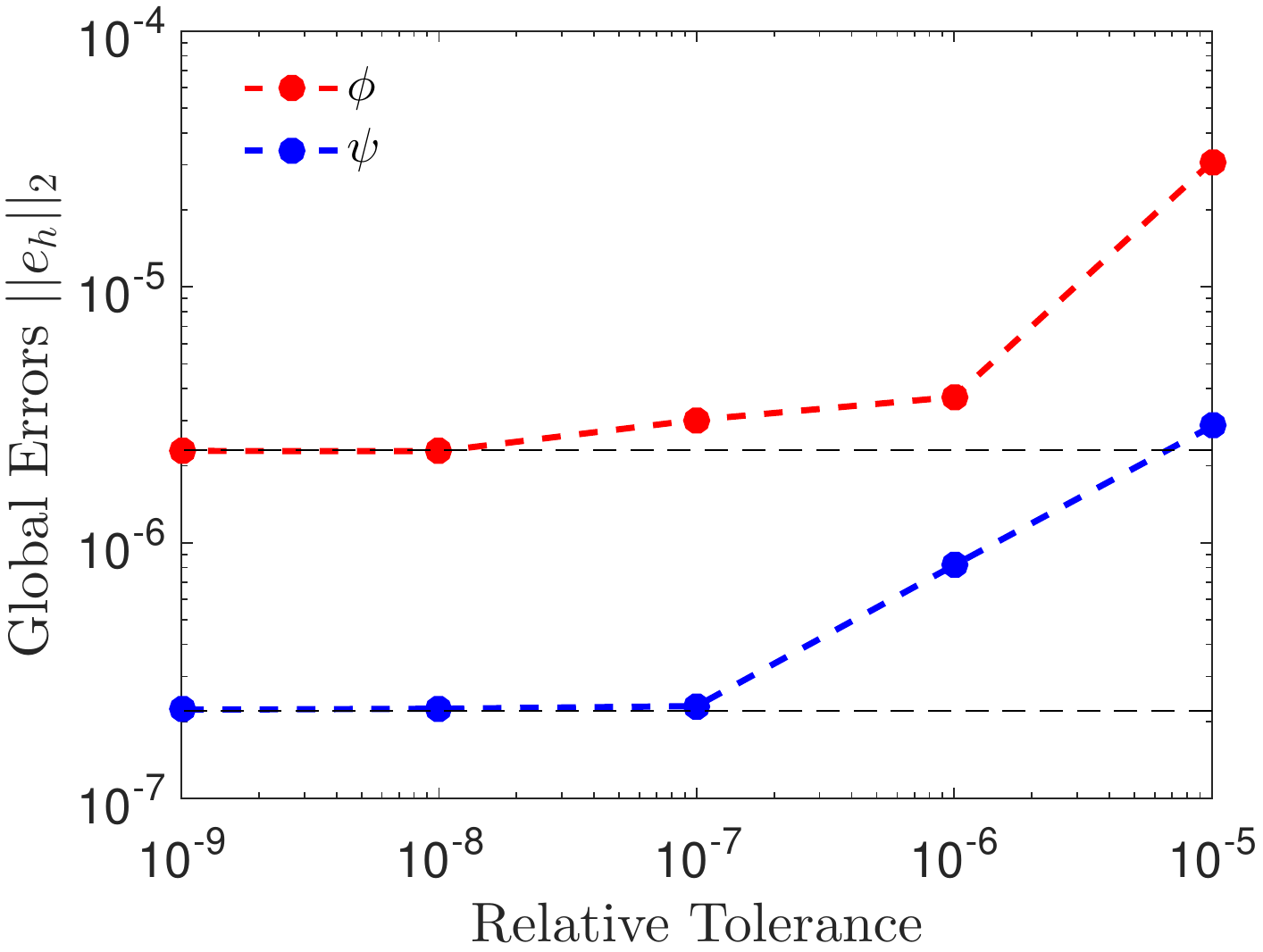}}
    \caption{Local (a) and global (b) errors of $\phi$ and $\psi$ as a function of the relative tolerance of the iterative solver based on the PETSc library for the grid $N_R\times N_Z\times N_\varphi = 64 \times 64 \times 64$, and $L_R=37.5\,\rho_{s0}$ and $L_Z=50\,\rho_{s0}$. The horizontal dashed lines represent the discretization error when the direct solver based on the MUMPS library is used. Convergence is achieved for values of the relative tolerance below $10^{-7}$. Results are show in dimensionless units.}
    \label{fig:verif_petsc}
\end{figure}

We discuss now the verification of the iterative solver. The error affecting the solution of an iterative solver is the combination of the discretization error and the tolerance error, the last depending on the tolerance threshold. We consider the solution of Poisson and Ampère equations and we compare the results of the direct and iterative solvers based on the MUMPS and PETSc libraries.  
Fig.~\ref{fig:verif_petsc} shows the global and local errors of $\phi$, for Poisson equation, and $\psi$, for Ampère equation, as a function of the relative tolerance of the iterative solver. We note that the errors decrease with the relative tolerance until a value of $10^{-7}$. Below this value, the discretization error dominates, independently of the solver tolerance.

\subsection{Verification of the neutral model}\label{sec:verification_neut}

We verify here, for the first time, the implementation of the neutral model. 
We proceed in two steps. First, we verify the correct implementation of the routine used to perform integrals over a neutral trajectory and over the velocity space, which appear in the kernel functions, Eqs.~\eqref{eqn:kernel_first}--\eqref{eqn:kernel_last}. Second, we focus on the construction and inversion of the matrix $K$, Eq.~\eqref{eqn:k_matrix}, used to evaluate the neutral density.

\begin{figure}
    \centering
    \subfloat[]{\includegraphics[width=0.45\textwidth]{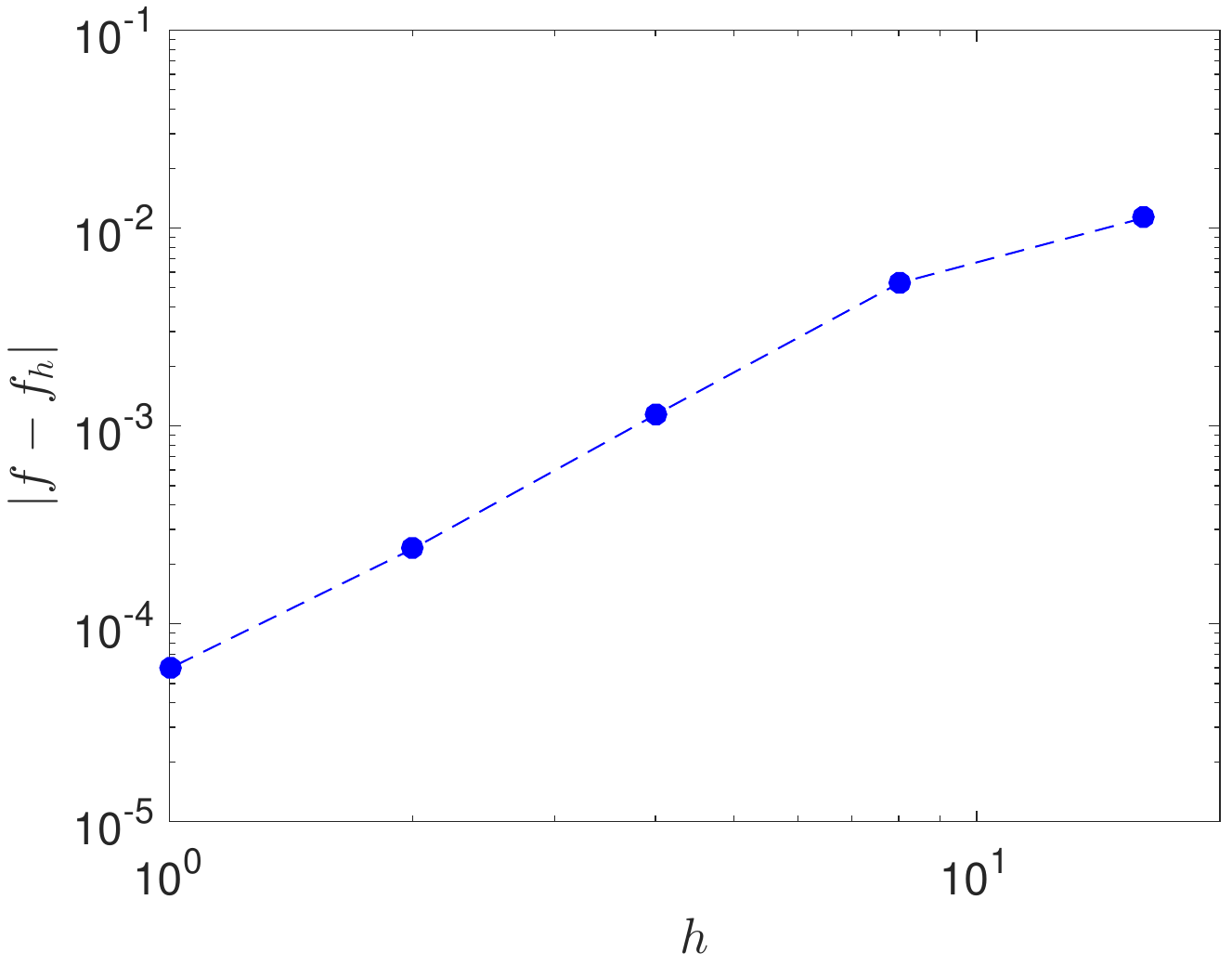}}\quad
    \subfloat[]{\includegraphics[width=0.45\textwidth]{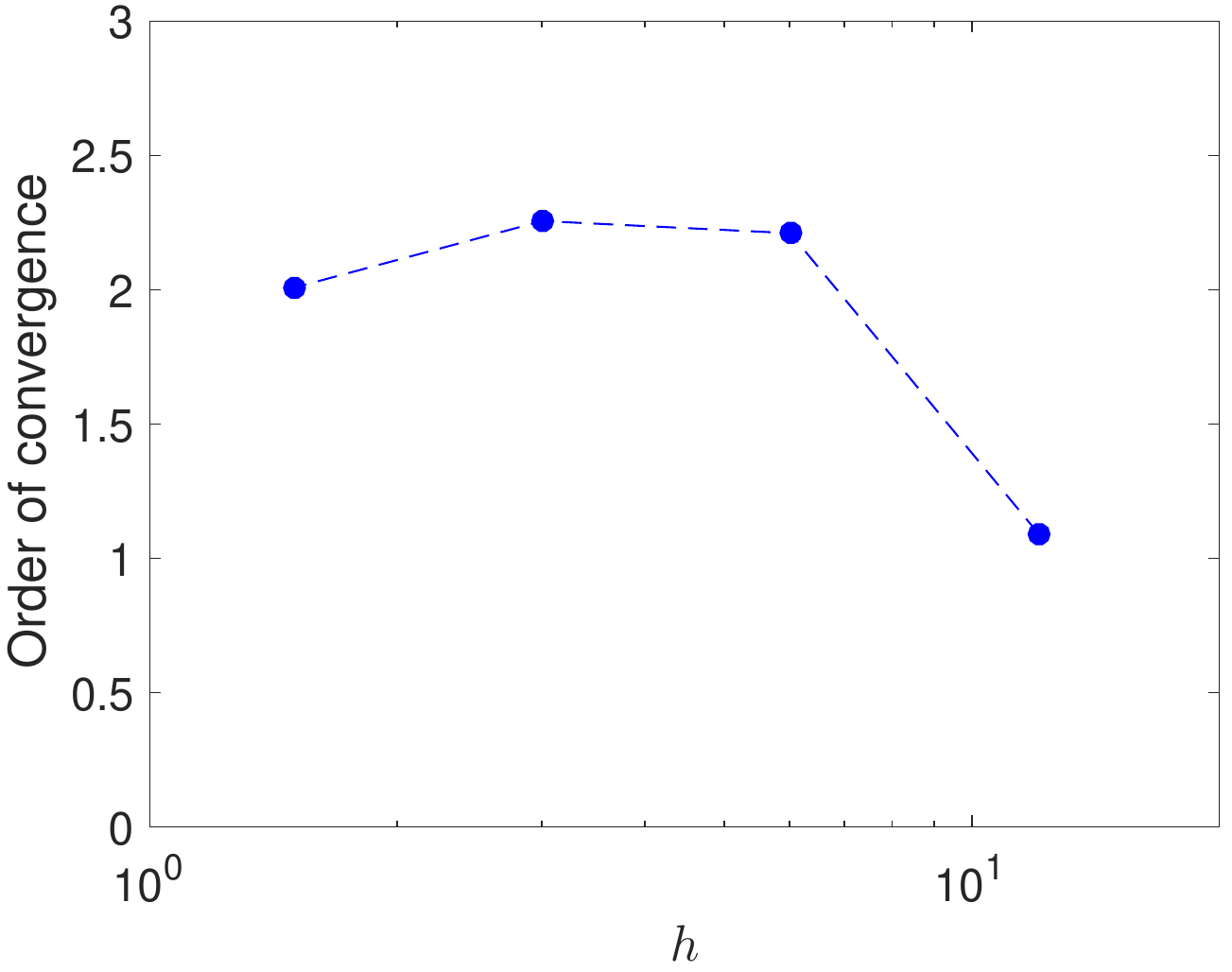}}
    \caption{Discretization error of the interpolated test function $f(x)=2x^2+4y^2$, discretized on an arbitrary domain $x\in [x_0,x_1]$ and $y\in [y_0,y_1]$, as a function of the discretization parameter $h=\mathrm{d}x/\mathrm{d}x_0=\mathrm{d}y/\mathrm{d}y_0$ at the position $x_0=-10$, $x_1=10$, $y_0=-7$ and $y_1=7$ (a), and the corresponding order of convergence (b).  The expected order of convergence to the analytical result is correctly retrieved.}
    \label{fig:interpolation_convergence}
\end{figure}

\begin{figure}
    \centering
    \subfloat[]{\includegraphics[width=0.45\textwidth]{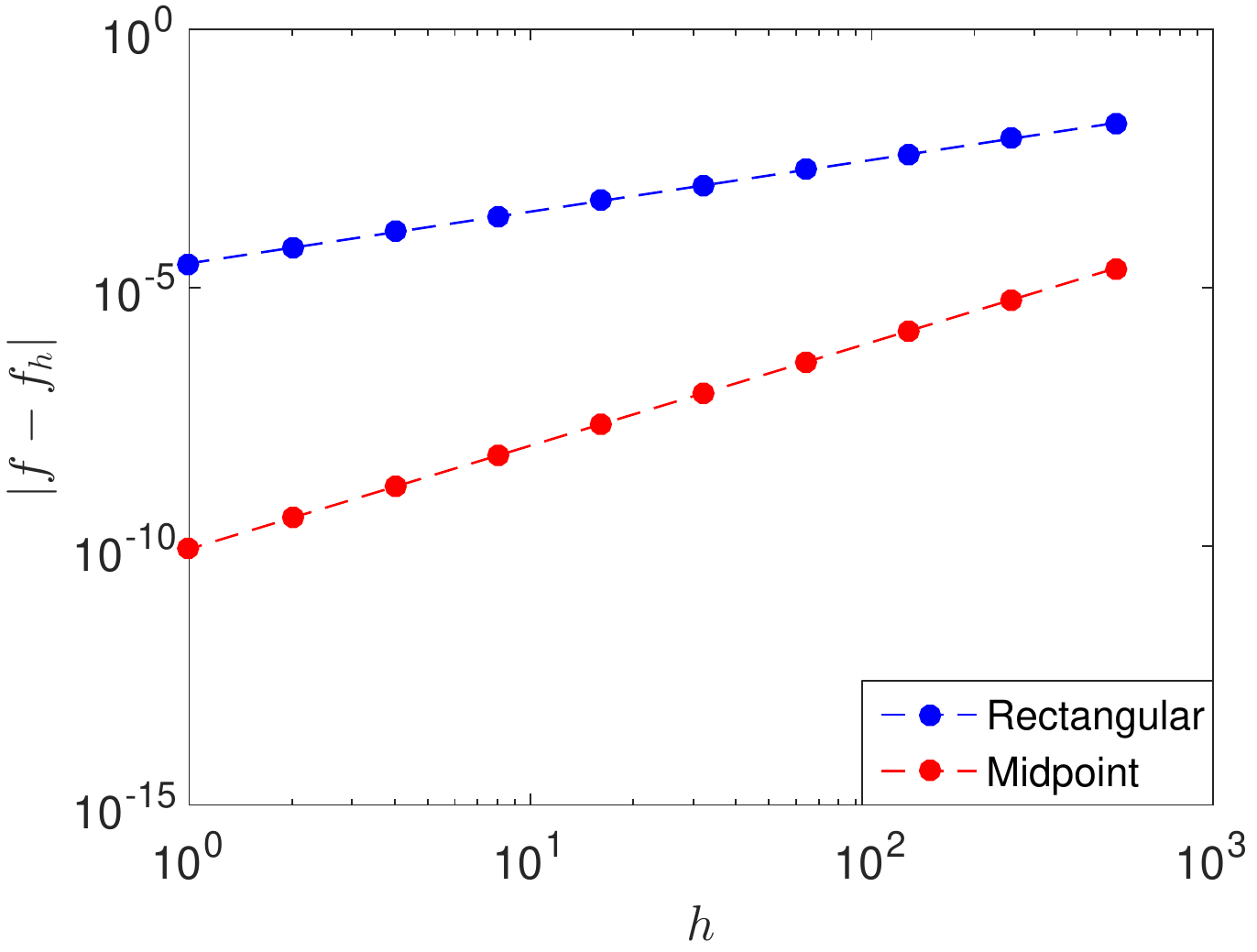}}\quad
    \subfloat[]{\includegraphics[width=0.45\textwidth]{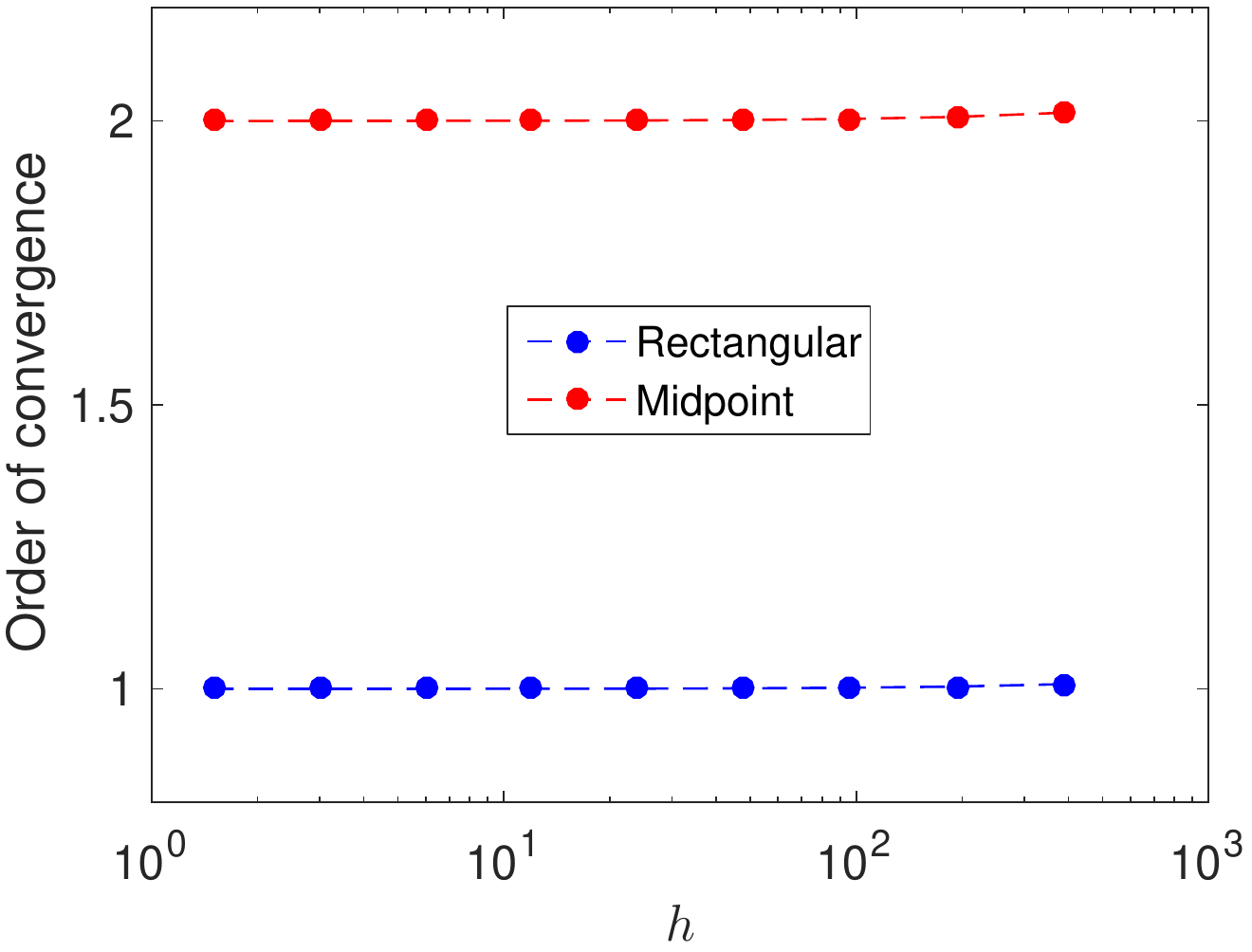}}
    \caption{Result of the unit test to verify the numerical integration routine used in GBS. The discretization error resulting from the numerical integration of the test function $f(x)=\cos x$ (a) and the corresponding order of convergence (b) are shown as a function of the discretization parameter $h=\mathrm{d}x/\mathrm{d}x_0$. The integral is performed between $x=1$ and $x=3$. The expected order of convergence to the analytical result is correctly retrieved for both the rectangular and the midpoint rules.}
    \label{fig:integration_convergence}
\end{figure}

To carry out the first verification step, we perform unit tests that allow us to verify individually the routines used to numerically integrate and to interpolate along a neutral trajectory.
The interpolating routine is tested on an analytical function $f$, considering various resolutions.
Fig.~\ref{fig:interpolation_convergence} shows the typical discretization error in a point of $f$ domain as a function of the grid resolution. The order of convergence is correctly retrieved as $h$ decreases.
The integration routine is tested by comparing the numerical and analytical result of the integral of an analytical function $f$ over an arbitrary interval. The discretization error as well as the order of convergence is shown in Fig.~\ref{fig:integration_convergence}.
The results of the unit tests point out the correct implementation of the routines used to evaluate integrals along a neutral trajectory.

\begin{figure}
    \centering
    \subfloat[]{\includegraphics[width=0.45\textwidth]{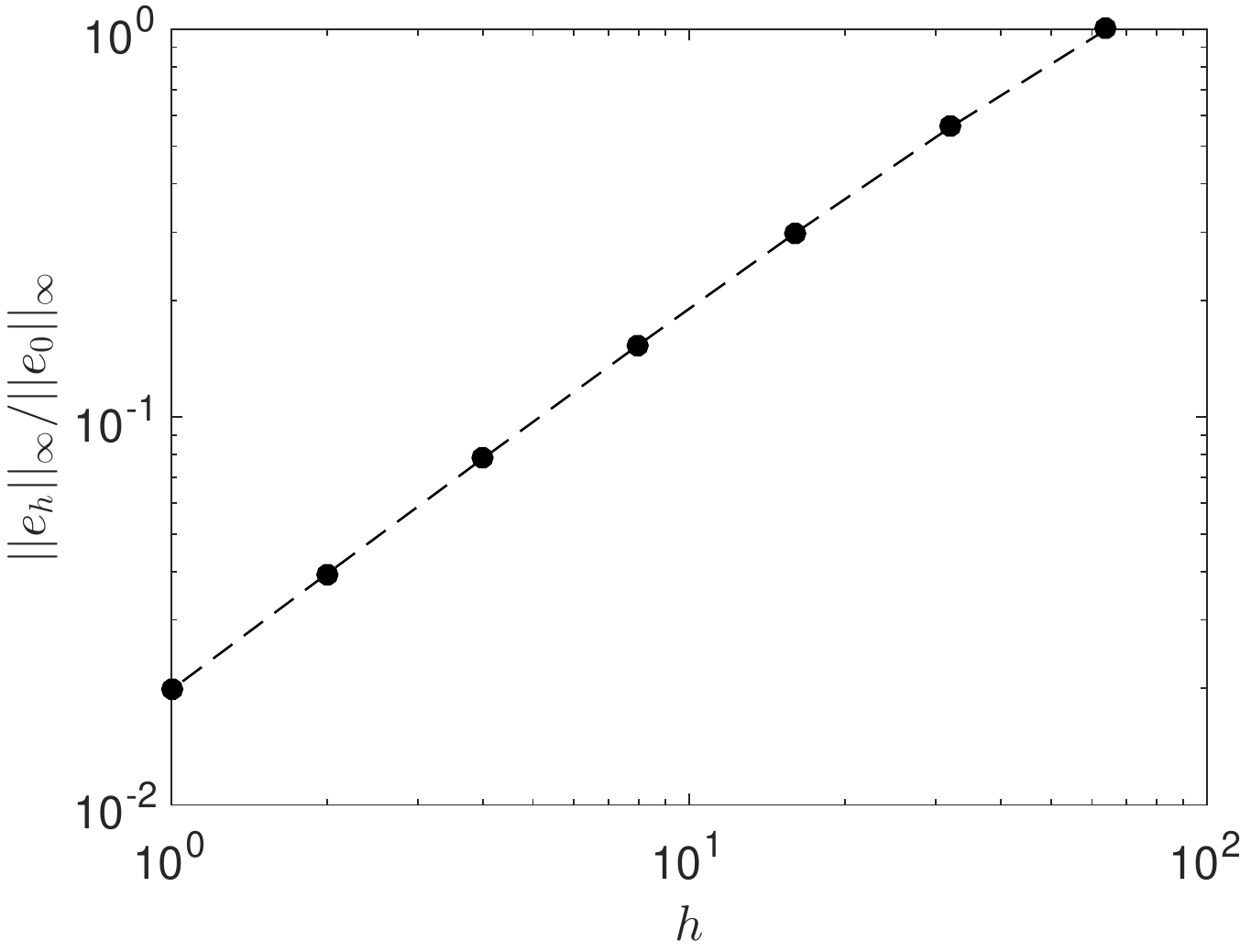}}\quad
    \subfloat[]{\includegraphics[width=0.45\textwidth]{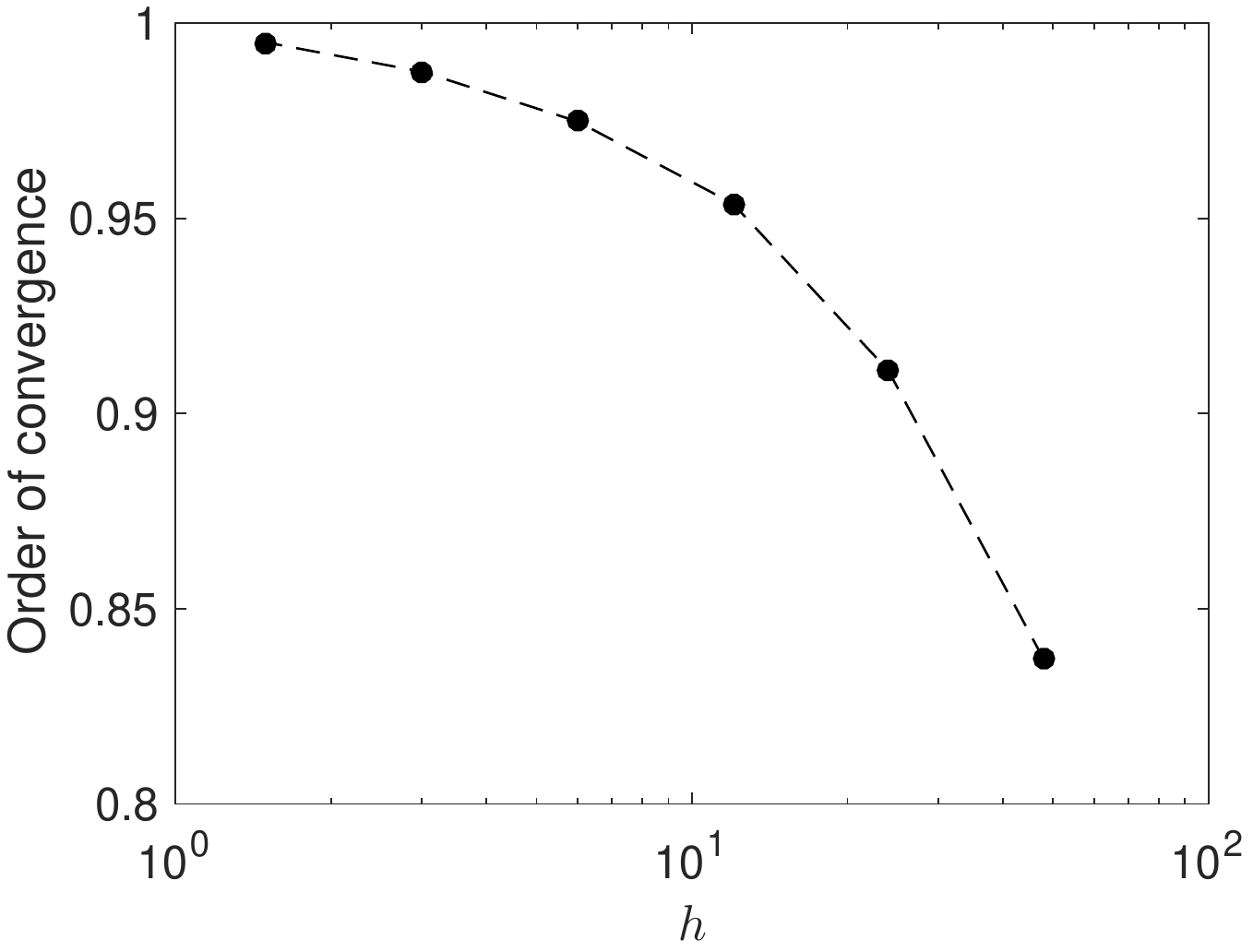}}
    \caption{Local discretization error of the neutral density as a function of the discretization parameters $h=\mathrm{d}v/\mathrm{d}v_0$ (a) and the corresponding order of convergence (b).
    The poloidal neutral spatial grid considered here is $N_R'\times N_Z' = 10\times 10$.
    The expected order of convergence of $n_\text{n}$ to the analytical solution $n_{\text{n},M}$ is correctly retrieved as $h$ decreases.}
    \label{fig:neutrals_mms}
\end{figure}

We focus now on the verification of the correct construction and inversion of the matrix $K$. For this purpose, we compare the numerical solution of the neutral density, $n_\text{n}$, to the analytical manufactured solution,
\begin{equation}
    \label{eqn:n_m}
    n_{\text{n},M} = A_\text{n} + B_\text{n} \sin(\alpha_\text{n} R) \cos(\beta_\text{n} Z)\,,
\end{equation}
where $A_\text{n}$, $B_\text{n}$, $\alpha_\text{n}$, and $\beta_\text{n}$ are arbitrary constants. 
In addition, we consider the manufactured neutral flux as vanishing, i.e.  $\Gamma_{\text{out,n},M}=0$.
The analytical source term required by the MMS is directly computed by substituting $n_{\text{n},M}$ and $\Gamma_{\text{out,n},M}$ in Eq.~\eqref{eqn:neutral_density},
\begin{equation}
    \label{eqn:source_manu}
    S_M(\mathbf{x}_\perp) = n_{\text{n},M}(\mathbf{x}_\perp) - \int_D n_{\text{n},M}(\mathbf{x}_\perp')\nu_\text{cx}(\mathbf{x}_\perp')K_{p\rightarrow p}(\mathbf{x}_\perp,\mathbf{x}_\perp')\mathrm{d}A'\,,\\
\end{equation}
where the contribution from the ion flux is not included to decouple the neutral and plasma modules.
We note that Eq.~\eqref{eqn:source_manu} requires the analytical evaluation of the kernel function $K_{p\rightarrow p}(\mathbf{x}_\perp,\mathbf{x}_\perp')$, Eq.~\eqref{eqn:kernel_first}.
For simplicity, we choose $\nu_\text{iz}(\mathbf{x}_\perp)=0$, $\chi_{\perp\text{in}}(\mathbf{x}_\perp,\mathbf{v}_\perp)=0$, $\nu_\text{cx}(\mathbf{x}_\perp)=$~const,  and
\begin{equation}
    \Phi_{\perp}(\mathbf{x}_\perp,\mathbf{v}_\perp) = \frac{2 m_i}{\sqrt{\pi}T_{i0}}r_\perp \exp\Bigl(\frac{\nu_{cx}r_\perp}{v_\perp}-\frac{m_i v_\perp^2}{T_{i0}}\Bigr)\,.
\end{equation}
The choice of these functions allows for the straightforward analytical evaluation of $K_{p\rightarrow p}(\mathbf{x}_\perp,\mathbf{x}_\perp')$, resulting in $K_{p\rightarrow p}(\mathbf{x}_\perp,\mathbf{x}_\perp')=$~const. 
As a consequence, all the elements of the matrix associated to $K_{p\rightarrow p}$ arising from the discretization are equal.
The integral over the poloidal plane in Eq.~\eqref{eqn:neutral_density}, evaluated through the matrix inversion (see Eq.~\eqref{eqn:neutral_system}), is discretized using the rectangular integration rule. Therefore, the numerical solution of the neutral density is expected to converge to the manufactured solution with the order of convergence $p=1$ as $h$ decreases. 
The local discretization error of the neutral density and the corresponding order of convergence are shown in Fig.~\ref{fig:neutrals_mms} as a function of $h$. 
The expected order of convergence of the neutral density to the analytical solution is retrieved, hence verifying the correct implementation in GBS of Eq.~\eqref{eqn:neutral_system}.

\section{Parallelisation scalability tests}\label{sec:scal}

The use of finite differences on a uniform $(R,\varphi,Z)$ grid allows for an effective implementation of GBS on parallel high-performance computers. Domain decomposition is applied to all three spatial dimensions and implemented with the Message Passing Interface (MPI) for both the plasma and neutral equations. 
Communication between different processors is carried out by means of ghost cell passing, using standard MPI functions. 
In addition, the global MPI communicator is split into two independent ones in order to evolve the plasma and neutral equations in parallel.
After every neutrals calculation, the neutral and the plasma modules synchronize, i.e. the neutral terms appearing in the plasma equations and the plasma quantities in the neutral module are updated (see Fig.~\ref{fig:workflow}). 
The splitting of the MPI communicators has some important advantages with respect to the serial approach: (i) the neutral density is evolved continuously, updating the neutral sources in the plasma equations at the desired frequency by choosing the number of MPI tasks for the plasma and neutral modules, (ii) the time to solution is reduced, and (iii) the code scalability is improved allowing us to run on a larger number of computing nodes, which is essential on large systems due to the large amount of memory required by the neutral module.
We show here that this approach leads to a very efficient strong and weak scalability. 

\begin{figure}
    \centering
    \subfloat[]{\includegraphics[width=0.45\textwidth]{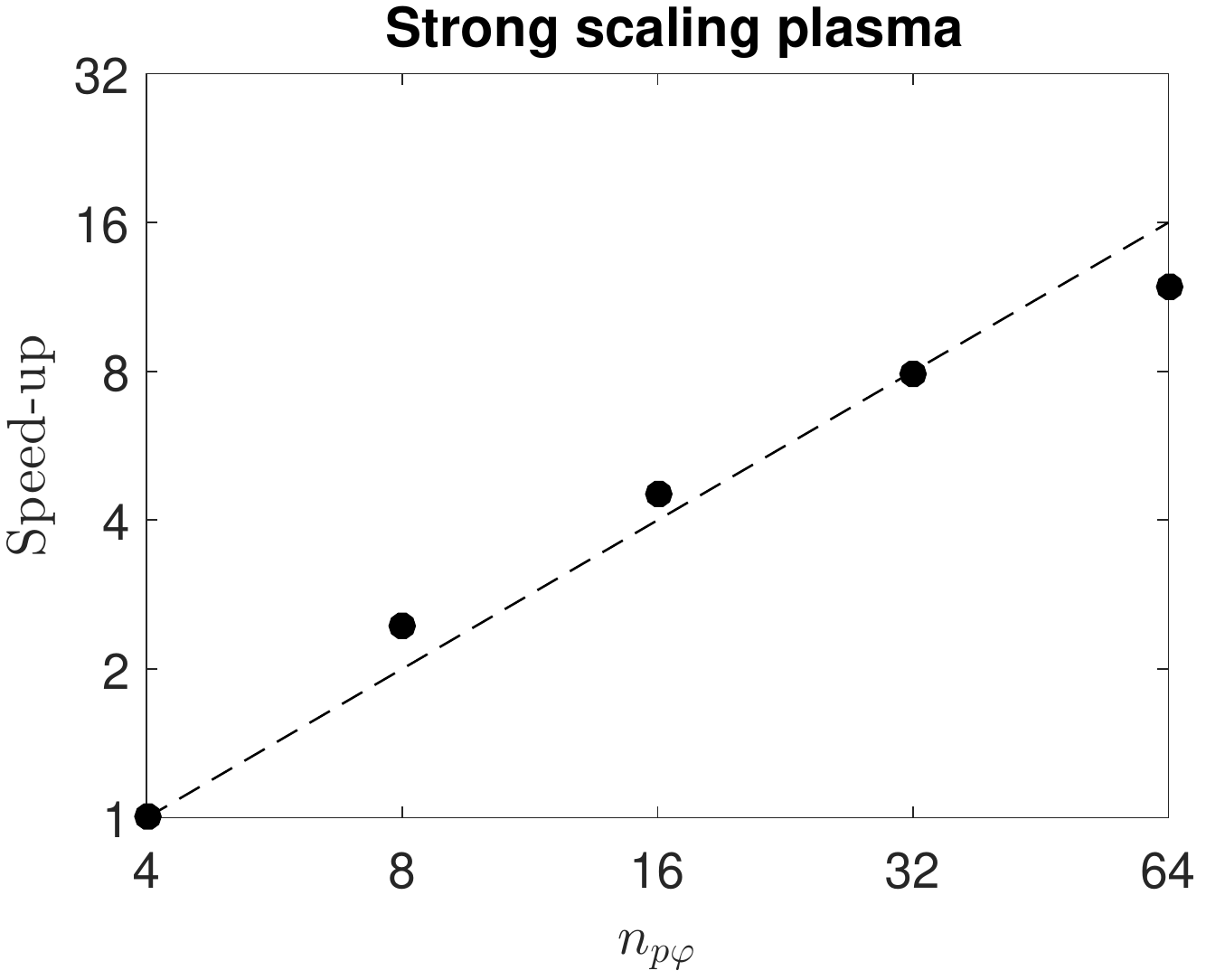}}\quad
    \subfloat[]{\includegraphics[width=0.45\textwidth]{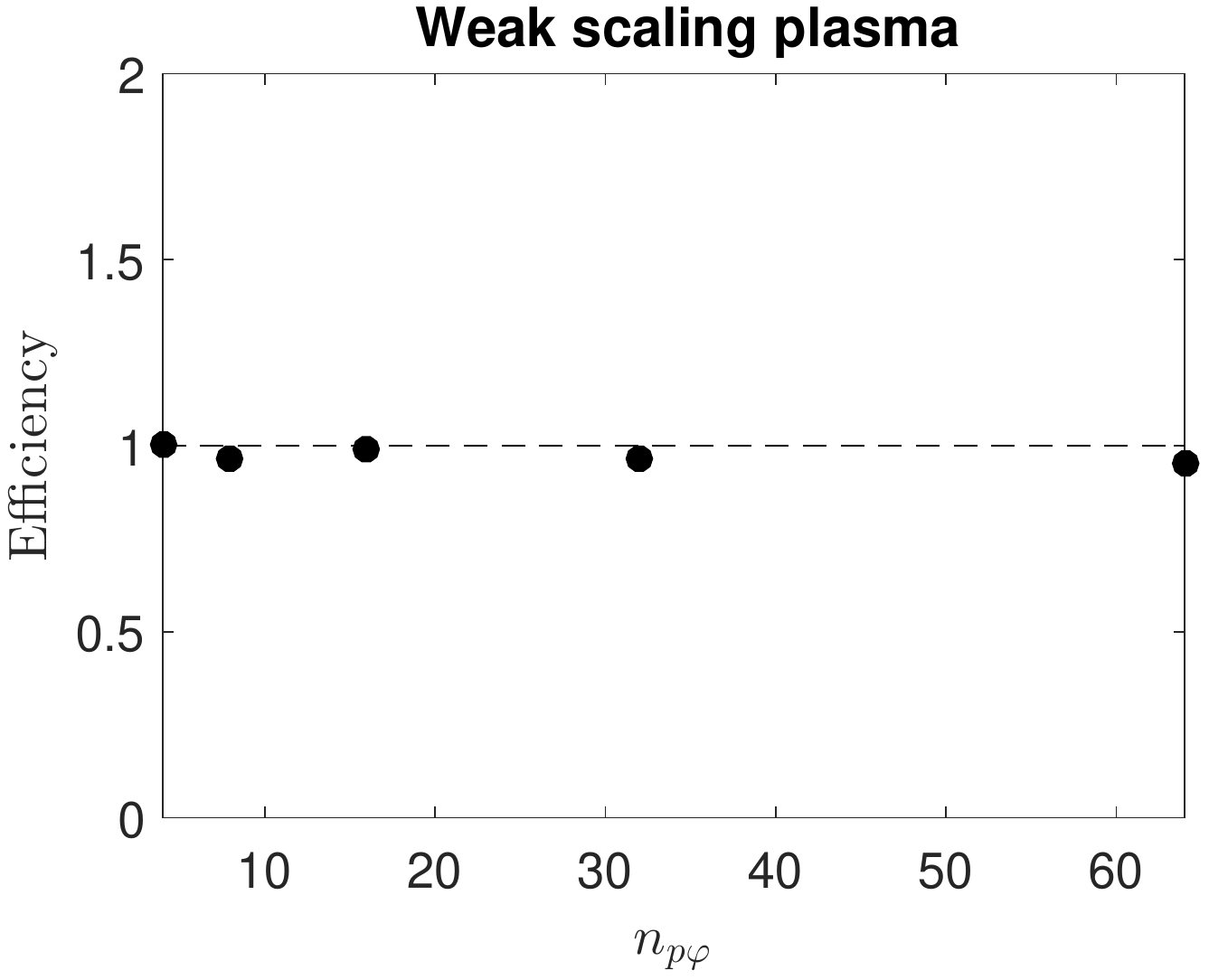}}
    \caption{Strong (a) and weak (b) scalability tests performed on the multi-core partition of Piz Daint (Cray XC40 equipped with two 18-core Intel Xeon E5-2695 v4 CPUs at 2.10GHz). The strong scalability is carried out on a grid of size $N_R\times N_Z\times N_\varphi=300\times 600 \times 128$, with one node (36 cores) on the $RZ$-plane.  
    GBS scales with an almost perfect efficiency up to $n_{p\varphi} = 32$, while efficiency decreases to 0.75 at $n_{p\varphi} = 64$, corresponding to $n_{pR}\times n_{pZ} \times n_{p\varphi} = 2304$ cores (64 nodes).
    The weak scalability test is carried out on grids of size $N_R\times N_Z\times N_\varphi=300\times 600 \times 2 n_{p\varphi}$, where $n_{p\varphi}$ is the number of MPI tasks in the $\varphi$ direction.}
    \label{fig:scalability}
\end{figure}

Being independent, the parallelization properties of the plasma and neutral modules are analyzed separately.
Focusing, first, on the plasma module, we consider a typical grid size of a TCV simulation, with $n_{pR}$, $n_{pZ}$ and $n_{p\varphi}$ the number of MPI tasks in the $R$, $Z$ and $\varphi$ direction, respectively. 
To avoid the demanding communications on poloidal planes required by the Poisson and Ampère equations, the strong scalability test is carried out by only increasing the MPI resources allocated to the $\varphi$ dimension, i.e. by solving whole $RZ$-planes on a computing node.
As shown in Fig.~\ref{fig:scalability}~(a), GBS scales with an almost perfect efficiency up to $n_{p\varphi} = 32$, while efficiency decreases to 0.75 at $n_{p\varphi} = 64$ (two toroidal planes per computing node), pointing out the high scalability of GBS especially on the toroidal direction, mainly thanks to the fact that the Poisson and Ampère equations are solved independently on each poloidal plane.
The weak scalability of the plasma module is carried out on grids of size $N_R\times N_Z\times N_\varphi=300\times 600 \times 2 n_{p\varphi}$, where $n_{p\varphi}$ is the number of MPI tasks in the toroidal direction.  The results are shown in Fig.~\ref{fig:scalability}~(b), confirming the good scalability properties of the plasma module in GBS.

\begin{figure}
    \centering
    \subfloat[]{\includegraphics[width=0.45\textwidth]{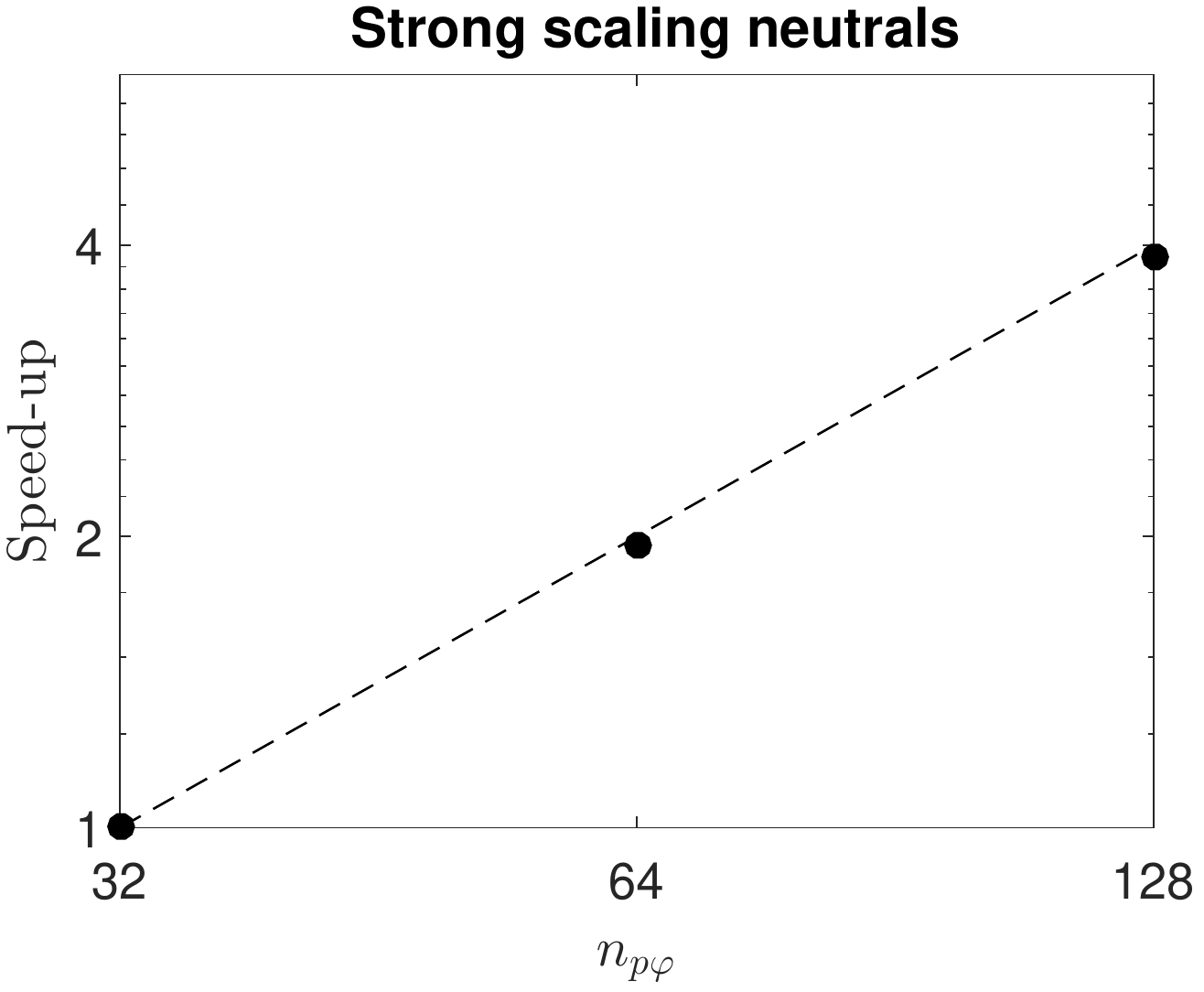}}\quad
    \subfloat[]{\includegraphics[width=0.45\textwidth]{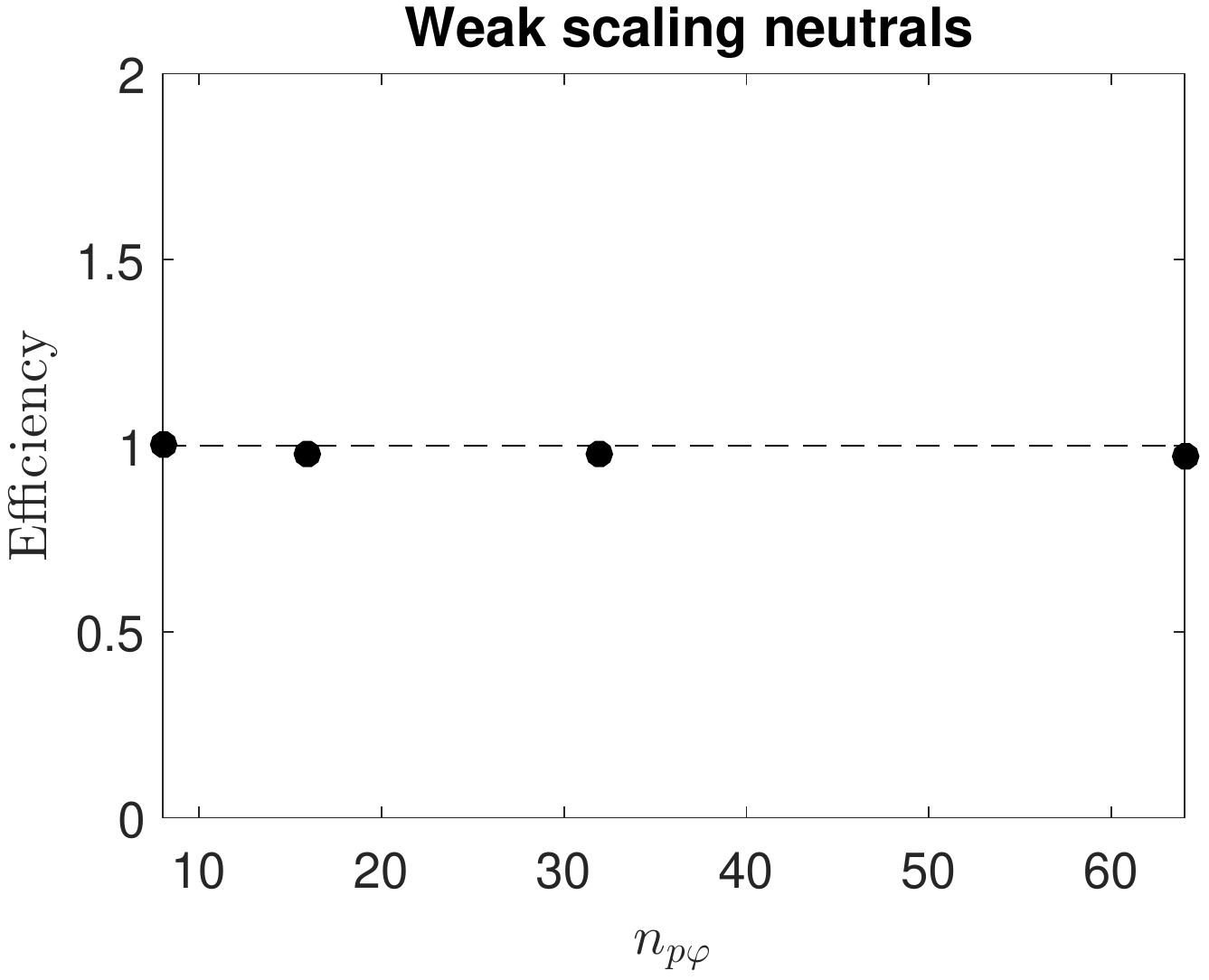}}
    \caption{Strong (a) and weak (b) scalability tests of the neutral module performed on the multi-core partition of Piz Daint (Cray XC40 equipped with two 18-core Intel Xeon E5-2695 v4 CPUs at 2.10GHz). The strong scalability is carried out on a neutral grid of size $N_R'\times N_Z'\times N_\varphi'=100\times 200 \times 128$, with one node (36 cores) on the $RZ$-plane, and is tested starting from 32 nodes, which is the minimum number of nodes required by memory constraints. The speed-up is almost ideal up to $n_{p\varphi}=128$, corresponding to a total of 4608 cores (128 nodes). The weak scalability is carried out on neutral grids of size $N_R'\times N_Z'\times N_\varphi'=300\times 600 \times 2 n_{p\varphi}$, where $n_{p\varphi}$ is the number of MPI tasks along $\varphi$.}
    \label{fig:scalability_neutrals}
\end{figure}

The strong scalability of the neutral module is carried out on a neutral grid typical of medium size simulations such as TCV.
Similarly to the plasma scalability, we consider one computing node in the $RZ$-plane, while increasing $n_{p\varphi}$, starting from the minimum number of nodes allowed by memory requirements.
In fact, the solution of the neutral system, Eq.~\eqref{eqn:neutral_system}, requires the allocation of matrices of sizes $[N_R'\times N_Z'+2(N_R'+N_Z')]^2$ in each poloidal plane, thus being highly memory consuming\footnote{ For example, on Piz Daint, plasma and neutral simulations carried out on 16 nodes require a memory of approximately 60 GB/node, which is above the memory per node available in Piz Daint.}.
As shown in Fig.~\ref{fig:scalability_neutrals}~(a), the speed-up is almost ideal up to $n_{p\varphi}=128$.
The weak scalability is performed on neutral grids of size $N_R'\times N_Z'\times N_\varphi'=300\times 600 \times 2 n_{p\varphi}$. Fig.~\ref{fig:scalability_neutrals}~(b) shows an almost perfect efficiency as the system size increases.

\section{Convergence properties}\label{sec:conv}

We test here the convergence of GBS results with respect to the plasma and neutral grid refinement. We consider the magnetic field given by the analytical flux function in Eq.~\eqref{eqn:psi} and we compare the results of three simulations with increasing plasma grid resolution and no coupling with neutrals, and of three simulations with increasing neutral grid resolution and same plasma grid resolution. 
The results are analyzed when the simulations reach a quasi-steady state, resulting from the balance between the plasma and heat sources and losses at the wall.
We also study the convergence with respect to the frequency of the neutral computation. 

\begin{figure}
    \centering
    \subfloat[]{\includegraphics[height=0.25\textheight]{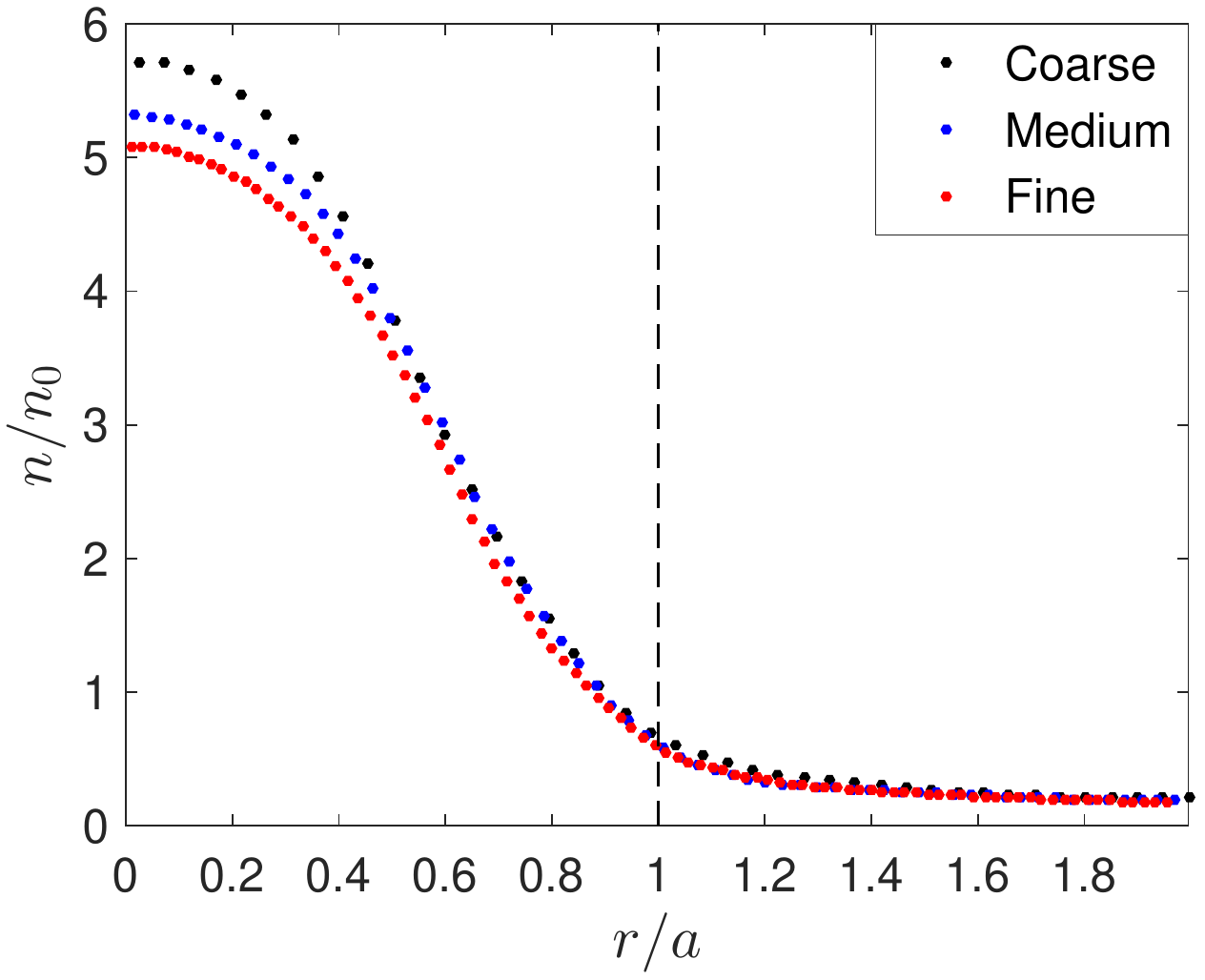}}\quad
    \subfloat[]{\includegraphics[height=0.25\textheight]{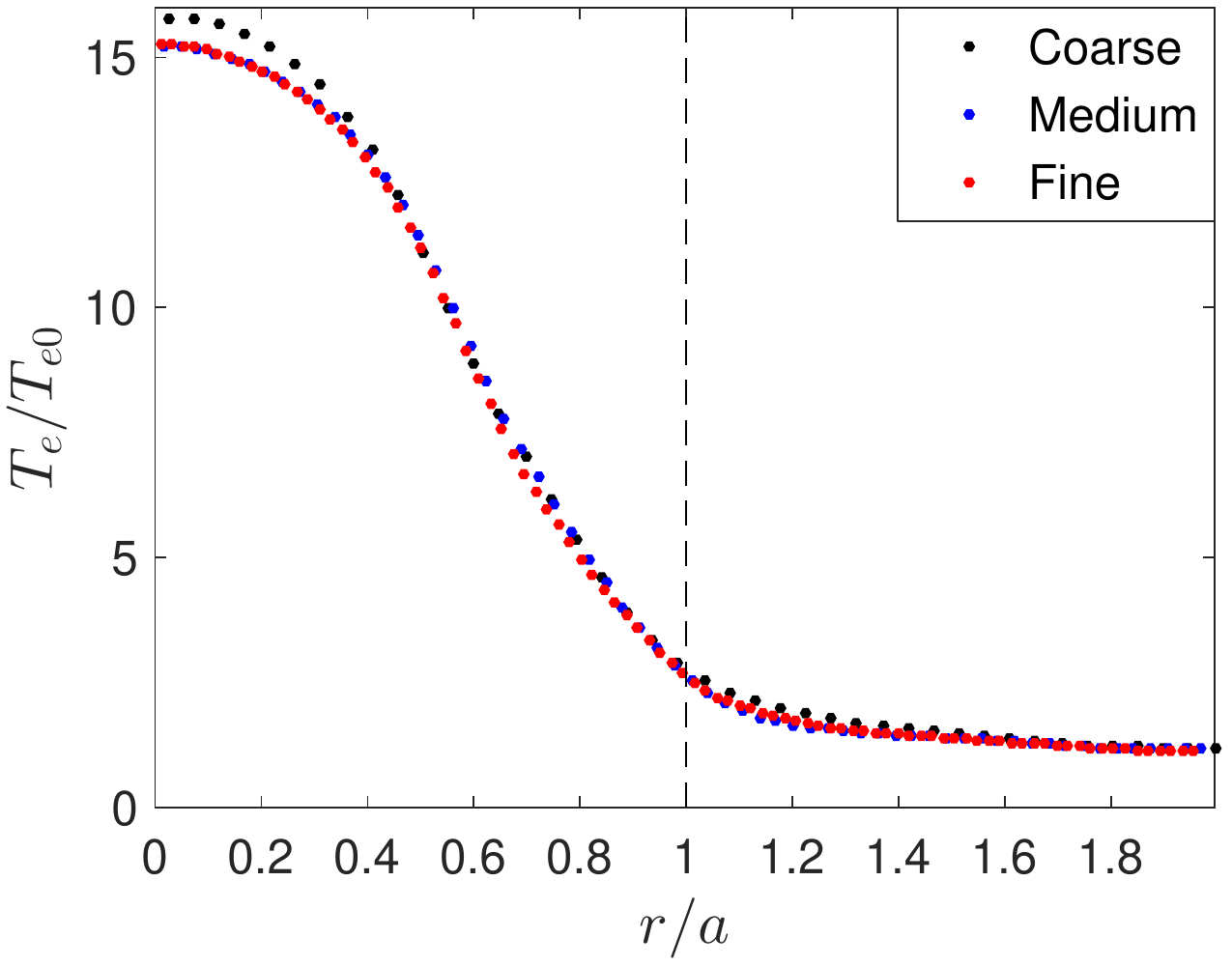}}\\
    \subfloat[]{\includegraphics[height=0.25\textheight]{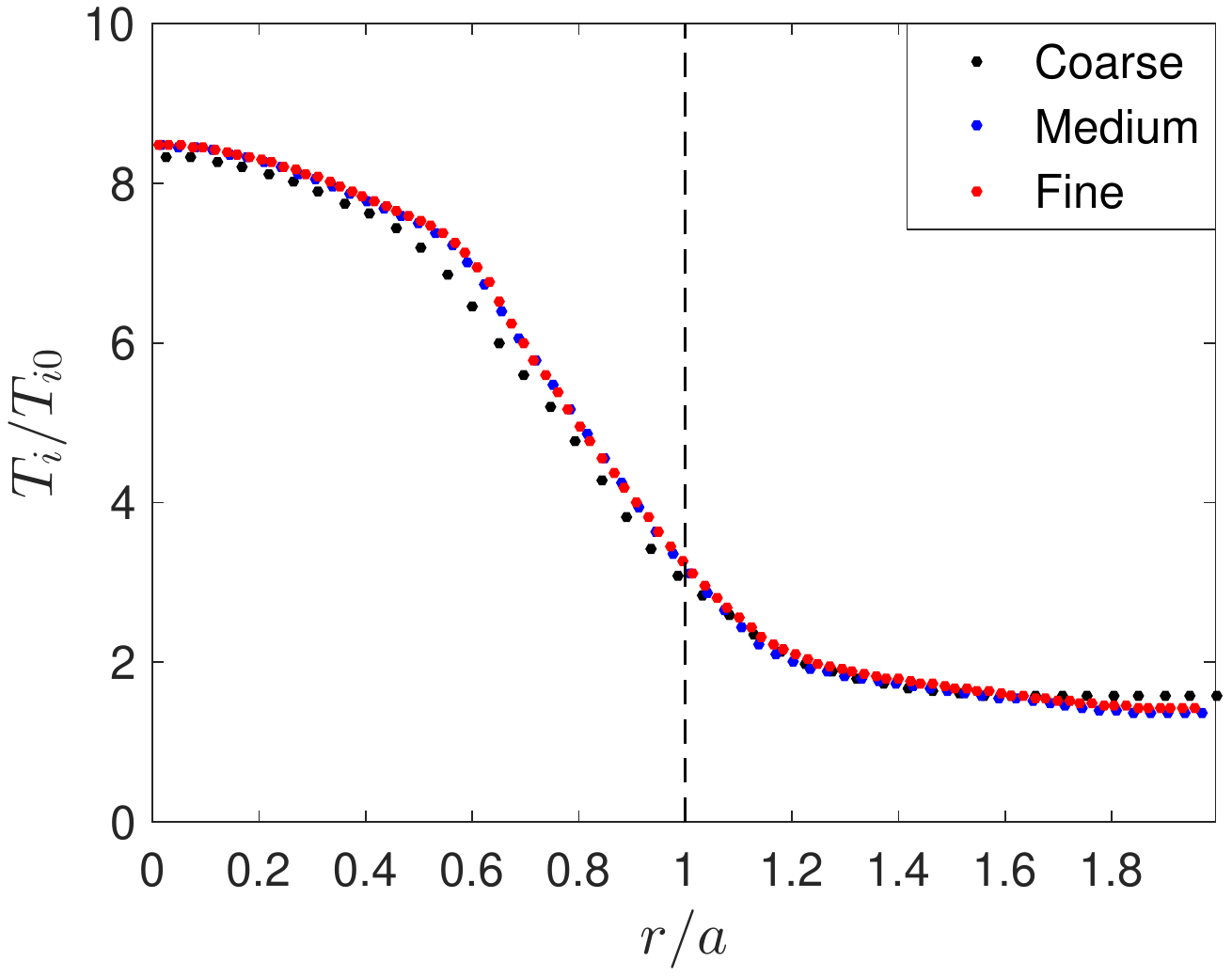}}\quad
    \subfloat[]{\includegraphics[height=0.25\textheight]{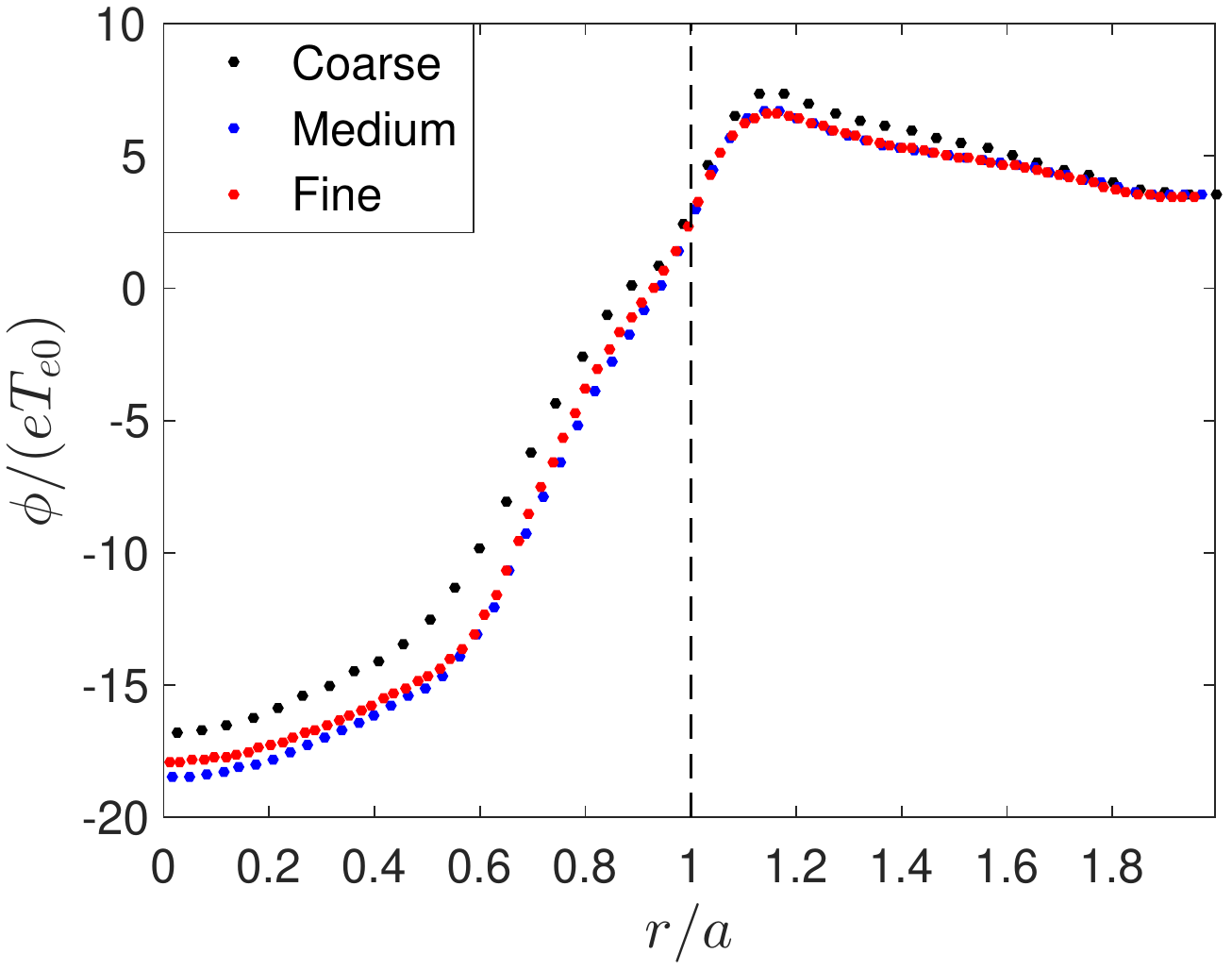}}\\
    \caption{Time and toroidally averaged radial profiles of density (a), electron temperature (b), ion temperature (c), and electrostatic potential (d) at the outer midplane of GBS simulations at different plasma grid resolutions ($N_R\times N_Z\times N_\varphi = 80\times 110 \times 24$, coarse grid; $N_R\times N_Z \times N_\varphi = 120\times 160 \times 32$, medium grid; $N_R\times N_Z \times N_\varphi = 180\times 240\times 48$, fine grid). The dashed vertical line represents the position of the separatrix.}
    \label{fig:conv_profiles}
\end{figure}

The simulations considered for studying the convergence with respect to the plasma grid are characterized by a coarse ($\Delta R = \Delta Z = 3.75~\rho_{s0}$ and $R_0\Delta \varphi = 65.4~\rho_{s0}$),  a medium ($\Delta R = \Delta Z = 2.50~\rho_{s0}$ and $R_0\Delta \varphi =49.1~\rho_{s0}$) and a fine ($\Delta R = \Delta Z = 1.67~\rho_{s0}$ and $R_0\Delta \varphi =32.7~\rho_{s0}$) resolution grid. 
In order to investigate the convergence, we consider the time and toroidally averaged radial profiles at the outer midplane of $n$, $T_e$, $T_i$ and $\phi$. 
As shown in Fig.~\ref{fig:conv_profiles}, the radial profiles clearly converge to the ones obtained from the simulation with the finest resolution. 
Assuming that $\rho_{s0}$ is the characteristic spatial length along the direction perpendicular to the magnetic field for phenomena occurring in our simulations, we conclude that $\Delta R = \Delta Z \simeq 2.50~\rho_{s0}$ is the minimum spatial resolution of the plasma grid on the poloidal plane that guarantees convergence of simulation results.
In order to guarantee stability of the simulations, the resolution along $\varphi$ is chosen according to the ratio of the poloidal to the toroidal component of the magnetic field, $R_0 \Delta \varphi \lesssim B_0 \Delta Z/B_Z \sim B_0 \Delta R/B_R$.

\begin{figure}
    \centering
    \subfloat[]{\includegraphics[width=0.3\textwidth]{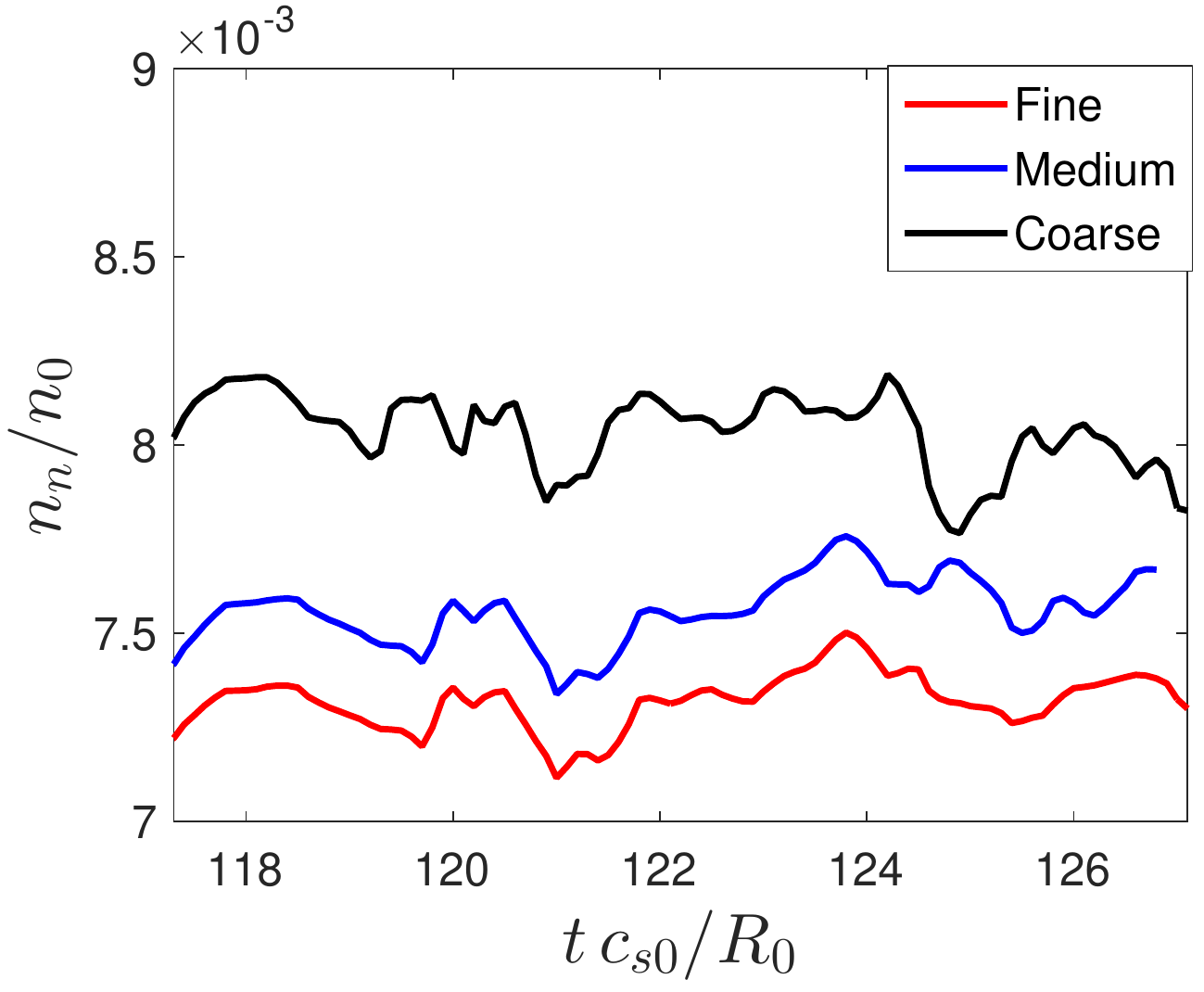}}\quad
    \subfloat[]{\includegraphics[width=0.3\textwidth]{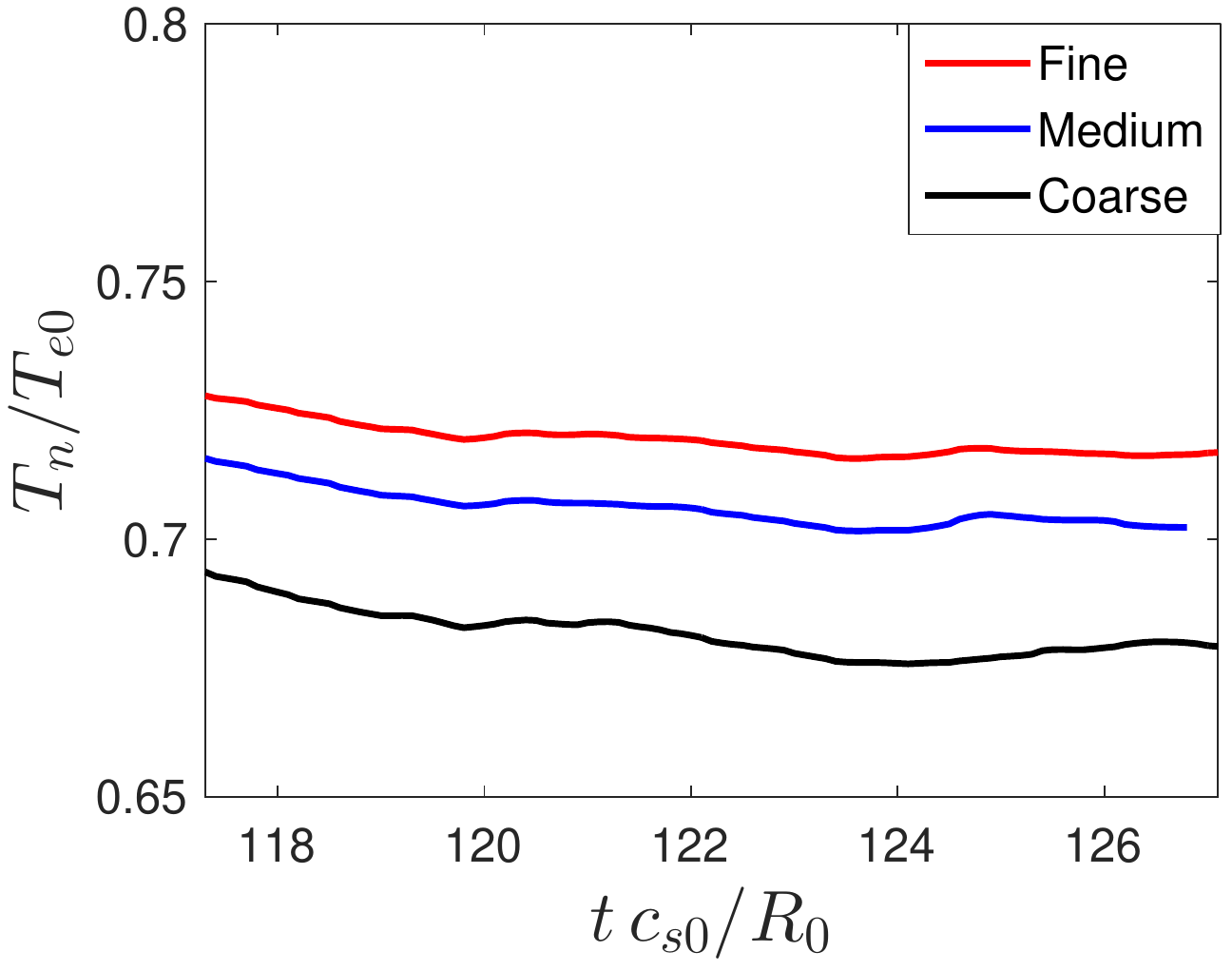}}\quad
    \subfloat[]{\includegraphics[width=0.3\textwidth]{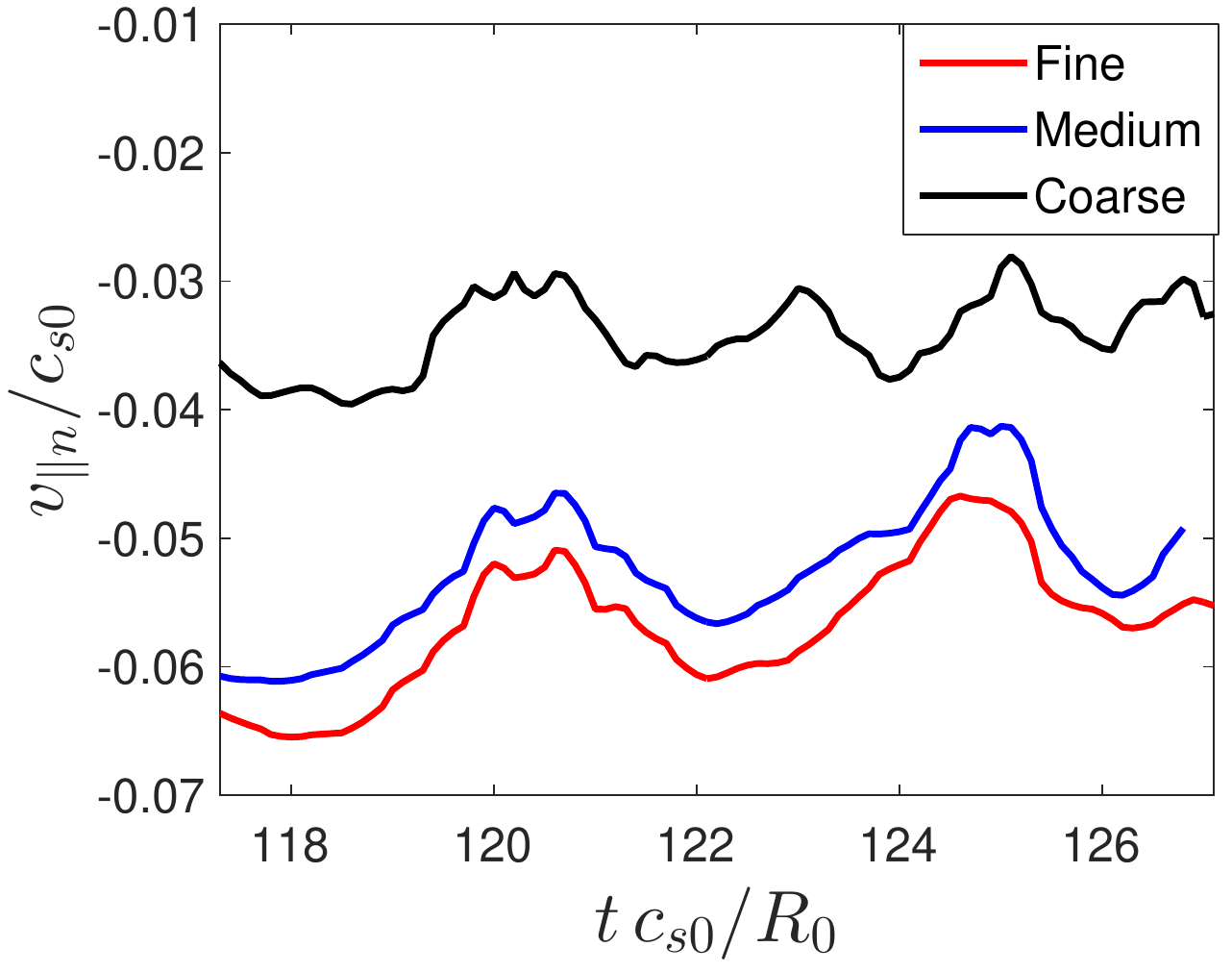}}
    \caption{Time traces of the spatially averaged neutral density (a), neutral temperature (b) and neutral parallel velocity (c) at different neutral grid resolutions ($N_R'\times N_Z' = 24\times 28$, coarse; $N_R'\times N_Z' = 42\times 56$, medium; $N_R'\times N_Z' = 84\times 112$, fine).}
    \label{fig:neu_av_conv}
\end{figure}

\begin{figure}
    \centering
    \subfloat[]{\includegraphics[width=0.45\textwidth]{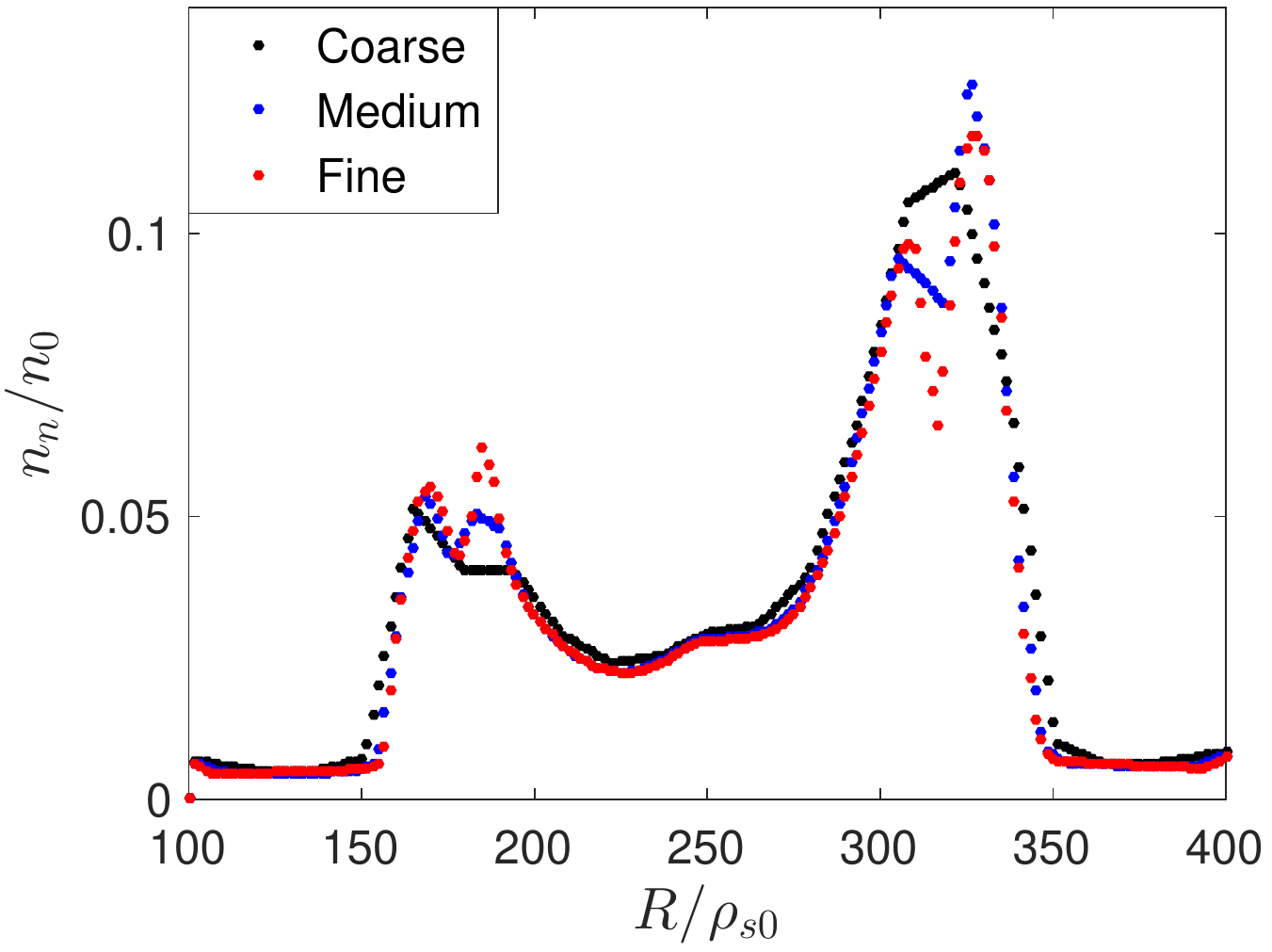}}\quad
    \subfloat[]{\includegraphics[width=0.45\textwidth]{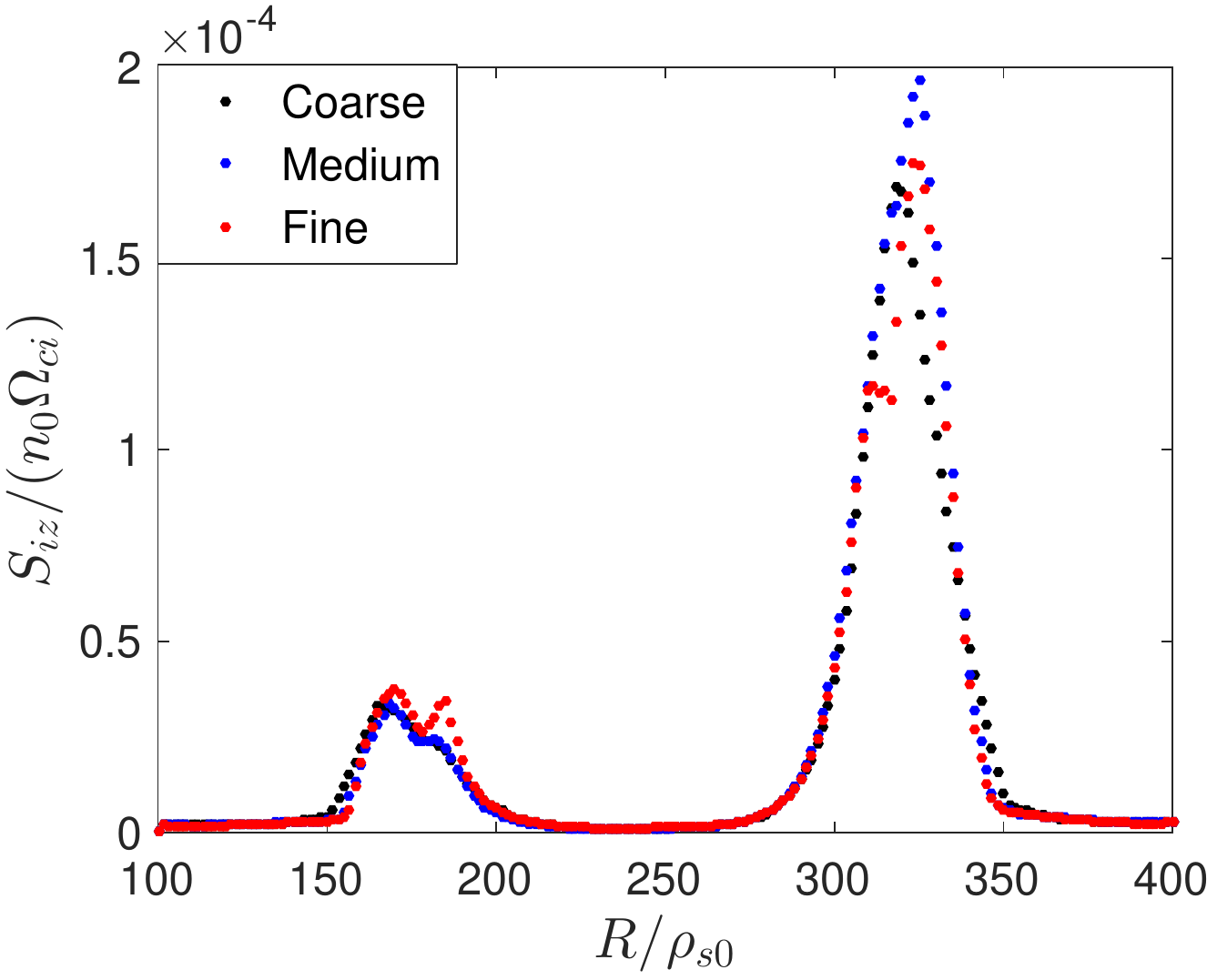}}
    \caption{Time and toroidal average of the neutral density (a) and ionization source (b) at the bottom wall at different neutral grid resolutions ($N_R'\times N_Z' = 24\times 28$, coarse; $N_R'\times N_Z' = 42\times 56$, medium; $N_R'\times N_Z' = 84\times 112$, fine).}
    \label{fig:neutral_convergence}
\end{figure}

The convergence with respect to the neutral grid refinement is tested by comparing the results of three simulations with the same plasma resolution, $\Delta R = \Delta Z = 1.67~\rho_{s0}$ and $R_0\Delta \varphi =32.7~\rho_{s0}$,  and different neutral grid resolutions: coarse ($\Delta R' = \Delta Z' = 0.15~\lambda_n$),  medium ($\Delta R' = \Delta Z' = 0.075~\lambda_n$), and fine ($\Delta R' = \Delta Z' = 0.038~\lambda_n$) resolutions, where $\lambda_{n}$ is the mean free path of neutrals for ionization, evaluated by considering a value of $T_e$ and $n$ in the simulations near the separatrix ($T_e \simeq 20$~eV and $n\simeq 4\times 10^{19}$~m$^{-3}$).
The resolution along $\varphi$ is given by the plasma grid.
The neutral density is evaluated every $\Delta t = 0.08\ R_0/c_{s0}$.
The time traces of the spatially averaged neutral density, neutral temperature and neutral parallel velocity, $v_{\parallel \text{n}}= \mathbf{v}_\text{n} \cdot \mathbf{b}$, show a clear convergence to the results of the finest neutral grid (see Fig.~\ref{fig:neu_av_conv}).
We also analyze the time and the toroidal averages of the neutral density and ionization source, $S_{iz}$, at the bottom wall, where the two strike points are located. 
Fig.~\ref{fig:neutral_convergence} shows that there is no noticeable difference in the neutral density and ionization source profiles between simulations with different neutral grid resolutions, pointing out that the resolution $\Delta R' = \Delta Z' \simeq 0.075~\lambda_n$ is sufficient to guarantee the convergence of the simulation results.

\begin{figure}
    \centering
    \subfloat[]{\includegraphics[height=0.25\textheight]{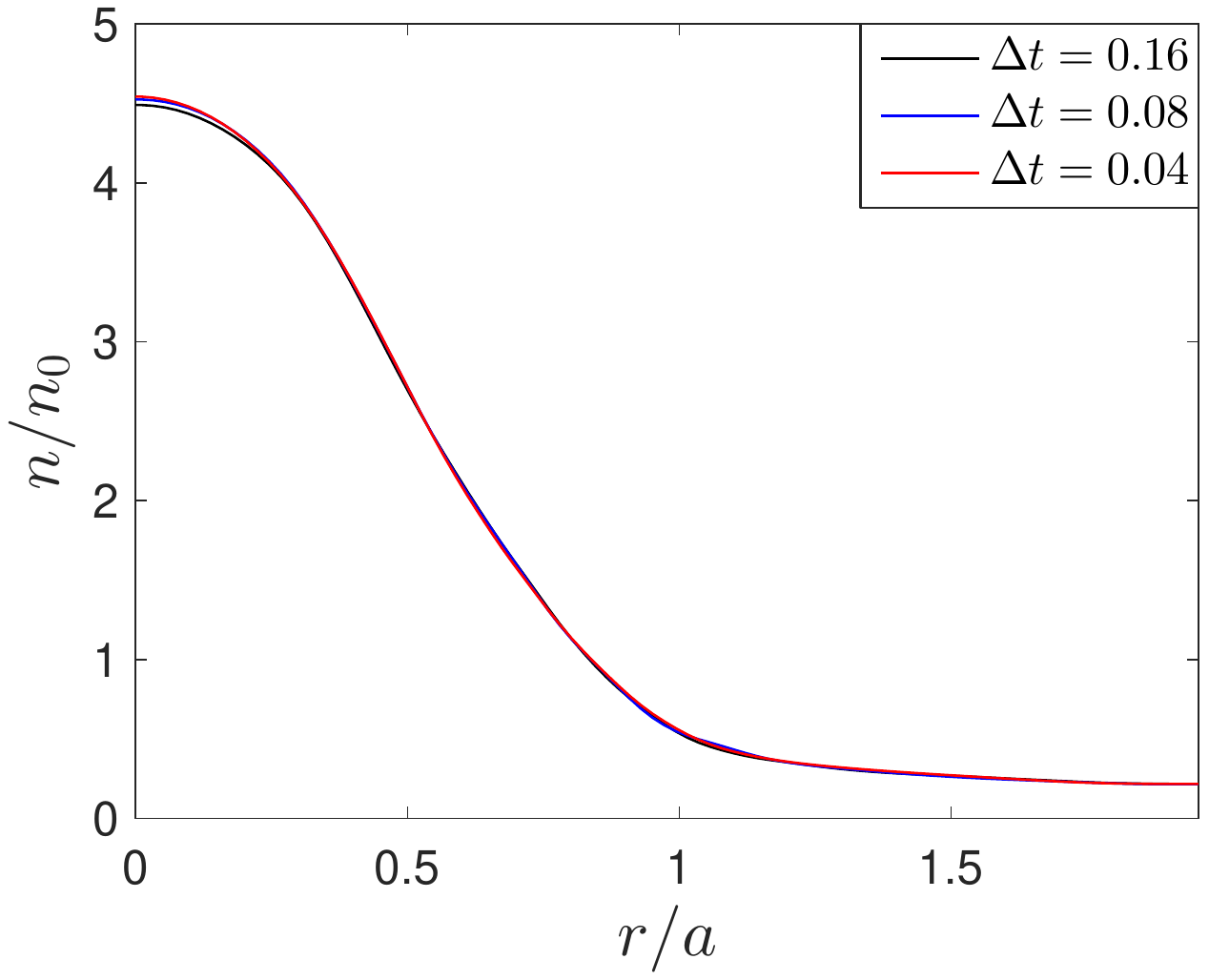}}\quad
    \subfloat[]{\includegraphics[height=0.25\textheight]{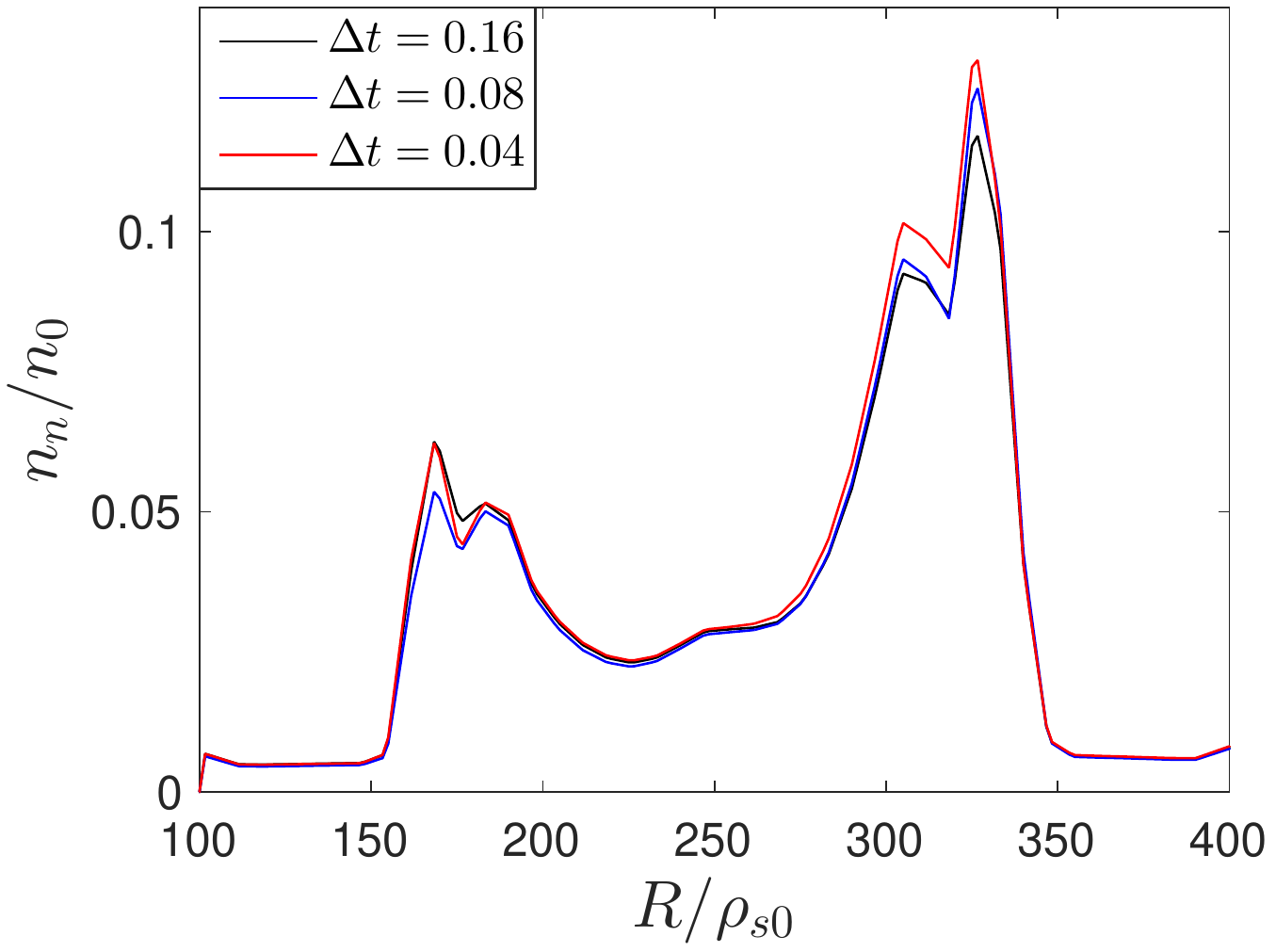}}
    \caption{Time and toroidally averaged radial density profile at the outer midplane (a) and neutral density at the bottom wall (b) of GBS simulations with the neutral density evaluated every $\Delta t=0.04\ R_0/c_{s0}$, $\Delta t=0.08\ R_0/c_{s0}$ and $\Delta t=0.16\ R_0/c_{s0}$.}
    \label{fig:frequency_convergence}
\end{figure}

Finally, we study the convergence of GBS results with respect to the frequency of the neutral calculation. We consider three simulations with a plasma grid resolution of $\Delta R = \Delta Z = 1.67~\rho_{s0}$ and $R_0\Delta \varphi =32.7~\rho_{s0}$, and a neutral grid resolution of $\Delta R' = \Delta Z' = 0.075~\lambda_n$. The neutral density is evaluated every $\Delta t= 0.04\ R_0/c_{s0}$, $\Delta t= 0.08\ R_0/c_{s0}$ and $\Delta t= 0.16\ R_0/c_{s0}$, respectively. 
Fig.~\ref{fig:frequency_convergence} shows that the time and toroidally averaged radial density profile at the outer midplane and neutral density at the bottom wall do not display any significant difference among the three simulations considered here.
We conclude that evaluating the neutral density every $\Delta t \simeq 0.1\ R_0/c_{s0}$  guarantees the correct convergence of the results. 
We note that, since in our simulations turbulence is mainly driven by ballooning modes with a maximum growth rate of $\gamma_B=\sqrt{2} c_s/\sqrt{L_p R_0}$~\cite{mosetto2013}, with ${L_p = |p_e/\partial_R p_e|}$  the equilibrium pressure gradient length at the separatrix on the outer midplane, the minimum neutral calculation frequency that guarantees the convergence of the simulation results is $\gamma_B\Delta t \simeq 0.7$.
Since the neutral and the plasma models can be evolved simultaneously, the number  of MPI tasks for the plasma and neutral modules can be chosen to guarantee this neutral calculation frequency.

\section{First simulation results in an experimentally-relevant scenario}\label{sec:application}

As an example of application of the GBS version presented here, we report on the results of the first GBS electromagnetic simulation that includes the coupling with neutral dynamics of a lower single-null discharge performed in the TCV tokamak (tokamak major radius $R_0=0.9$~m and  minor radius measured from the tokamak magnetic axis to the separatrix at the outer midplane $a=0.20$~m).
The magnetic equilibrium considered in the simulation is given by the equilibrium reconstruction of the TCV discharge \#65402 at 1~s (plasma elongation and triangularity at the separatrix $\kappa=1.71$ and $\delta=0.35$, respectively, toroidal magnetic field at the tokamak magnetic axis $B_0=0.9$~T, plasma current $I_p$=146~kA).
The discharge considered here is in forward field (ion-$\nabla B$ drift direction pointing toward the active X-point). The upstream density and electron temperature at the separatrix, taken as reference density and temperature in the simulation, are $n_0=0.6\times 10^{19}$~m$^{-3}$ and $T_{e0}=35$~eV, respectively.
With these reference values, the ion sound Larmor radius is $\rho_{s0}\simeq 1$~mm, the sound speed is $c_{s0}\simeq 4.1\times 10^4$~m/s, and the reference time is $t_0=R_0/c_{s0}\simeq 0.02$~ms.
In order to reduce the computational cost of the simulation, we consider here a domain corresponding to half size of TCV, i.e. $L_R = 300\,\rho_{s0}$ and $L_Z= 600\,\rho_{s0}$. 
The dimensionless simulation parameters are $\rho_*^{-1}=450$, $\tau=1$, $\eta_{0e}=3\times10^{-4}$, $\eta_{0i}=1$, $\chi_{\parallel e}=20$, $\chi_{\parallel i}=1$, $D_f=7$ for $f=\{n, T_e, T_i, \Omega, U_{\parallel e},v_{\parallel i}\}$, $m_i/m_e=3000$, $\beta_{e0}=2\times 10^{-6}$, and $\nu_0=0.05$. The amplitude of the electron temperature source is chosen so that $T_e/T_{e0}\simeq 1$ at the separatrix. 
This leads to a power source with an intensity, in physical units, of approximately 150~kW, which is close to the experimental estimated value of the power crossing the separatrix, $P_\text{sep} = 120$~kW.
No external ion temperature source is used in the simulation and the ion heating is provided by the equipartition term in Eq.~\eqref{eqn:ion_temperature}.
The value of neutral reflection coefficient is $\alpha_\text{refl}=0.2$. 
Although we consider a discrete gas puff in the private flux region, most of the neutral particles are generated from the ion recycling at the wall. A discrete pump at the bottom wall in the region $R> 500 \rho_{s0}$ is also considered. 

The plasma spatial grid is $N_R\times N_Z\times N_\varphi = 150\times 300 \times 64$ and the neutral spatial grid is $N_R'\times N_Z'\times N_\varphi' = 50\times 100 \times 64$. The time step is $\mathrm{d}t=10^{-5} R_0/c_{s0}$ and the neutral kinetic equation is solved every $\Delta t= 0.04\ R_0/c_{s0}$.
The simulation reaches the turbulent quasi-steady state after approximately 80~$R_0/c_{s0}$, starting from flat initial profiles of density and temperature. 

\begin{figure}
    \centering
    \subfloat[]{\includegraphics[height=0.3\textheight]{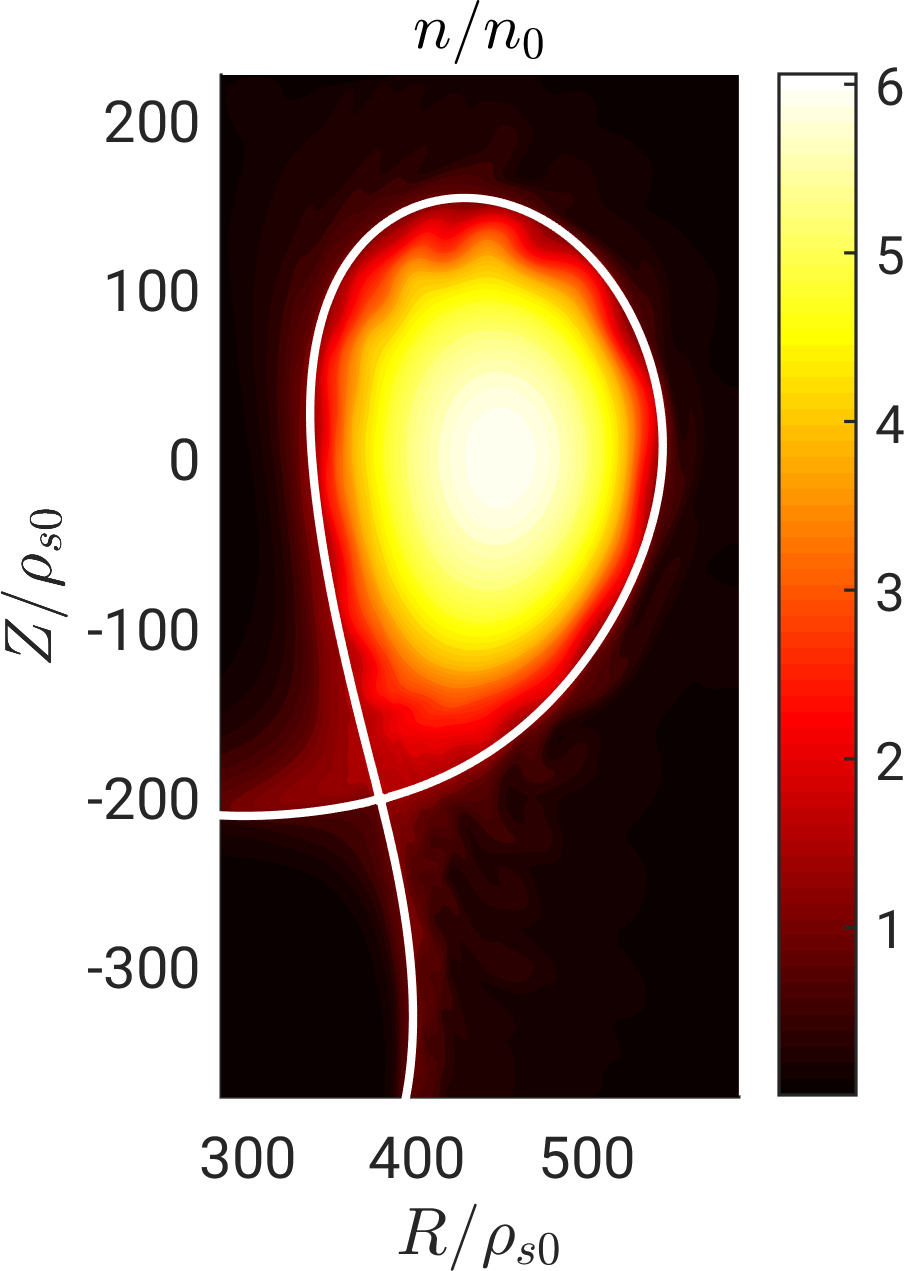}}\,
    \subfloat[]{\includegraphics[height=0.3\textheight]{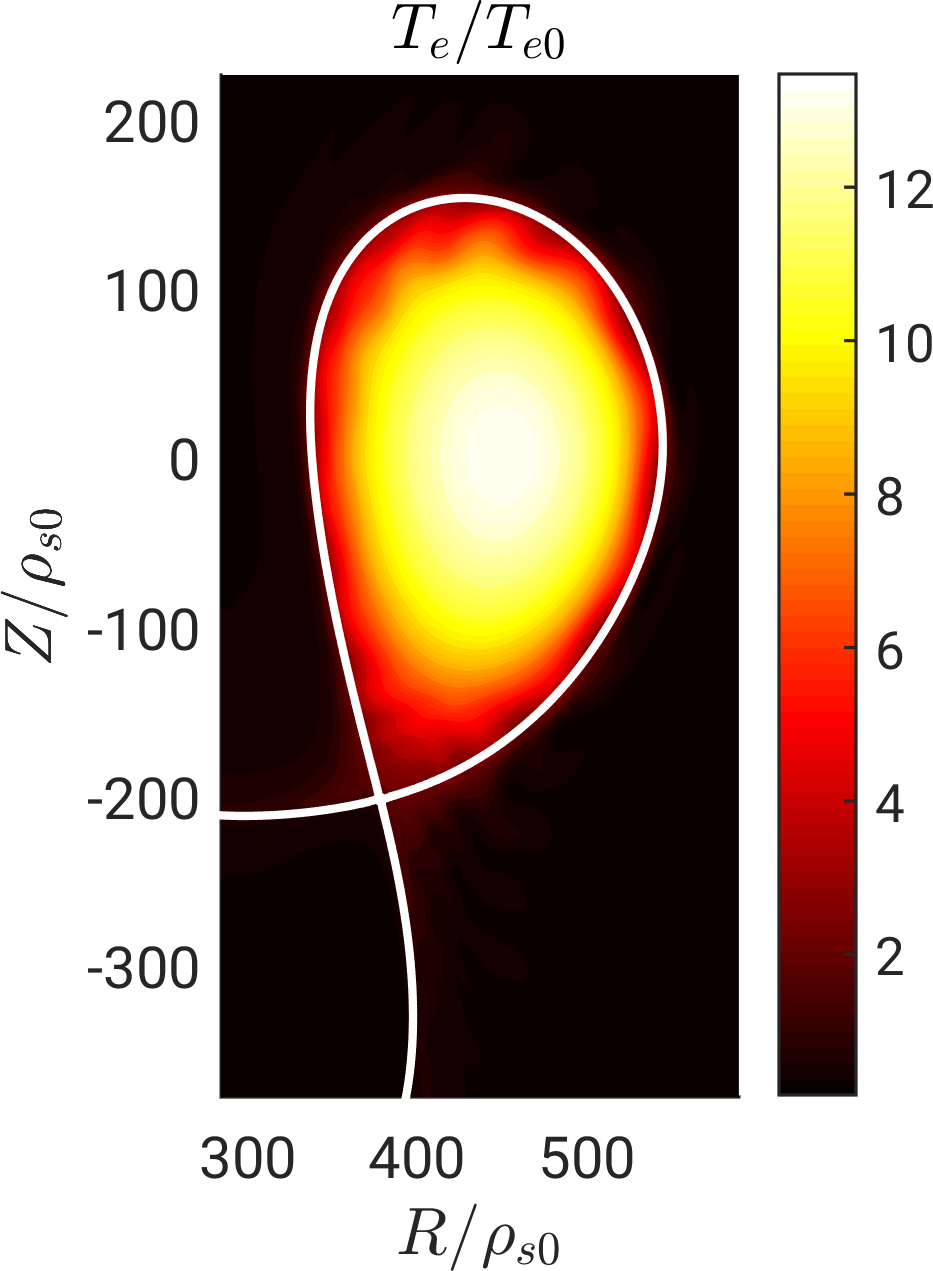}}\,
    \subfloat[]{\includegraphics[height=0.3\textheight]{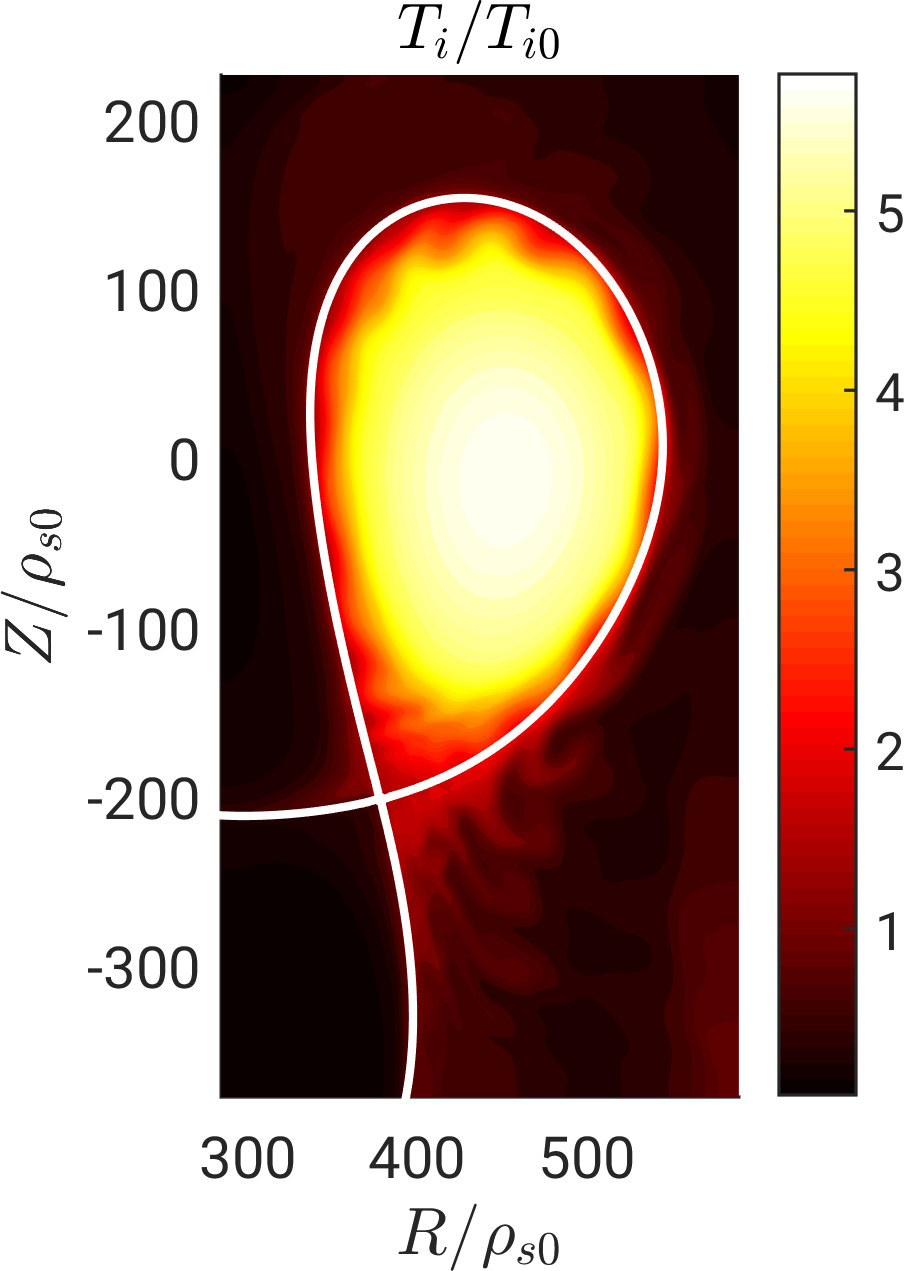}}\\
    \subfloat[]{\includegraphics[height=0.3\textheight]{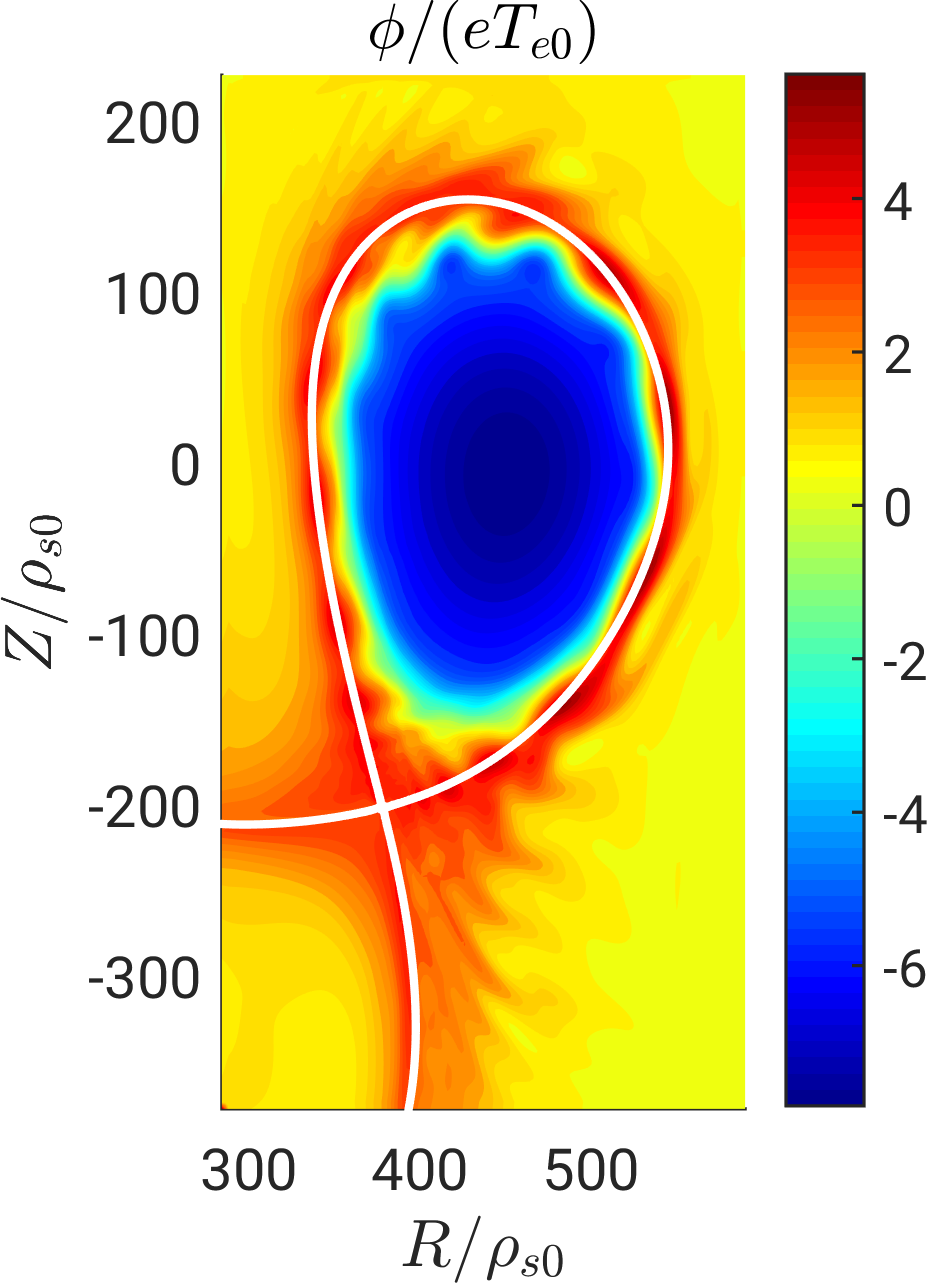}}\,
    \subfloat[]{\includegraphics[height=0.3\textheight]{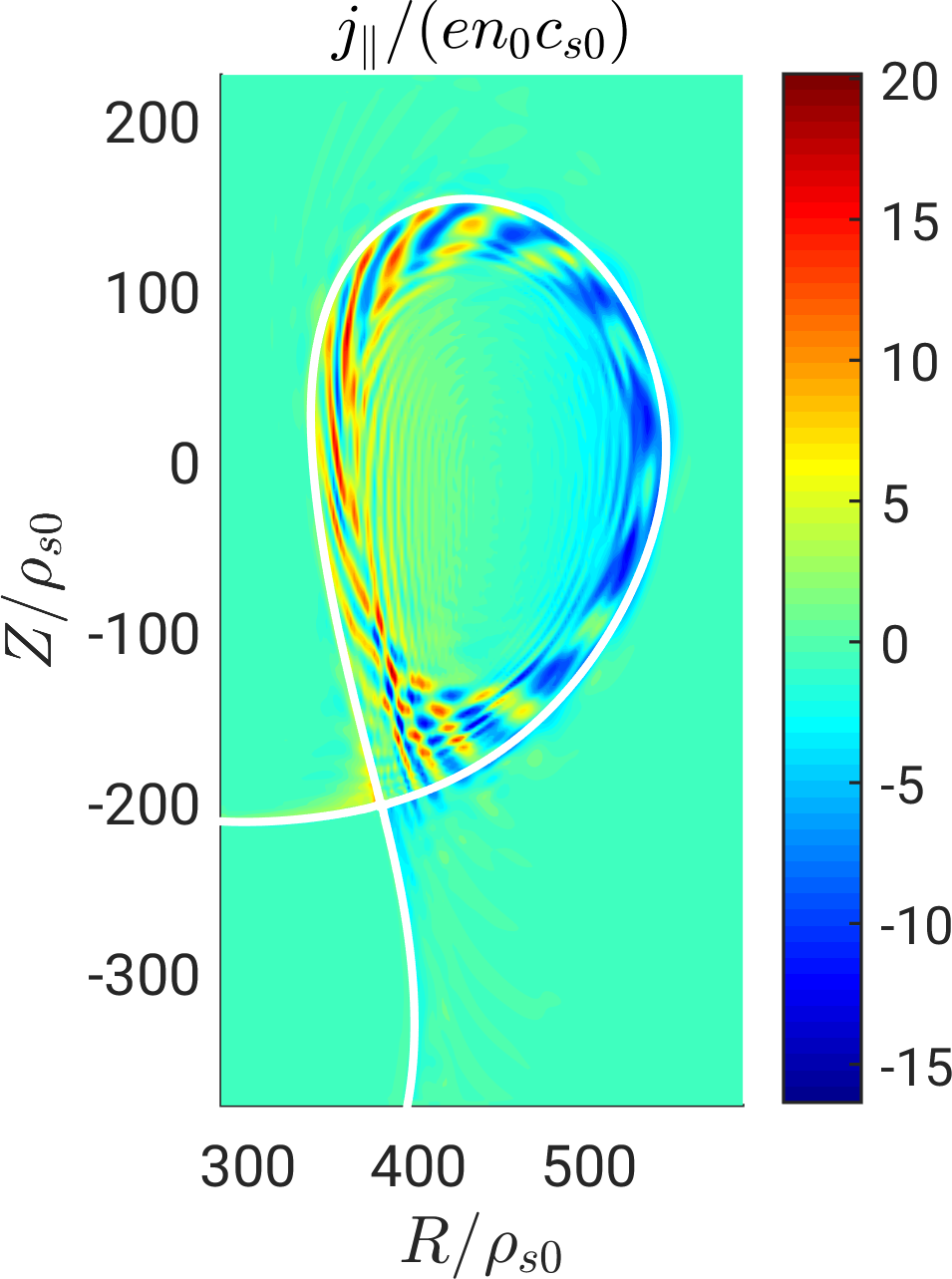}}\,
    \subfloat[]{\includegraphics[height=0.3\textheight]{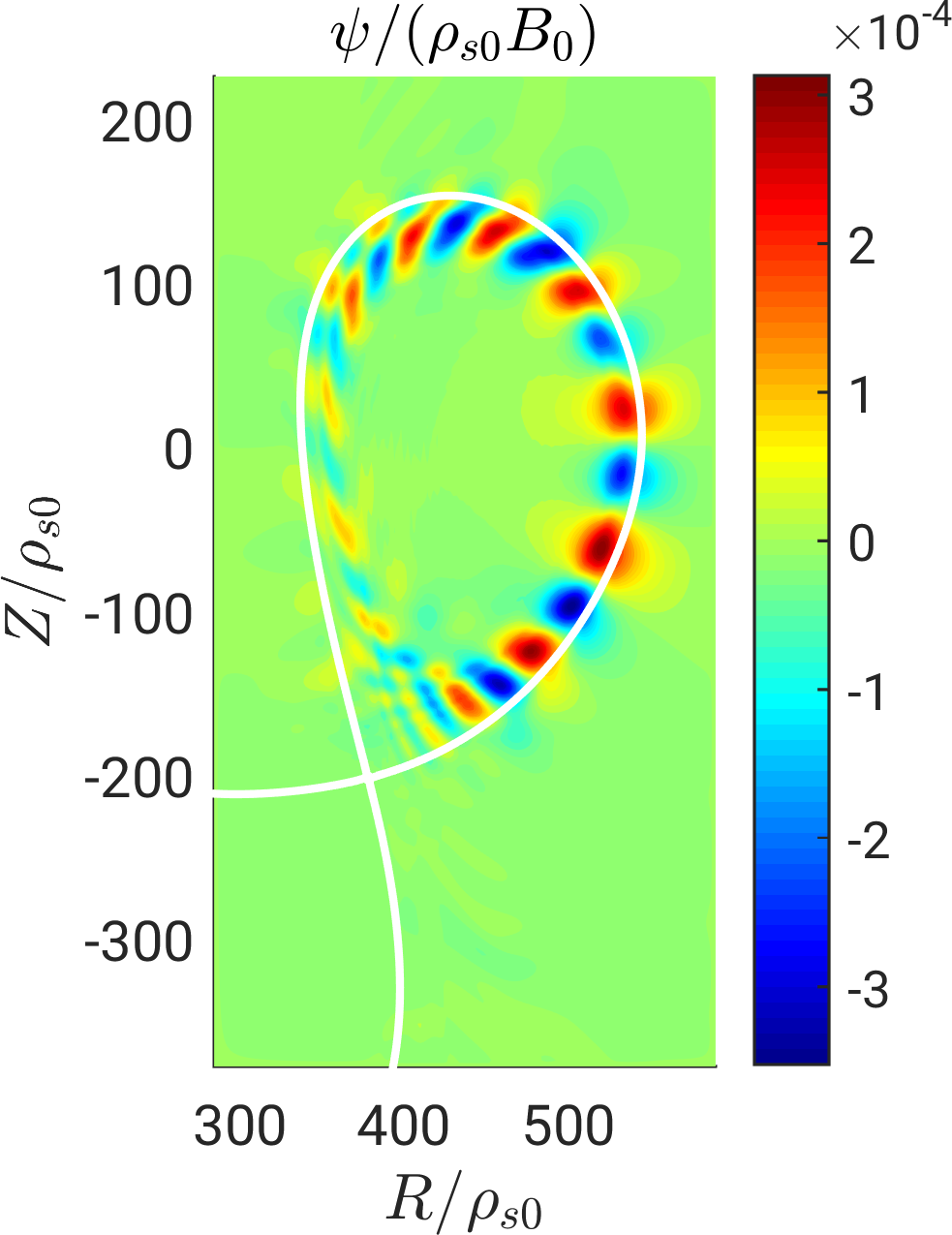}}
    \caption{Typical snapshots of the density (a), electron temperature (b), ion temperature (c), electrostatic potential (d), parallel current (e) and fluctuating vector potential (f) for the TCV simulation described in Sec.~\ref{sec:application}.}
    \label{fig:tcv_snapshot}
\end{figure}

We briefly discuss here the simulation results, showing that they are in agreement with typical experimental observations. 
Typical snapshots on a poloidal plane of $n$, $T_e$, $T_i$, $\phi$, $j_\parallel$ and $\psi$ are presented in Fig.~\ref{fig:tcv_snapshot}. 
As experimentally observed (see, e.g., Refs.~\cite{cima1995,freethy2016}), the tokamak core is characterized by low-amplitude density and temperature fluctuations, which increase from the core to the edge. A wave-like turbulent dynamics, experimentally observed~\cite{furno2008,xu2009}, is clearly visible in the edge (see Fig.~\ref{fig:tcv_snapshot}). When crossing the separatrix, the turbulent eddies experience a strong $\mathbf{E}\times\mathbf{B}$ shear and detach from the main plasma, forming filaments that propagate radially in the SOL, in agreement with experimental observations described, e.g., in Refs.~\cite{furno2008,xu2009}.
In addition, as it is also experimentally observed~\cite{goncalves2005,tanaka2009,walkden2017,tsui2018}, turbulence in the far SOL is characterized by intermittent events due to coherent plasma filaments~\cite{ippolito2011}.

The strong electric field gradient observed across the separatrix also reflects experimental observations~\cite{schirmer2006,mcdermott2009}. 
Because of the ambipolarity of the plasma flow at the sheath, the electrostatic potential in the SOL is positive and proportional to the electron temperature~\cite{stangeby2000}. Therefore, the electrostatic potential increases from the far SOL to the separatrix. On the other hand, the electrostatic potential inside the separatrix is negative with an electric field proportional to the ion pressure gradient. 
As a consequence, a strong electric field gradient and poloidal $\mathbf{E}\times\mathbf{B}$ shear form in the region across the separatrix.
The $\mathbf{E}\times\mathbf{B}$ shear can play an important role in suppressing turbulence at the tokamak edge, as discussed in detail in Ref.~\cite{giacomin2020}, where GBS simulations are used to investigate the turbulent transport regimes in this region.  Here we show that the $\mathbf{E}\times\mathbf{B}$ shear persists also when the neutral dynamics is included.

\begin{figure}
    \centering
    \subfloat[]{\includegraphics[height=0.3\textheight]{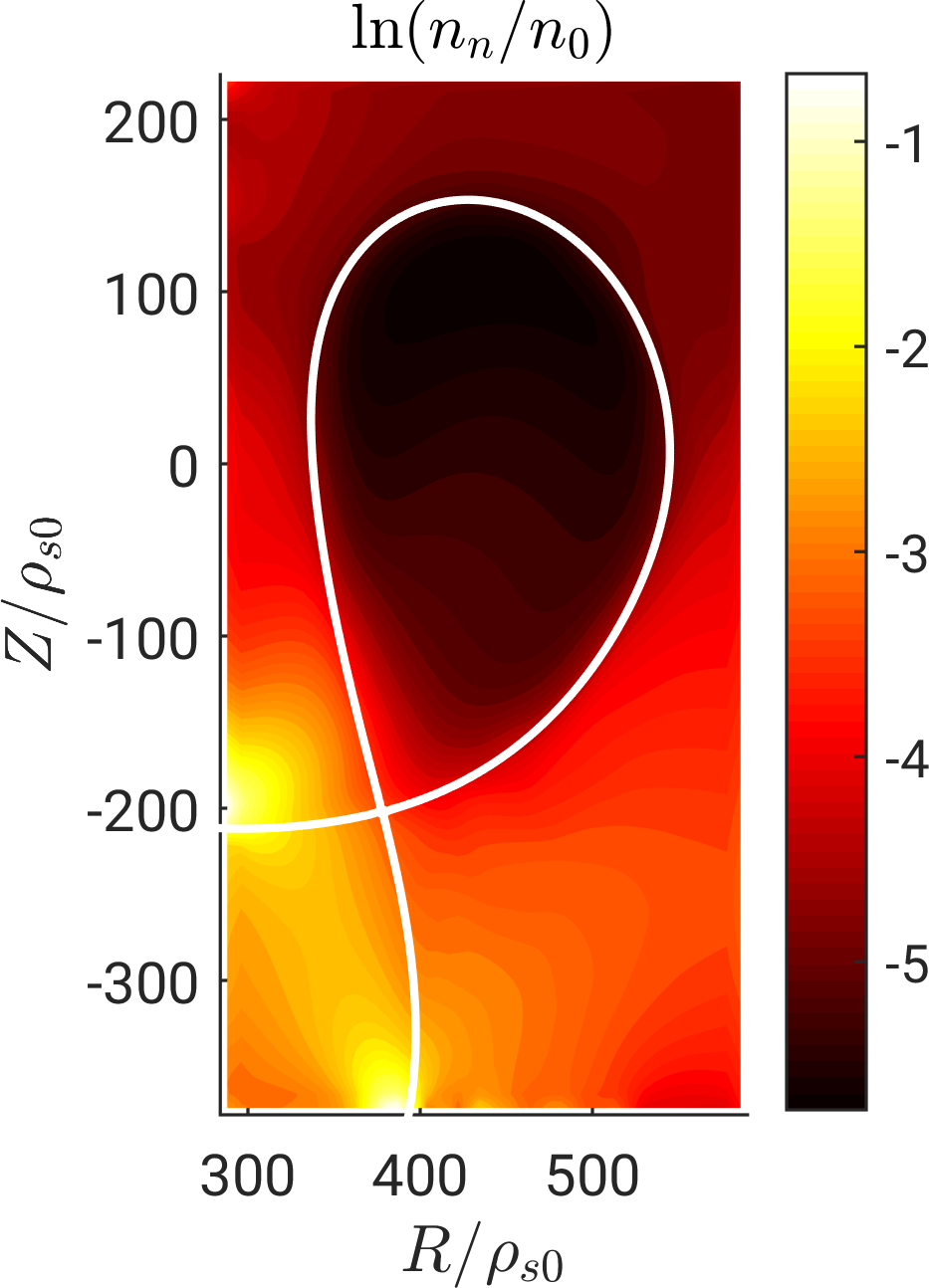}}\,
    \subfloat[]{\includegraphics[height=0.3\textheight]{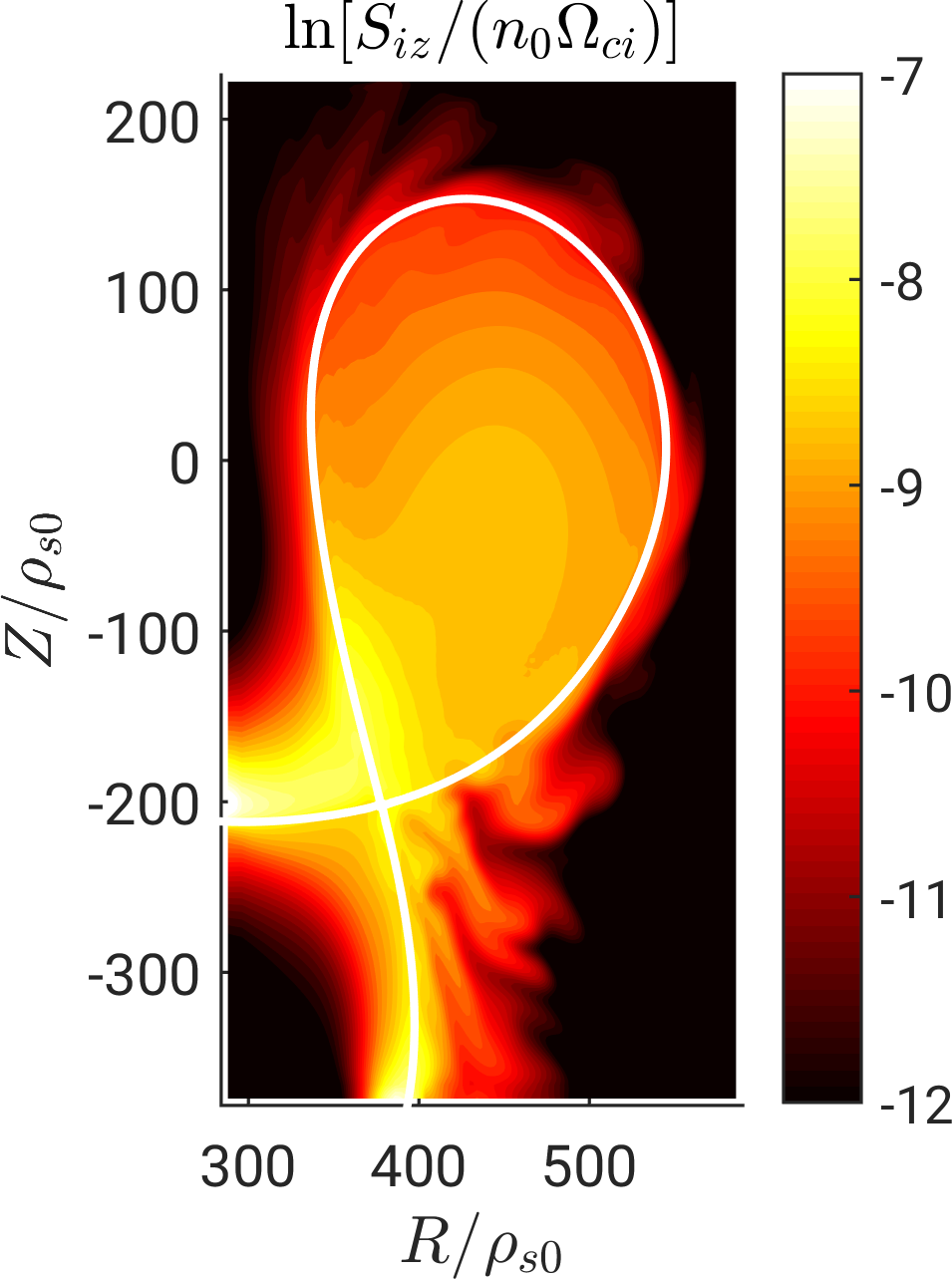}}\,
    \subfloat[]{\includegraphics[height=0.3\textheight]{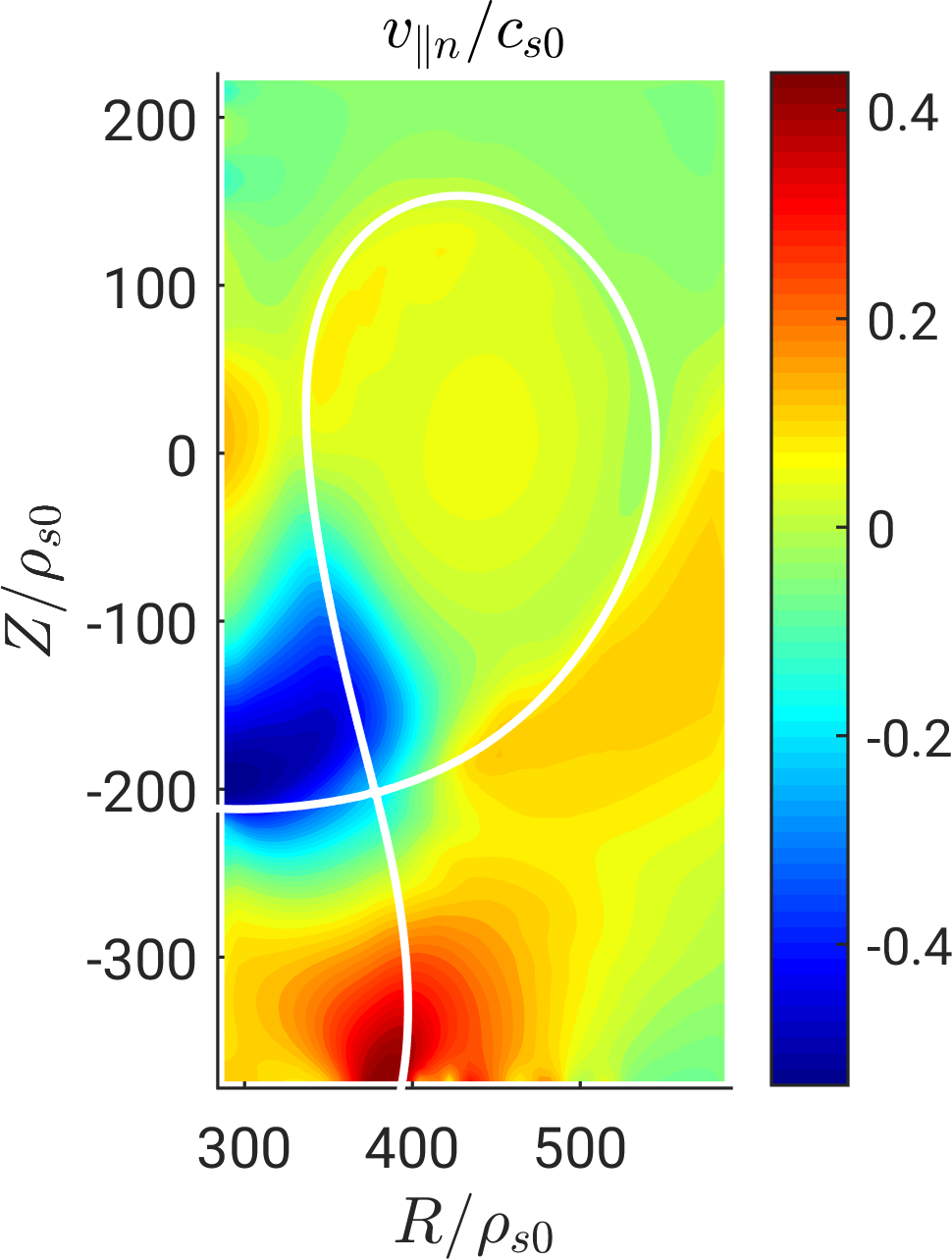}}
    \caption{Typical snapshots of the neutral density (a), ionization source (b) and neutral parallel velocity (c) for the TCV simulation described in Sec.~\ref{sec:application}.}
    \label{fig:tcv_neutrals}
\end{figure}

Typical snapshots of $n_\text{n}$, $S_{iz}$ and $v_{\parallel \text{n}}$ are shown in Fig.~\ref{fig:tcv_neutrals}. 
As expected, the neutral density is larger in the proximity of the target plates, where most of the recycling takes place, although a non-negligible fraction of neutrals can be found in the core, mainly in the region close to the X-point.
In fact, a significant fraction of neutral particles reach the separatrix from the target on the high-field side before being ionized (see Fig.~\ref{fig:tcv_neutrals}~(b)). 
In our simulation, core ionization accounts for approximately 25\% of the total ionization source, which is in agreement with the fact that the TCV discharge considered here is in the low-density and low-recycling regime. 
We note that $v_{\parallel \text{n}}$ is positive in the outer and negative in the inner divertor regions (see Fig.~\ref{fig:tcv_neutrals}~(c)). Since the direction of the magnetic field points towards (away from) the outer (inner) target, the parallel neutral flux is directed towards the wall at both targets.

\section{Summary and conclusions}

The present paper describes a new version of GBS, a three-dimensional, flux-driven, global, two-fluid turbulence code, developed for the self-consistent simulation of plasma turbulence and kinetic neutral dynamics in the tokamak boundary.
In the version of GBS presented here, the simulation domain is extended to encompass the whole plasma volume, avoiding an artificial boundary with the tokamak core and retaining the core-edge-SOL turbulence interplay.
Both the plasma and neutral implementations are carefully optimized leading to a significant speed-up of the code.
In particular, a new iterative solver based on the PETSc library is implemented and optimized for the solution of the Poisson and Ampère equations, allowing us to efficiently carry out electromagnetic simulations at TCV size, while avoiding the use of the Boussinesq approximation.
Also the neutral module is refactored and optimized by implementing an iterative solver based on the PETSc library.

The implementation of the plasma and neutral models is then carefully verified by means of the MMS, including, for the first time, the verification of the electromagnetic terms and of the kinetic neutral model. 
The verification of the neutral model is completed by a set of unit tests to verify the routines used to compute the integrals over the neutral trajectories. 

Our tests show the efficient scalability of GBS on parallel high-performance computers, mainly thanks to the numerical scheme used to discretize the differential operators. The uniform Cartesian grid allows for a massive parallelisation through MPI domain decomposition.
The splitting of the plasma and neutral communicators improves the parallelisation of GBS.

The convergence properties with respect to the plasma and neutral grid refinement are tested in typical turbulence simulations, showing that convergence of the simulation results is achieved with a plasma and neutral grid spacing of approximately 2.5~$\rho_{s0}$ and 0.075~$\lambda_n$, respectively. 
The convergence with respect to the neutral calculation frequency is also studied, showing that evaluating the neutral density every $\Delta t = 0.1 R_0/c_{s0}$ is sufficient to guarantee the convergence of the simulation results. 

Finally, the results of the first GBS electromagnetic simulation of a TCV lower single-null discharge, which include the self-consistent evolution of neutral dynamics, are presented. 
A preliminary analysis of the simulation results show that GBS results agree well with experimental observations. 

In the future, we plan to address three aspects to extend the flexibility and applicability of GBS. First, GBS will be adapted to account for the complex geometry of the tokamak wall, which may include baffles, such as those installed in TCV~\cite{reimerdes2021}, or a complex divertor region. Second, by leveraging the relatively simple numerical scheme used in the version of GBS presented here, GBS will be ported to GPU, allowing for efficient simulations of large scale magnetic fusion devices, such as ITER and DEMO. 
Third, GBS plasma model equations will be extended by adding further moments of the electron and ion distribution functions~\cite{jorge2017}, allowing us to retain kinetic effects that are expected to play an important role in the hot boundary of large tokamaks.

\section*{Acknowledgments}
The simulations presented herein were carried out at the Swiss National Supercomputing Center (CSCS) under the project IDs s882 and s1028. 
This work was supported by the EUROfusion -- Theory and Advanced Simulation Coordination (E-TASC).
This work, supported in part by the Swiss National Science Foundation, was carried out within the framework of the EUROfusion Consortium and has received funding from the Euratom research and training programme 2014 - 2018 and 2019 - 2020 under grant agreement No 633053. The views and opinions expressed herein do not necessarily reflect those of the European Commission.

\bibliographystyle{unsrt}
\bibliography{library}

\end{document}